\documentclass[useAMS,usenatbib]{mn2e}
\usepackage{amsmath}
\usepackage{txfonts}
\usepackage{graphicx}
\usepackage{aas_symbols}
\usepackage{graphicx}
\usepackage{epstopdf}

\newcommand{\beq}{\begin{equation}}
\newcommand{\eeq}{\end{equation}}
\newcommand{\bea}{\begin{eqnarray}}
\newcommand{\eea}{\end{eqnarray}}
\newcommand{\gae}{\gtrsim}
\newcommand{\lae}{\lesssim}
\usepackage[usenames,dvipsnames,svgnames,table]{xcolor}
\usepackage[draft]{hyperref}
\definecolor{darkblue}{rgb}{0.0,0.0,0.3}
\hypersetup{colorlinks,
            linkcolor=darkblue,urlcolor=darkblue,
            anchorcolor=darkblue,citecolor=darkblue}
\DeclareSymbolFont{cmletters}{OML}{cmm}{m}{it}
\DeclareMathSymbol{v}{\mathalpha}{cmletters}{"76}
\begin{document}

\title[Dissipation in AGN Jets]{Three-dimensional Simulations
  of AGN Jets: Magnetic Kink Instability versus Conical Shocks}

\author[R. Barniol Duran, A. Tchekhovskoy \& D. Giannios]{Rodolfo
  Barniol Duran$^{1}$\thanks{Email: rbarniol@purdue.edu (RBD),
    atchekho@berkeley.edu (AT), dgiannio@purdue.edu (DG)}, Alexander
  Tchekhovskoy$^{2,3,4}$\footnotemark[1]\thanks{Einstein Fellow and TAC Fellow}, Dimitrios Giannios$^{1}$\footnotemark[1] \\
$^{1}$Department of Physics and Astronomy, Purdue University, 525 Northwestern Avenue, West Lafayette, IN 47907, USA \\
$^{2}$Departments of Astronomy and Physics, Theoretical Astrophysics Center, University of California Berkeley, Berkeley, CA 94720-3411, USA \\
$^{3}$Lawrence Berkeley National Laboratory, 1 Cyclotron Rd, Berkeley,
CA 94720, USA \\
$^{4}$Center for Interdisciplinary Exploration \& Research in Astrophysics (CIERA) and Department of
Physics and Astronomy, Northwestern University, \\
Evanston, IL 60208, USA}

\date{Accepted; Received; in original form 2016}

\pubyear{2016}

\maketitle

\begin{abstract}
  Relativistic jets in active galactic nuclei (AGN) convert as much as
  half of their energy into radiation. To explore the poorly
  understood processes that are responsible for this conversion, we
  carry out fully 3D magnetohydrodynamic (MHD) simulations of
  relativistic magnetized jets. Unlike the standard approach of
  injecting the jets at large radii, our simulated jets
  self-consistently form at the source and propagate and accelerate
  outward for several orders of magnitude in distance before they interact with
  the ambient medium. 
  We find that this interaction can trigger strong energy dissipation
  of two kinds inside the jets, depending on the properties of the
  ambient medium.  Those jets that form in a new outburst and drill a
  fresh hole through the ambient medium fall victim to a 3D magnetic
  kink instability and dissipate their energy primarily through magnetic
  reconnection in the current sheets formed by the instability. On the
  other hand, those jets that form during repeated cycles of AGN
  activity and escape through a pre-existing hole in the ambient medium
  maintain their stability and dissipate their energy primarily at MHD
  recollimation shocks. In both cases the dissipation region can be 
  associated with a change in the density profile of the ambient gas. 
  The Bondi radius in AGN jets serves as such a location.
\end{abstract}


\section{Introduction}
\label{sec:introduction}

Black holes (BHs) reprocess infalling gas into outflows and
radiation. Of particular importance are relativistic collimated
outflows, which we will refer to simply as jets. Relativistic
motions have been inferred in various astrophysical systems including
long- and short-duration gamma-ray bursts \citep[GRBs,
e.g.,][]{frail01,pk02, tayloretal2004}, active galactic nuclei
(AGN, e.g.,
\citealt{birettaetal99,jorstad_agn_jet_2005,2011ApJ...742...27L,meyeretal2013}),
and black hole X-ray binaries \citep[e.g.,][]{fend04a}. The
relativistic motion of jets implies that they
start out highly magnetized, with very few baryons (otherwise they would have insufficient power
to accelerate to the observed Lorentz factors), and are likely powered by the rotation of the
central supermassive BHs \citep[e.g.,][]{sashaetal2011}. Observationally, jet power is comparable to
the accretion power
\citep{1991Natur.349..138R,2014Natur.515..376G,2015MNRAS.449..316N},
suggesting that the jets are an energetically important component of
the system. Indeed, jets can affect the evolution of BHs and their
host galaxies, e.g., via extracting the black hole spin energy
\citep[e.g.,][]{2004ApJ...602..312G,sasha15}, heating the ambient gas and suppressing its
infall, with important implications for the cooling flow problem in
the context of galaxy clusters and
star formation \citep[e.g.,][]{2006ApJ...645...83V,2012MNRAS.427.2998I,2012ApJ...746...94G,2015ApJ...811...73L,2016arXiv160606734F}.

Even though jets are widely thought to be powered at least in part by
magnetic fields, most studies of the dissipation zone (i)~assume
axisymmetry or ignore magnetic fields
\citep[e.g.,][]{2000ApJ...531L.119A,2007ApJ...665..569M,2007ApJ...671..171V,2008A&A...488..795R,2008A&A...491..321M,2010A&A...519A..41P,2011ApJ...742...25Y,kohleretal2012,2012ApJ...751...57D, lopezcamaraetal2013,mizunoetal2015},
(ii)~prescribe jet injection at large distances
\citep[e.g.,][]{nakamuraetal2007,2010MNRAS.402....7M, 
  mignoneetal2013,2014ApJ...781...48G}, or (iii) consider a segment of
a pre-existing jet
\citep{mizunoetal2012,2014ApJ...784..167M,porthandkomi2015,singhetal2016}. Assumption
(i) ignores known 3D magnetic instabilities, the most serious
of which is the magnetic kink instability \citep[e.g.][]{begelman1998, lyubarskii1999, narayanetal2009},
which can result in the dissipation trigger for the jet radiation
\citep[e.g.,][]{begelman1998, nakamuraandmeier2004,
  gianniosandspruit2006} or can even globally disrupt the jets
\citep{sashaandomer16}.  Assumption (ii) breaks the
self-consistent connection of the jets to the central engine and
prevents them from establishing a natural value of the \emph{magnetic
  pitch}, or the ratio of poloidal and toroidal magnetic field
strength, that is crucial for jet stability as we discuss below \citep[see also][]{omerandsasha16}. Assumption (iii) considers a segment of
the jets and limits the ability to address
global jet stability properties.

Indeed, jets propagate through a vast range of distances from the
event horizon of the BH ($\sim 10^{-7} [M_{\rm BH}/10^9M_\odot]$~kpc, where $M_{\rm BH}$ 
is the black hole mass)
to outside the galaxy ($\sim10^3$~kpc), giving wide latitude --- ten orders of
magnitude in distance  --- for magnetic instabilities to develop.
Along the way, the jets radiate as much as 10--50\% of their luminosity as photons at
characteristic scales, where major
dissipation events take place (the ``dissipation zone''), such as the blazar zone and the knots
in AGN and the prompt emission in GRBs
\citep[e.g.,][]{pk02,2012Sci...338.1445N,2014Natur.515..376G}. The cause of energy dissipation is actively debated. It
can be triggered by internal processes (e.g., internal shocks, MHD
instabilities) and external interactions (e.g., recollimation shocks,
external shocks), or the interplay between the two.  
Since, within the MHD paradigm, jets are expected to reach the 
dissipation zone magnetically dominated, dissipation of magnetic 
energy is the most likely source of jet radiation 
\citep[e.g.,][]{spruitetal2001, 2003astro.ph.12347L, 2011MNRAS.416.2193N,
sironietal2015}. 

In this paper, we study the effect of the external medium on the jet
dissipation zone using relativistic magnetohydrodynamic (MHD)
simulations.  In order to capture the development of non-axisymmetric
magnetic instabilities, we include magnetic fields and carry out the
simulations in full 3D. We also consider the launching of the jet
essentially from the surface of the central compact object.  With our
approach, the \emph{magnetic pitch} at the base of the jet is at its
natural value self-consistently set by the rotation of the central
object, instead of being chosen {\it a priori}. Because the magnetic
pitch controls jet stability to the 3D magnetic kink
\citep{2000A&A...355..818A,narayanetal2009,2014ApJ...781...48G}, our
approach allows us to study the stability of relativistic jets from
first principles.  

Kink instability in a magnetized jet is similar to that in a
  spring: the spring flies sideways as we squeeze it. Whether the jet
  is susceptible to the kink instability depends on whether the
  instability has enough time to grow. If the instability growth
  time-scale is shorter than the time it takes for the fluid to travel
  from the base to the tip of the jet, the jet will become
  unstable. If the jet is strongly compressed and therefore
  toroidally-dominated, the instability grows faster. And because jet
  compression is determined by the interaction with the ambient medium,
the properties of the ambient medium that the jet propagates through can
have a profound effect on jet stability 
\citep{omerandsasha16}. In a new outburst, a jet drills a hole through
the ``pristine'' ambient medium and develops a highly pressurized
region at the tip of the jet -- the \emph{jet head} -- that does the
drilling and pushes the ambient medium out of the way. We refer to
such jets as \emph{headed} jets. The push against the external medium
compresses the headed jets into a tighter magnetic helix. This
increases their toroidal magnetic pressure and destabilizes them
against the 3D magnetic kink instability \citep{omerandsasha16}. A
more lucky jet, for which a hole was pre-drilled along the jet propagation 
path by previous AGN activity and has not had the time to close,  
does not develop a head. We refer to such jets as
\emph{headless} jets; they tend to be more stable. For instance, 3D
simulations of such jets did not find significant dissipation due to
3D instabilities \citep{2006ApJ...641..103H,mckinneyandblandford2009,sashaetal2011}.

The distinction between the two types of jets is not always
clear-cut. For instance, short-GRB jets, believed to be produced by
compact binary mergers, could initially be headless but eventually run
into the material ejected early in the merger process (e.g.,
\citealp{murguiaetal2014}) and develop instabilities.  Conversely, core-collapse
GRBs initially have to drill their escape route through the progenitor
star, but once they emerge from it, they become headless. A clear
picture of what determines the jet stability in general is still
lacking.  In this paper, we attempt to elucidate on this topic by
clearly separating headless and headed jets and their interaction with
the external medium.

The density profile of the ambient gas through which the jet 
propagates depends on the specifics of the astrophysical system. 
Here we use observations of jets in nearby AGN to guide us in including the 
effects of the ambient density into our models. 
Recent 
X-ray observations with \emph{Chandra} inferred a shallow profile $\propto r^{-1}$, potentially
accompanied by a jump, in the external medium profile of the M87 galaxy, near the
edge of the central black hole's sphere of influence, or the Bondi
radius, $r_{\rm B}\sim0.1$~kpc \citep{russelletal2015}.
\cite{sashaandomer16} considered the interaction of uncollimated
mildly relativistic jets with the external medium and hypothesized
that this interaction starts around~$r_{\rm B}$.  To make their
simulations computationally feasible, they initiated the jets at
$r_{\rm B}$ and followed their propagation to galaxy scales,
$\sim10^3r_{\rm B}\sim 100$~kpc. However, by the time real jets reach
$r_{\rm B}$, they are already well-collimated and move at relativistic
velocities. For instance, the M87 jet collimates into a parabola
(e.g., \citealp{nakamuraandasada2013} and references therein) and
accelerates to Lorentz factors $\sim6$ (e.g., \citealp{birettaetal99,
  meyeretal2013}) over $\sim 6$ orders of magnitude distance before it
reaches $r_{\rm B}$. In this paper, we -- for the first time -- study the
effects of acceleration and collimation of jets on their interaction with
breaks, jumps, and other features in the ambient medium that are
possibly present at~$r_{\rm B}$.

As the first attempt to attack this 3D multi-scale problem of jet
acceleration, collimation, and interaction with the ambient medium, we
make several simplifications. We reduce the length of the
acceleration zone down to $2$ orders of magnitude and consider 
jets of high-power (which propagate the fastest); we will also ignore
gravity and assume a monopolar magnetic field geometry at the central
compact object (see Sec.~\ref{simulation_setup}).
In Section \ref{simulation_setup} we
describe our numerical method and simulation setup. In
Section~\ref{sec:modell-jet-inter} we present our simulation results
for headed and headless jets.  In Section~\ref{sec:discussion} we
discuss the astrophysical implications of our results and in
Section~\ref{sec:conclusions} we conclude.

\section{Simulation setup} \label{simulation_setup}

We carry out time-dependent 3D relativistic MHD numerical
simulations using the {\sc harm} code \citep{gammieetal2003,
  nobleetal2006, sashaetal2007, mckinneyandblandford2009,
  sashaetal2011}. 
  We adopt a simulation setup inspired by 
  \cite{sashaandomer16, omerandsasha16}.
 We use modified spherical polar coordinates $(r, \theta, \varphi)$ and
a numerical grid that spans the range $r_{\rm in} \le r \le r_{\rm out}$,
$0\le \theta \le \pi$ and $0 \le \varphi \le 2\pi$. 
All our simulations use $r_{\rm in}=r_0$, and $r_{\rm out} = 10^5 r_{\rm in}$.
We initiate our jets at $r = r_0$, which is a few times the radius of the central
compact object. So long as the jet is sub-Alfvenic at $r_0$, the value of $r_0$ does 
not affect the simulation outcome. Using
a large value of $r_{\rm out}\gg r_{\rm in}$ ensures that the jets do not reach
the outer boundary in a simulation time. 

To isolate the intrinsically 3D
effects, we will carry out both 2D and 3D simulations.  In our 2D simulations the jets are pinned to the
rotational axis by the axisymmetry, but in 3D they are free to deviate
from the axis. Because of this, for our 2D simulations, we orient the
rotational axis along the polar axis, $\theta = 0$. However, for our 3D
simulations, we direct the rotational axis along the $x$-axis, $\theta=\pi/2$, $\varphi = 0$ (e.g.,
\citealp{molletal08}, see details in \citealp{omerandsasha16}) to
allow complete freedom of 3D jet motion and avoid jet interaction with the polar singularity. 
For simplicity of presentation, when showing the results of
both 2D and 3D simulations,  regardless of the actual direction of the
rotational axis, we will refer to it as
the $z$-axis and orient it vertically in our figures.

Initially, we fill the domain with cold gas that we refer to as
the ``external density'', ``ambient density'' or simply
``density''. We describe its spatial distribution below. The gas
  in our simulations is 
initially cold, apart from a very small amount of thermal energy supplied 
by the density ``floors'', which have no effect on our
results. Namely, if the values of density and/or internal energy drop
below their floor values, $\rho_{\rm floor} = b^2/50$ and
$u_{g, \rm floor}=b^2/250$ respectively, where $b$ is the total magnetic 
field strength in the fluid frame, we reset them to their floor
values. The inner
boundary is a perfectly conducting magnetized sphere. We will refer to
it as the central object. We neglect gravity.

We thread the central object and the
computational domain with a
monopole (radial) magnetic field; using a dipole magnetic field
produces similar jet properties
\citep{omerandsasha16}. We fix the radial (normal) magnetic field strength of the
central object and allow the other two magnetic field components to
relax. The
simulations start by instantaneously spinning up the sphere
to a constant angular frequency $\Omega = 0.8c/r_{\rm in}$. The
electric field is zero in the frame instantaneously comoving with the sphere. The
rotation coils up the initially radial magnetic field lines into
helices and generates twin magnetized outflows from the surface
of the central object that propagate mainly along the rotational
axis.  The rotation is uniform within $50^{\circ}$ of
the rotational axis and smoothly goes to zero at $70^{\circ}$: this
ensures that the twin outflows do not interact with the coordinate
singularity. In other words,  
this ensures that in 3D there is no rotation of the central object
at $\theta = 0$, so there is no need to treat the boundary conditions at $\theta=0$
in any special way. The initial magnetization at the base of our jets is $\sigma_0 =
2p_{\rm mag,0}/\rho_0 c^2 \approx 25$, where $p_{\rm mag,0}=b_0^2/2$ is the magnetic
pressure and $\rho_0$ is the density at the base of the jet. 
The initial internal energy of the jets is
  $u_0 = p_{\rm mag,0}/62.5$. Thus, the initial plasma beta at the
  base of the jets is
  $\beta_0 = (\gamma-1)u_0/p_{\rm mag,0} \approx 0.005 \ll 1$, since
  the jet is cold, apart from a small amount of thermal energy
  supplied through the boundary conditions and the floors in our
  simulations solely for numerical stability purposes. This small
  amount of thermal energy has no effect on our results.  Once the
  jets are launched by the rotation of the central compact object,
  they send a shock wave into the ambient medium. This blast wave
  heats up the ambient medium, and the thermal pressure of the medium
  confines and collimates the jets.

Our radial grid is uniformly spaced in $\log r$ out to a radius
$10^4 r_{\rm in}$, where the grid becomes progressively sparse. The
angular grid is modified to concentrate the grid cells toward the
rotational axis by following a parabolic shape
$\delta\theta \propto r^{-\nu/2}$ (where $\delta\theta$ is measured
away from the rotational axis), where we usually adopt $\nu = 0.85$. The
jets form and initially propagate along the rotational axis of the
central object. In all figures, the central object is centred at the
coordinate origin, $r=0$, and we will usually show only one jet
that propagates from the center along the positive
$z$-axis. 

\begin{table}
\begin{center}
\begin{tabular}{lccccccc|c}
\hline
Model & Density & $\alpha$ & $\beta$ & $r_t/r_0$ & $\chi$ & $\delta$ & $z_0/r_0$ & $P_{\rm jet}$\\
name  & (equation)  & & & & & & & (${\rm erg\; s^{-1}}$)\\
\hline
A   & (\ref{rho_A}) & 3 &  &  & & & &\\
A2 & (\ref{rho_Ax}) & 3 & 1 & 100  & 1    & & & $10^{47}$\\
A2x3 & (\ref{rho_Ax}) & 3 & 1 & 100   & 3   & & & $3 \times 10^{46}$\\
A1 & (\ref{rho_Ax}) & 3 & 1 & 10   & 1  & & & $10^{46}$\\
\hline
B & (\ref{rho_B}) & 3 &  &     &  &    6 &  &\\
B2   & (\ref{rho_Bx}) & 3 & 1 &  100  &  3 &    6 & 80 &\\
\hline
\end{tabular}
\end{center}
\caption{Initial ambient density profiles chosen for each of the simulations, according to equations 
(\ref{rho_A})--(\ref{rho_Bx}), see Fig. \ref{Model_schematics}.  
For simulations A2, A2x3 and A1 the density profile has a break at $r_t$, which 
we associate with the Bondi radius, $r_{\rm B}$. Using the M87 jet 
as a baseline, which has $P_{\rm jet} \approx 10^{44}$ erg/s, $r_{\rm B} \sim 0.1$ kpc
and number density $n(r_{\rm B}) \sim 1$ cm$^{-3}$ (e.g., \citealp{russelletal2015}), we 
calculate a dimensionless parameter 
$\xi \equiv P_{\rm jet}/(n_{\rm B} m_p c^3 r_{\rm B}^2)$, 
where $m_p$ is the proton mass and $n_{\rm B} \equiv n(r_{\rm B})$ is 
the ambient number density at the Bondi radius. We compute the physical
power of our simulated jets by
scaling the simulated density at $r_t$ to the characteristic value, $n_{\rm B}
= 1\ {\rm cm}^{-3}$, while keeping the value of $\xi$ constant. All jet powers 
presented in this paper correspond to one jet.}
\label{table0}
\end{table}

\section{Modelling jet interaction with the external medium}
\label{sec:modell-jet-inter}

\subsection{Headed Jets}
\label{sec:headed-jet-interacts}

\subsubsection{Modeling the external medium}
\label{sec:model-ambi-medi}

\begin{figure*}
\includegraphics[width=6.2cm, angle=0]{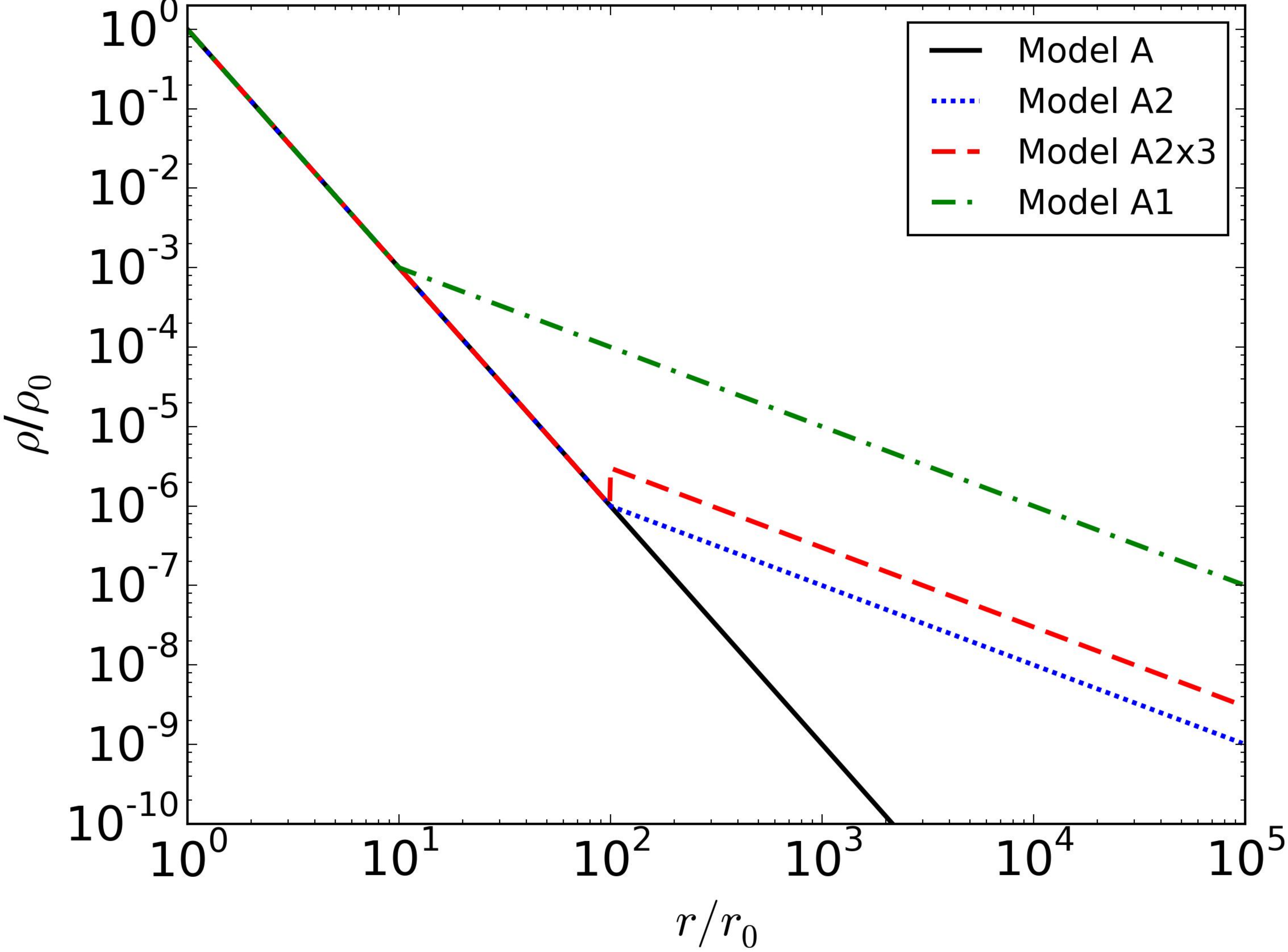} 
\includegraphics[width=5.5cm, angle=0]{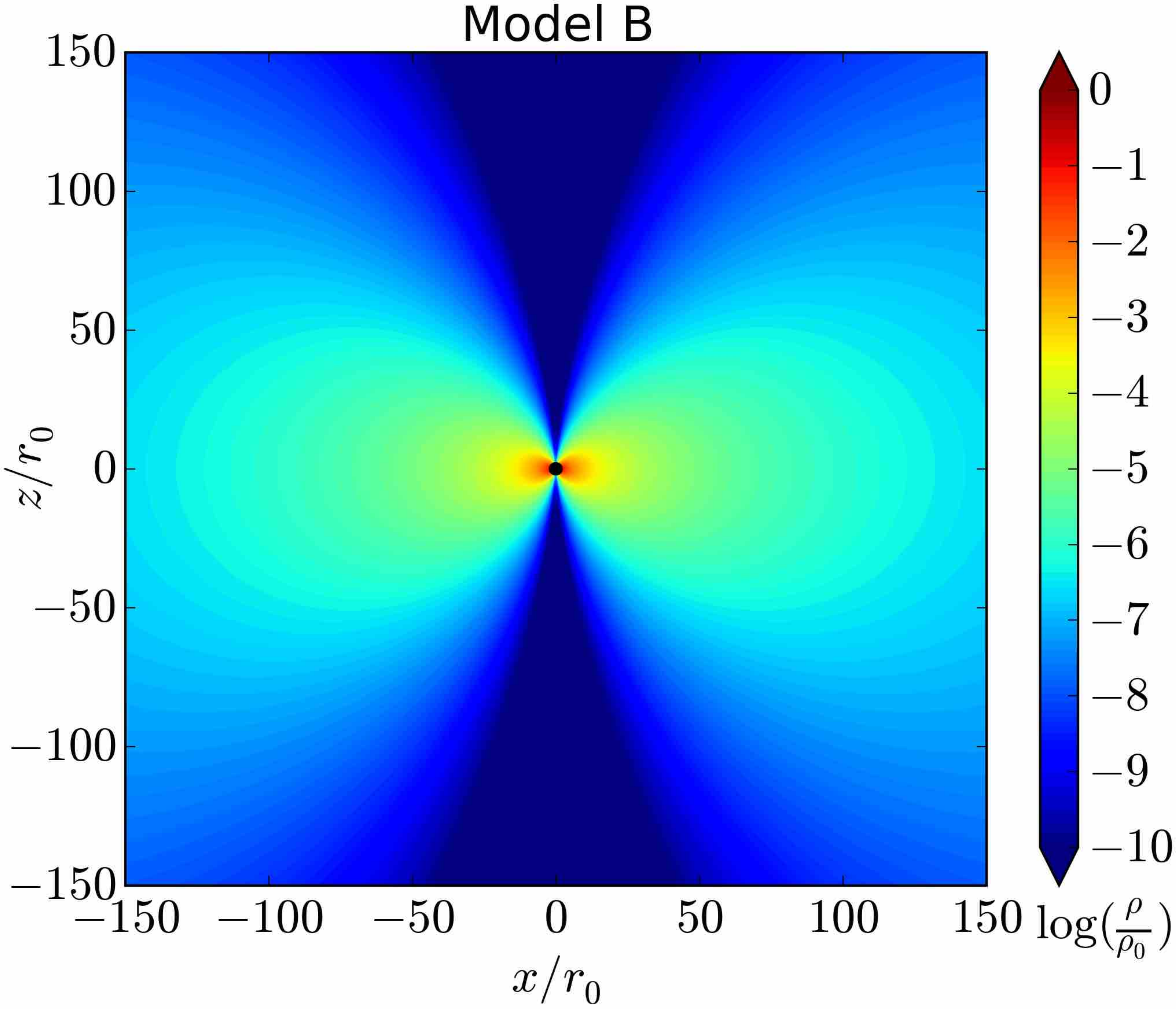} 
\includegraphics[width=5.5cm, angle=0]{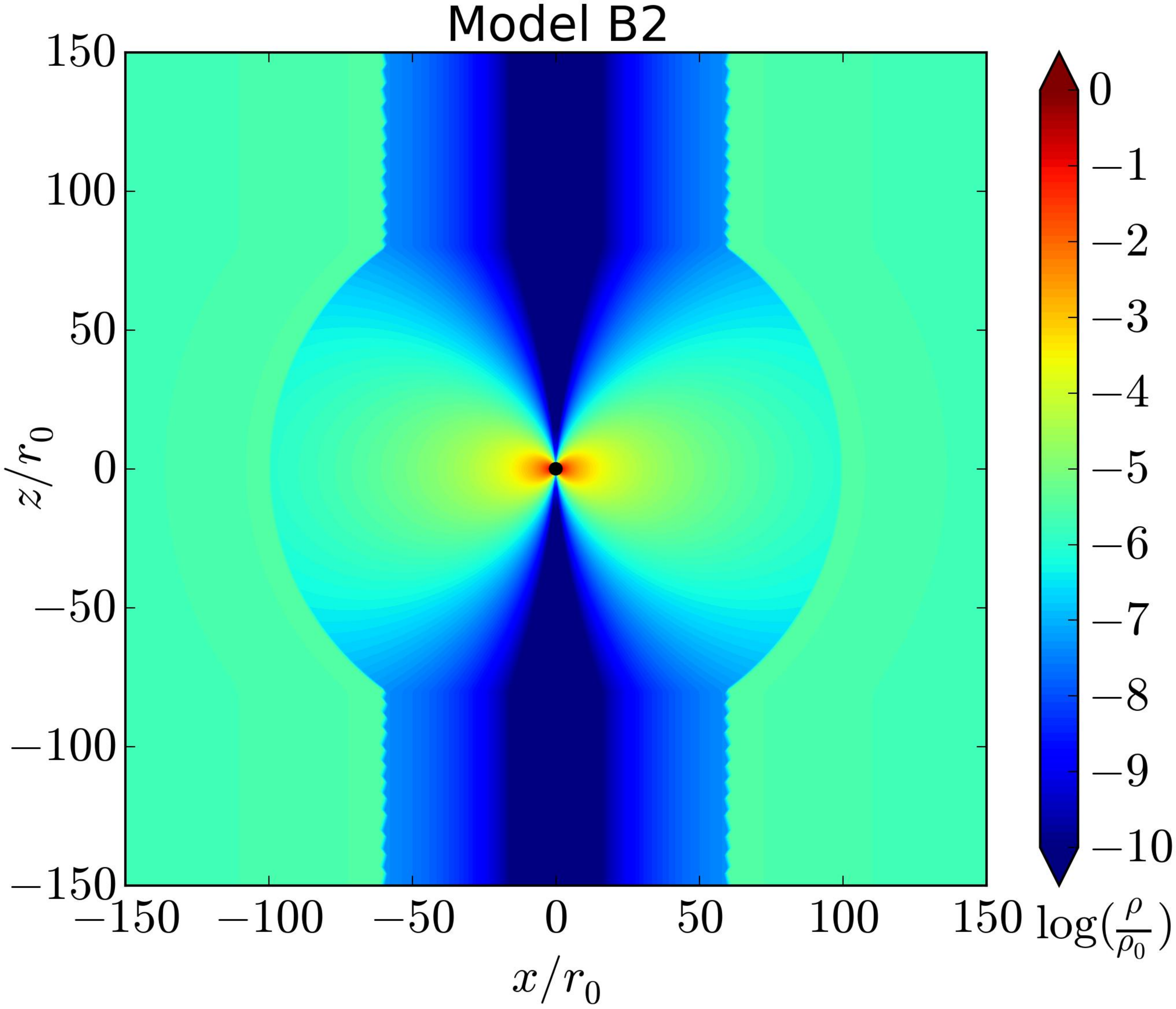} 
\caption{Models for the initial external medium we consider in this
  work. The left panel shows the density profiles in models A,
  which are spherically symmetric (see equations
  \ref{rho_A} and \ref{rho_Ax}).  The middle and right panels show
  density colour maps of $\log_{10}(\rho/\rho_0)$ for models B and B2,
  respectively (see equations~\ref{rho_B} and~\ref{rho_Bx}). The initial and boundary conditions of models B and B2 are
  axisymmetric (around the $z$-axis).}   
\label{Model_schematics} 
\end{figure*}

Relativistic jets are launched very close to the BH and are likely
powered by the magnetized rotation of the BH. Jets accelerate as
they collimate off the accretion disc wind 
(e.g., \citealp{2006MNRAS.367..375B, komietal2007, sashaetal2008}).  In this
work, we focus on the asymptotic behaviour of the jets, and we would like
to include the collimating effect of the accretion disc wind.  In order to
keep the simulation cost manageable, we opt to avoid the need to
resolve the turbulent motions characteristic of the accretion
disc. Because of this, we adopt a simplified approach that allows us
to focus on the physics of the jets. To
include the collimating effect of the accretion disc outflow, we
immerse the central compact object into a spherically-symmetric
ambient medium, with a power-law density profile\footnote{
In this model and in all models throughout the paper we 
fix the initial density and the initial magnetic field 
strength at $r_0$ to be $\rho_0=4500$ and $B_0 = (4\pi)^{1/2}$ in 
arbitrary units, respectively.},
\begin{equation} \label{rho_A}
\rho_{\rm A} = \rho_0 \left(\displaystyle\frac{r}{r_0}\right)^{-\alpha}
 \qquad \text{(Model A).}
\end{equation}
For our first model, which we refer to as model A, we choose the
density distribution given by eq.~\eqref{rho_A} with a rather steep density
profile, $\alpha = 3$ (see Table \ref{table0} and the left panel of
Fig.~\ref{Model_schematics}).  The rapid drop off of density with
distance roughly mimics the effect of the accretion disc wind.  
As we will see below, the ambient gas collimates the
jet into a parabolic shape that resembles that inferred for the
M87 jet \citep{asadaandnakamura2012} and 3D accretion disc--jet
simulations \citep[e.g.,][]{2006ApJ...641..103H,sashaetal2011}. The steep 
density profile also allows the jet to accelerate to high velocities without having 
to cover a huge dynamical range, which is computationally expensive. This motivates our choice
of $\alpha$. We carry out simulations of model A both in 2D and 3D, at different
resolutions, as described in Tables~\ref{table1} and
\ref{table3}. These tables also describe our simulation naming
convention. 

\begin{table}
\begin{center}
\begin{tabular}{lc}
\hline
Naming & Resolution \\
convention  & ($N_r \times N_{\theta} \times N_{\phi}$) \\
\hline
X-2D   & $256 \times 96 \times 1$\\
X-2D-hr   & $512 \times 192 \times 1$\\
X-2D-vhr   & $1024 \times 384 \times 1$\\
\hline
X-3D   & $256 \times 96 \times 192$\\
X-3D-hr   & $512 \times 192 \times 384$\\
\hline
\end{tabular}
\end{center}
\caption{The naming convention we use for our simulations: ``X''
  stands in for the model names followed by a ``2D'' or 
  ``3D'' suffix denoting the dimensionality of the simulation and by an
  optional resolution suffix: ``hr" for high resolution and ``vhr" for very high resolution.}
\label{table1}
\end{table}

\begin{table}
\begin{center}
\begin{tabular}{ccc}
\hline
Model & Simulation & $t_f$\\
name  & identifying suffix& $(r_0/c)$\\
\hline
A   & -2D, -2D-hr, -2D-vhr, -3D & 2700\\
A2   & -2D, -2D-hr, -2D-vhr, -3D & 3000\\
A2   & -3D-hr & 1500\\
A2x3   & -2D, -2D-hr, -2D-vhr, -3D & 3400\\
A1   & -2D, -2D-hr, -2D-vhr, -3D & 2800\\
B  & -2D, -2D-hr, -2D-vhr, -3D & 1500\\
B  & -3D-hr & 1200\\
B2   & -2D, -2D-hr, -2D-vhr, -3D & 2200\\
B2  & -3D-hr & 700\\
\hline
\end{tabular}
\end{center}
\caption{List of the simulations we have carried out. Their names
  consist of a model name (left column) followed by a suffix (middle
  column). The simulations have been run until a time $t_f$ in units
  of $r_0/c$ (right column).
We carried out a total of 27 different simulations.
We give the parameters of each model in Table~\ref{table0}
and the resolution of our simulations in Table~\ref{table1}. 
}
\label{table3}
\end{table}

\begin{figure*}
\includegraphics[width=16.60cm, angle=0]{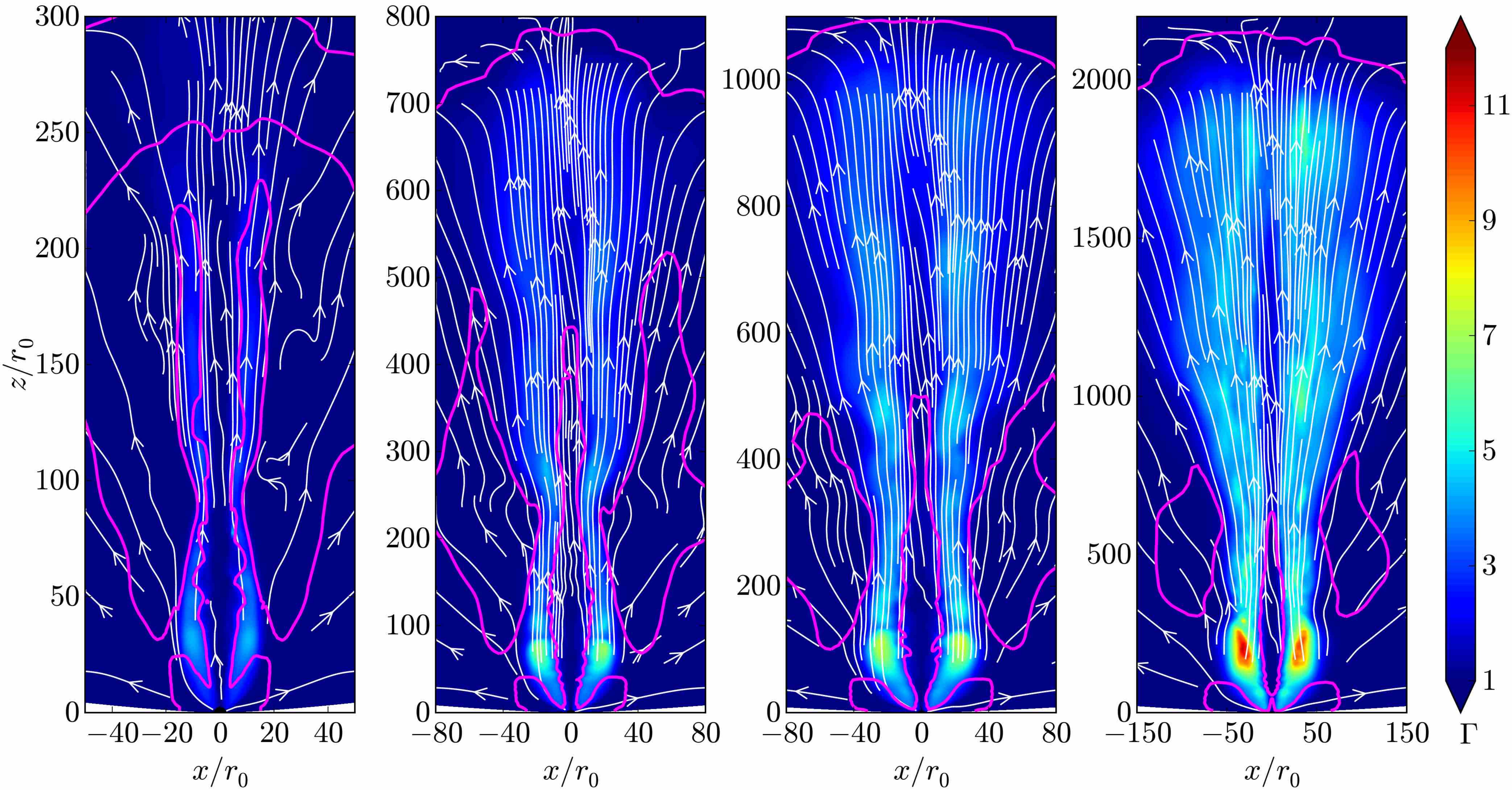}
\caption{Cuts through the rotational axis of the Lorentz factor in
  our 3D simulation A-3D, at different times (from left to right,
  $t \approx 800 r_0/c$, $t \approx 1300 r_0/c$,
  $t \approx 1700 r_0/c$ and $t \approx 2700 r_0/c$; note different
  scales on the plots), with red (blue) colour showing high (low)
  values (please see the colour bar).  The directed white lines show the
  velocity streamlines traced out in the image plane.
  In model A, the jet propagates into a
  density profile $\propto r^{-3}$.  Throughout this work, we only show one
  jet propagating along the $z$-axis. The central compact object is
  located at $(0,0)$. The jet accelerates up to $\Gamma \sim 10$, as seen
  in the right panel. As the jet drills through the ambient medium it recollimates, 
  which causes a drop in the Lorentz factor (seen at $z \sim 300 r_0$ 
  in the right panel). A weak recollimation point, initially at $z \sim 80 r_0$ in the left panel, 
  propagates outward at $\sim 0.3 c$. This jet mostly retains its
  axisymmetry, indicating global stability against 3D magnetic kink modes. 
  The magenta lines show the position of the fast
  magnetosonic surface. The jet (brightly coloured regions in these
  plots) accelerates to super-fast magnetosonic velocities by $z \sim
  20 r_0$. Therefore, information 
  about any potential changes beyond this distance in the external medium cannot propagate back to the compact
  object, just as expected in AGN jets in nature.}
\label{Sequence-A} 
\end{figure*}

\begin{figure*}
\includegraphics[width=16.60cm, angle=0]{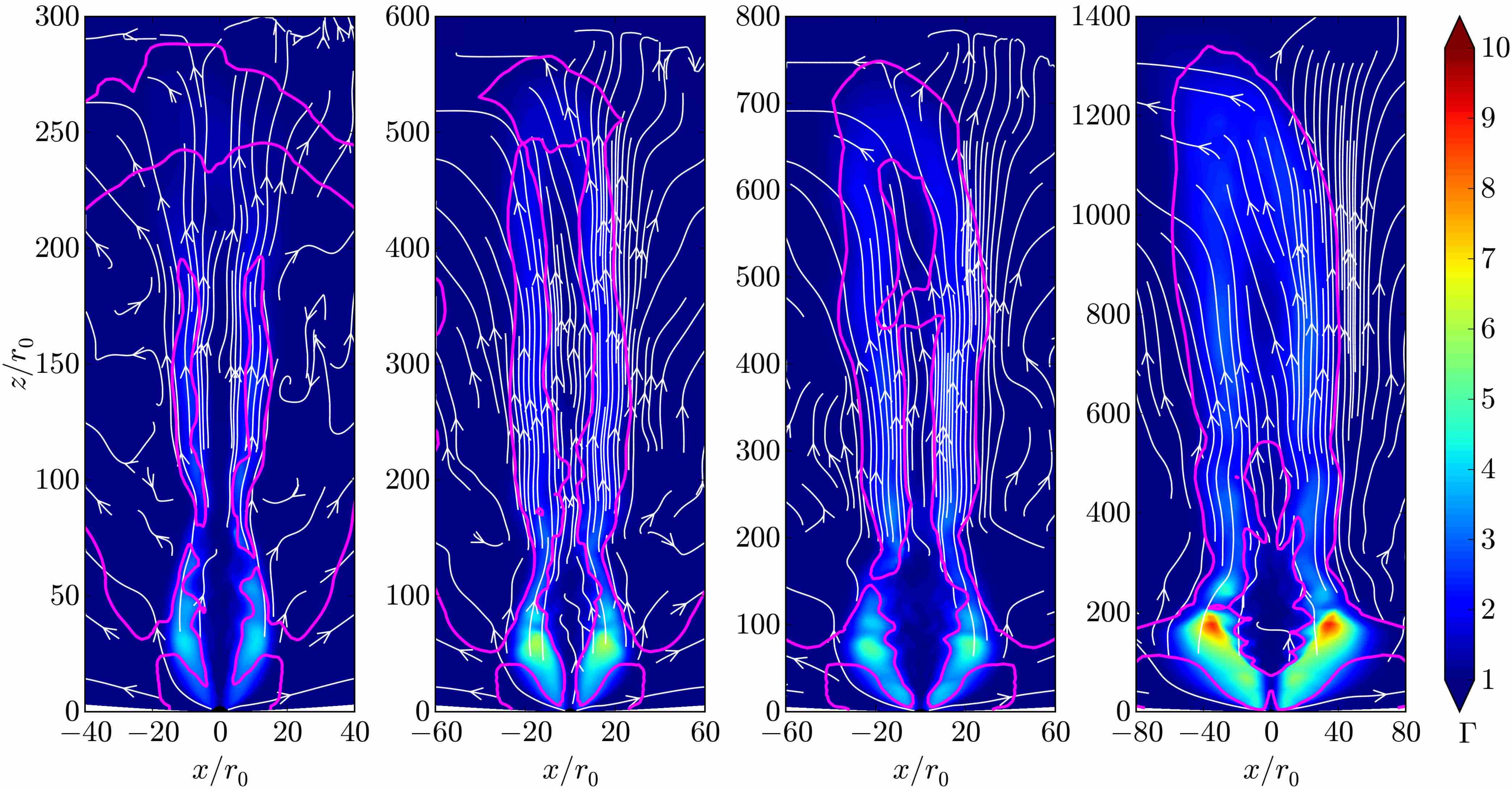}
\caption{Cuts through the rotational axis of the Lorentz factor colour maps in
  our 3D simulation A2-3D, at different times (from left to right,
  $t \approx 800 r_0/c$, $t \approx 1700 r_0/c$,
  $t \approx 2200 r_0/c$ and $t \approx 3000 r_0/c$; note different scales on
  the plots, all panels share the same colorbar). The A2-3D jet
   runs into a density profile $\propto r^{-3}$ that flattens to $\propto r^{-1}$ at $100r_0$. 
   This density break 
  slows down the jet and makes the recollimation point (much more noticeable 
  than in Fig. \ref{Sequence-A}) move slower than in simulation A-3D.
  The magenta lines indicate the position of the fast magnetosonic surface
  and the white arrows show velocity streamlines. The jet velocity 
  is super-fast magnetosonically beyond $z \sim 20 r_0$: no information 
  about the flattening in the density profile propagates back to the compact 
  object. Unlike in the simulation A-3D without the density break,
  the A2-3D jet here shows strong deviations from axisymmetry. Thus, the
  density break tends to destabilize jets. }
\label{Sequence-A2} 
\end{figure*}

Fig.~\ref{Sequence-A} shows a sequence of snapshots of our 3D
simulation of model A, which we refer to as simulation A-3D (see
Table~\ref{table1}). We show 2D colour maps of the jet Lorentz factor. When the jet runs into the ambient medium, which represents the
accretion disc wind, it collimates.  As the jet pushes the gas out of
the way, the recollimation point (or the ``pinch" of the jet) moves
away from the center at $v\sim 0.3c$, reaching $z\sim600r_0$ by
$t = 2700r_0/c$, as seen in the different panels of
Fig.~\ref{Sequence-A}. The shape of the jet changes beyond the
recolllimation point from parabolic-like to conical. As this sequence
of snapshots shows, when presented with a continuous power-law density
ambient medium, the jet self-similarly expands and does not show
significant deviations from axisymmetry. The expansion causes the flow
inside the recollimation point to accelerate, up to a Lorentz factor
of $\Gamma \sim 10$, as seen in the right panel (the more the jet
expands, the higher $\Gamma$ it reaches). Once the jet passes through
the recollimation point, it slows down and the Lorentz factor
drops down to $\Gamma \sim 5$.

The magenta lines in Fig.~\ref{Sequence-A} show the position of the
\emph{fast magnetosonic surface}, or simply the fast surface. 
The jets start out moving at sub-fast
magnetosonic velocity. However, as the jets expand and accelerate,
they eventually outrun the fast waves, which happens at the fast
surface located at $z \sim 20 r_0$, as seen in the left panel of
Fig.~\ref{Sequence-A}: beyond
this distance, the fast waves can no longer communicate any
information backwards. Because of this, when the jets undergo a
recollimation or run into an obstacle in the ambient medium, this
information is not communicated backwards along the jet. In fact, if
the jet remained sub-fast throughout, then the information about any
changes in the external medium could reach the central compact object
and change the jet properties there. However, this would be an
artefact of the reduced dynamic range of a simulation. In reality,
since $r_{\rm B}$ is extremely far from the compact
object,
we expect the fast surface to be located at a distance well inside the
Bondi radius. As we will see below, all our jets (except in model A1) accelerate to super-fast
magnetosonic speeds before they encounter any changes in the external
medium, as AGN jets do in nature.

As jets propagate outward in an AGN, they eventually 
encounter the ambient gas of the host galaxy. 
While it is generally uncertain how the radial
profile of the accretion wind disc transitions to the interstellar
medium (ISM) outside the sphere of influence of the black hole,
X-ray observations of M87 \citep{russelletal2015} suggest a flattening of the
density profile potentially accompanied with a factor of few
jump in density, as we discuss below. Motivated by this, we model this transition via a
break and a jump:
\begin{equation} \label{rho_Ax} 
\rho = \begin{cases} 
\rho_{\rm A}(r) & r \le r_t \\
\chi \left(\displaystyle\frac{r}{r_t}\right)^{-\beta} \rho_{\rm A}(r_t) & r > r_t,
\end{cases}
 \qquad \text{(Model A\#)}
\end{equation}
where the fiducial density profile $\rho_{\rm A}(r)$ is given by
eq.~\eqref{rho_A}, $r_t$ is the distance to the break in the density
profile (from $\propto r^{-\alpha}$ to $\propto r^{-\beta}$) 
and $\chi$ is the magnitude of the jump in density: for $\chi=1$ there is
no jump in density at the break, whereas for $\chi \ne 1$ there 
is a jump by a factor of $\chi$.  

The X-ray observations of M87 \citep{russelletal2015} suggest that a change
in the density profile occurs within a factor of a few of 
$r_{\rm B}$, with the asymptotic density slope $\beta\simeq 1$.
Motivated by this, we choose $\beta = 1$ and identify the
location of the break, $r_t$, with the Bondi radius, $r_{\rm B}$. 
That $r_{\rm B}/r_0
\sim 10^5{-}10^6$ is characteristic of M87 and many other AGN, implies that we
would need to place the density break $5{-}6$ orders
of magnitude away from the compact object, resulting in an enormous
dynamical range and making the cost of the numerical simulations
prohibitively high. As a compromise, in our model A2
we reduce the dynamical range to 2 orders of magnitude and take
$r_t = 100 r_0$, see the left panel of Fig. \ref{Model_schematics}.  

\begin{figure*}
\includegraphics[width=16.60cm, angle=0]{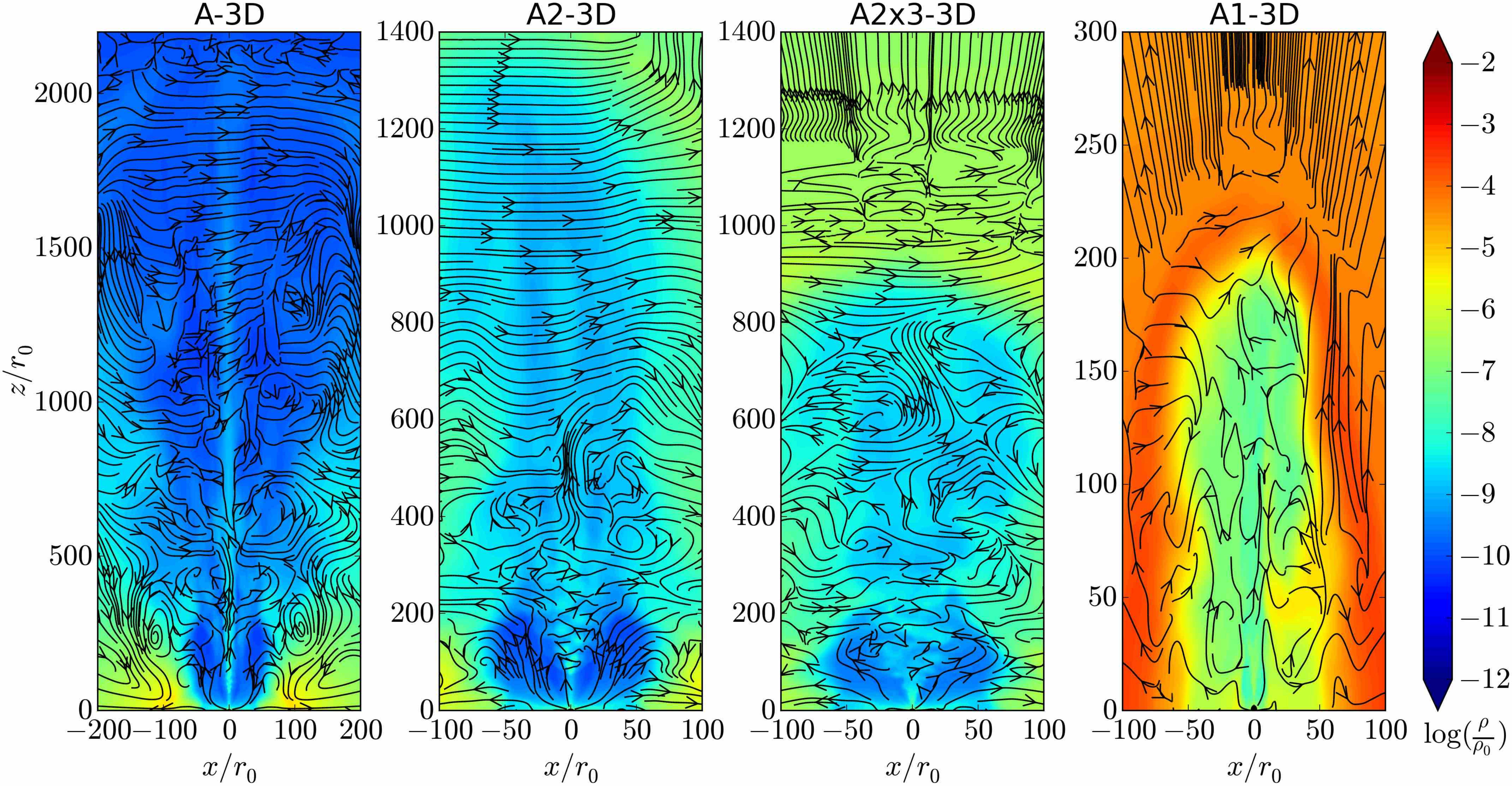}
\caption{Side by side comparison of density snapshots in our 3D simulations at similar times (note the differences
  in length scales), with red showing high and blue low values
(on a log scale; please see the colour bar). From left to right, we 
show vertical slices of density in models A ($t \approx 2700 r_0/c$), 
A2 ($t \approx 3000 r_0/c$), A2x3 ($t \approx 3400 r_0/c$) and 
A1 ($t \approx 2800 r_0/c$), respectively.  Black directed lines show magnetic field 
streamlines traced out in the image plane. The perturbed nature of magnetic lines reflects both the
emergence of irregular magnetic fields in the jets and the large-scale deviations of
the jets out of the image plane, both caused by the 3D
magnetic kink instability. Corresponding Lorentz factor contours of models 
A, A2, A2x3 and A1 can be found in Figs. \ref{Sequence-A}, \ref{Sequence-A2} and 
\ref{LF_A2x3_and_A1}. The jet 
in model A2x3 propagates much slower than in A2, since it runs into the
density higher 
by a factor of 3 at
$r = 100 r_0$. 
Note that the jet in model A1 propagates $\sim4$ times slower than in
the model A2x3 due to an even higher 
density it encounters (this is because the break in the density profile in this model is at $r=10 r_0$). 
The degree of deviation from axisymmetry increases from left
to right. We conclude that an increase in ambient density leads to a stronger magnetic kink 
instability and, as we will see in the text, stronger magnetic energy dissipation.}   
\label{Final_density_modelsA} 
\end{figure*}

\begin{figure}
\begin{center}
\includegraphics[width=12cm, angle=90]{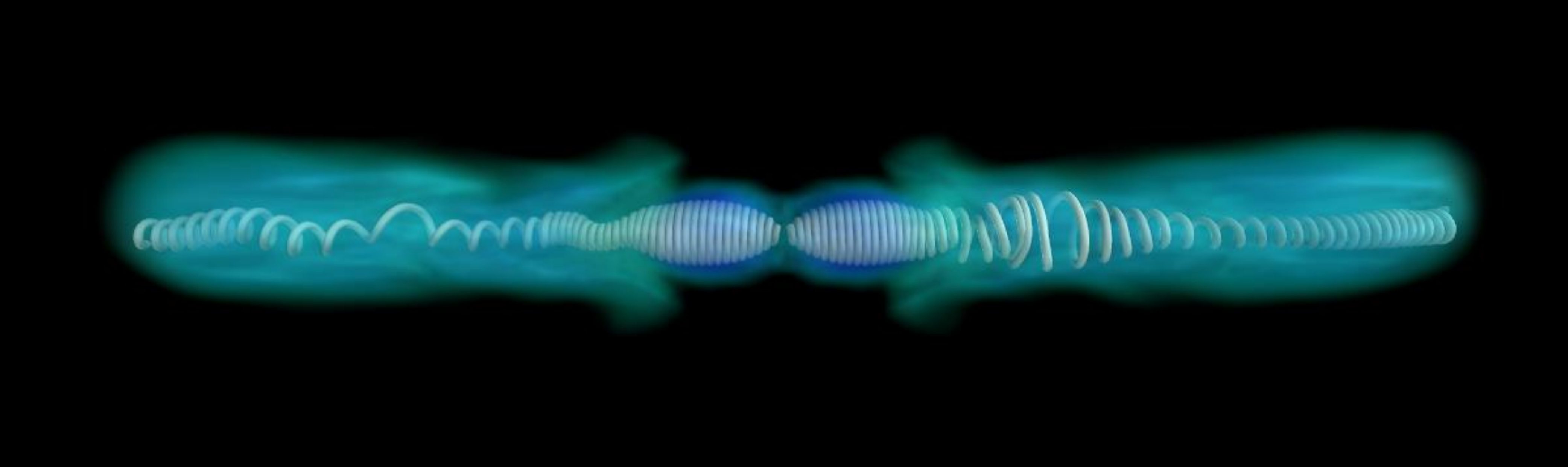}
\end{center}
\caption{3D rendering of density of our 3D simulation of model A2x3
at $t \approx 2200 r_0/c$. The compact object is located at the center of the
figure; this is the only figure that shows both 
jets. The jets extend to $z \sim 650 r_0$. 
We have chosen a representative magnetic field line (white), 
which is twisted as the compact object rotates.
The density breaks at $100 r_0$ to a shallower 
density profile, which causes the jets to recollimate and the toroidal 
magnetic field to build-up and become unstable to the 3D
magnetic kink instability. The irregular bends and asymmetries of the 
magnetic field lines are tell-tale 
signs of the instability. The instability dissipates toroidal magnetic
field into heat around the density break and leads 
to a less tightly wound magnetic field at large radii.}
\label{3D_figure} 
\end{figure}

\begin{figure}
\begin{center}
\includegraphics[width=8.30cm, angle=0]{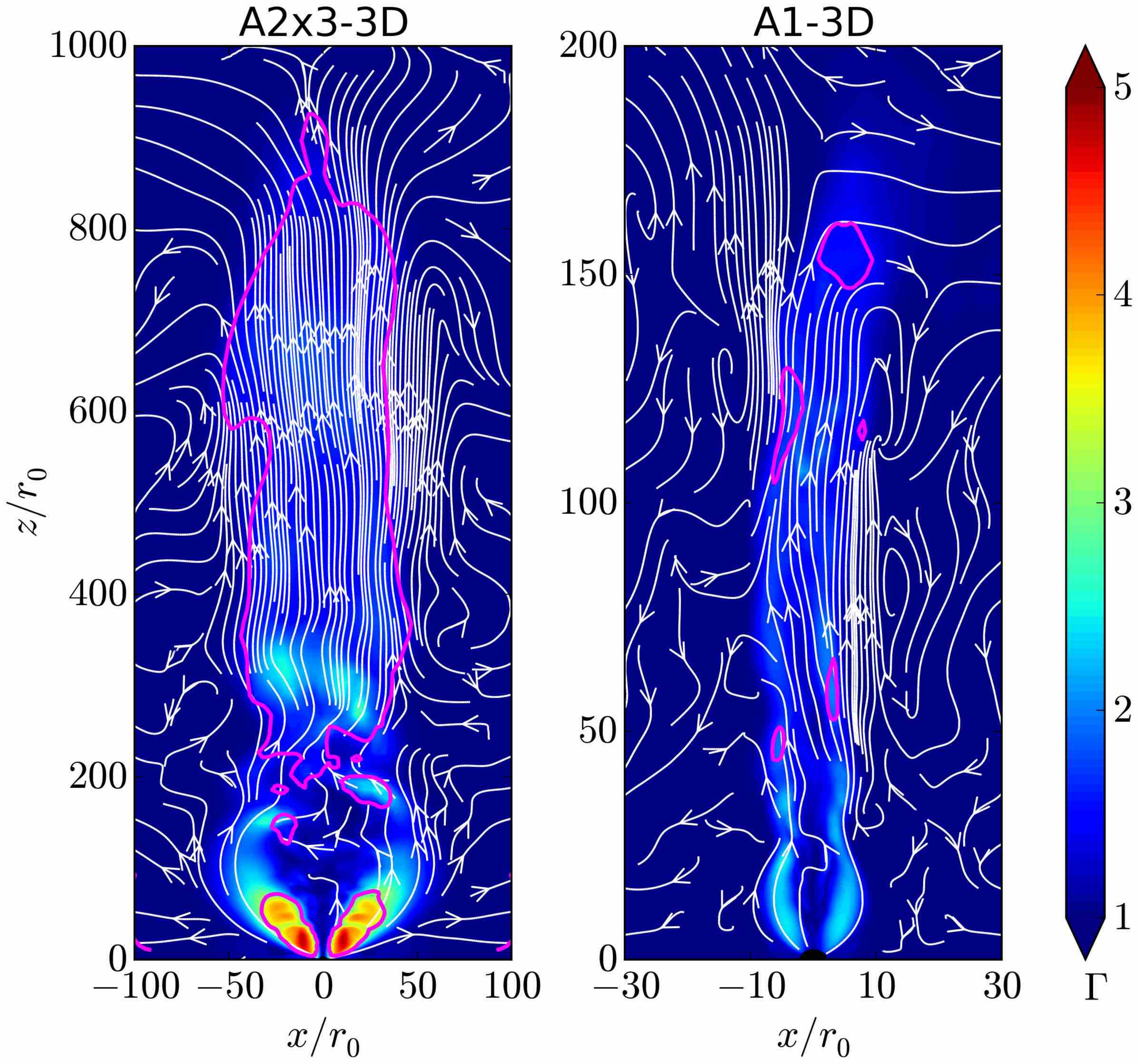}
\end{center}
\caption{Cuts through the rotational axis of the Lorentz factor colour map 
of our 3D simulations
A2x3-3D ($t \approx 3400 r_0/c$) and A1-3D ($t \approx 2800 r_0/c$) 
in the left and right panels, respectively. Both jets show deviations
from axisymmetry due to the magnetic kink instability. The magenta lines 
show the position of the fast magnetosonic surface, while the white
directed lines show the 
velocity streamlines traced out in the image plane. Note that these panels are zoomed-in 
versions of the density contours for these simulations shown in 
Fig. \ref{Final_density_modelsA}. The A2x3-3D jet turns sub-fast as it recollimates 
between $z \sim 100-200 r_0$ close to the location of the density profile break, 
whereas the A1-3D jet remains always sub-fast (except for small super-fast patches)
since the density break is placed 
very close to the central object, resulting in an insufficient dynamic
range for the jet to accelerate to a super-fast velocity. }
\label{LF_A2x3_and_A1} 
\end{figure}

\begin{figure}
\begin{center}
\includegraphics[width=5.8cm, angle=0]{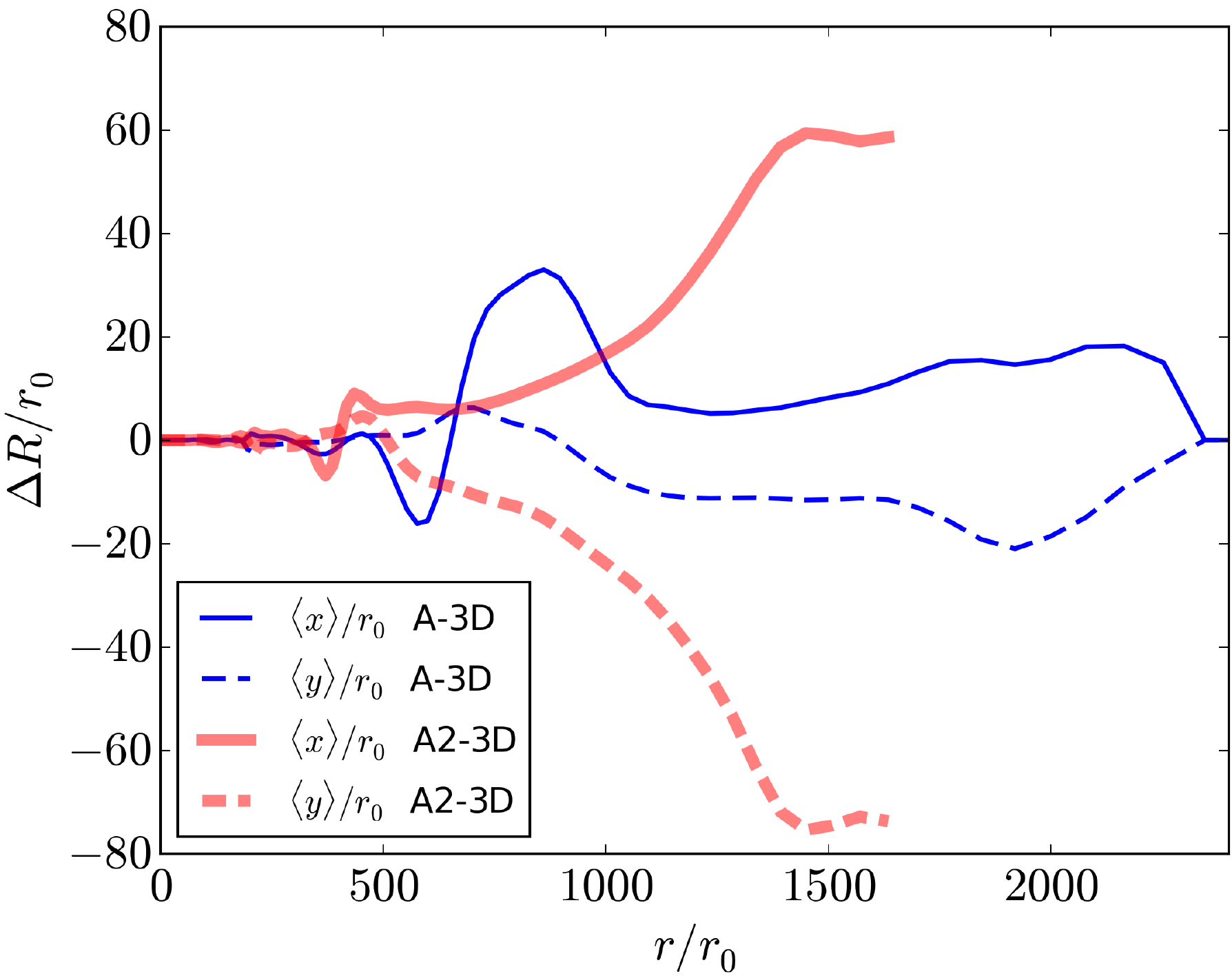} 
\includegraphics[width=5.8cm, angle=0]{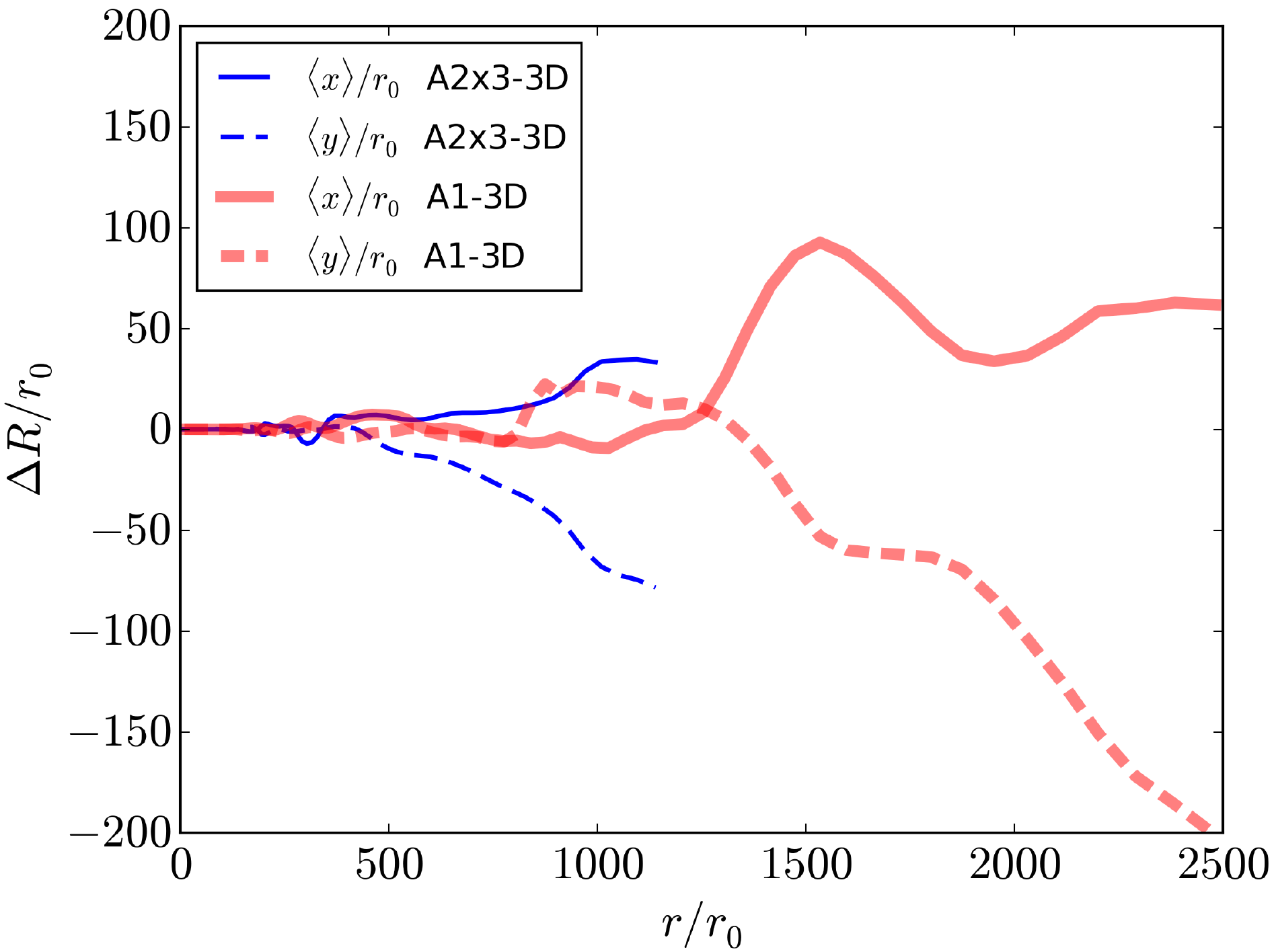} 
\end{center}
\caption{Non-axisymmetric modes in our headed jets, which drill into
  the ambient medium (A models, see Table~\ref{table1}), as measured
  through the deviation of jet ``center of energy flux'', $\langle x\rangle$
  and $\langle y\rangle$, from the rotational axis, $x=y=0$. These
  non-axisymmetric modes are only seen in 3D and signal the
  development of 3D instabilities. {\it Top panel:} The simulation 
  A2-3D (dashed red lines), which
  has a density break at $100 r_0$, shows larger deviations from axisymmetry 
  than simulation A-3D (blue lines), which does not have a density
  break. The measurements are performed at the end of each
  simulation ($t \approx 3000 r_0/c$ for A2-3D and
  $t \approx 2700 r_0/c$ for A-3D). The jet remains axisymmetric until
  the recollimation point; however, beyond it, the jet shows
  deviations from axisymmetry.  {\it Bottom panel:} Deviations from axisymmetry for
  the simulations A2x3-3D (blue lines) and A1-3D (dashed red lines) at the
  end of the simulations ($t \approx 3400 r_0/c$ and $t \approx 2800 r_0/c$,
  respectively). The presence of a jump in simulation A2x3 at
  $r_t = 100 r_0$ destabilizes the jet more than in simulation
  A2-3D, which is shown in the top panel and does not have the density
  jump. Simulation A1-3D also shows a large degree of asymmetry, as
  expected for jets of lower power (their power is 10x lower than for
  A2-3D and 3 times lower than for A2x3-3D). For ease of comparison across
  the models, we zoomed in on the simulation A1-3D by a factor of 10 times
  (i.e., just for this plot, we redefined $r_0 = 0.1r_{\rm in}$
  instead of $r_{\rm in}$ so that the density break is now at
  $r_t = 100 r_0$ instead of $10r_0$).  }
\label{Asymmetry_1} 
\end{figure}

Fig.~\ref{Sequence-A2} shows a sequence of 
snapshots of our 3D simulation of model A2, which we refer to as
simulation A2-3D (see Table~\ref{table3}). 
We show colour maps of the slice through the rotational axis of the Lorentz factor. For this simulation, 
the recollimation point is much more noticeable and violent
than in model~A; it also propagates much more slowly. It settles at $z\sim 200 r_0$, around 
the location of the external medium density break. Although it continues 
to move, the recollimation point moves only at $v\sim 0.1c$
towards the end of the simulation. Given that the external medium is 
not completely rigid, but it is allowed to be pushed by the jet, the recollimation point 
continuously moves out in our simulations. Similar to the A-3D jet, 
the magenta lines in Fig. \ref{Sequence-A2}
show that the A2-3D jet moves at a super-fast magnetosonic velocity beyond
$z \sim 20 r_0$. Also similar is that the jet 
accelerates to nearly $\Gamma \sim 10$
 and once it passes through the recollimation point, it slows down.
The presence of the break in the density profile at $r_t \sim 100 r_0$
not only makes the jet recollimate more effectively, but also leads to
a stronger deceleration of the jet, down to $\Gamma \sim 3$.

Beyond the recollimation
point the jet develops substantial asymmetries, characteristic of 3D,
non-axisymmetric magnetic kink instability. For simplicity of comparison, we show
the simulations A-3D and A2-3D side by side in the two leftmost panels of
Fig.~\ref{Final_density_modelsA}. This makes it clear that the
interaction of the jet with the break in the ambient density slows
down the propagation of both the jet head and the recollimation
point. The flattening of the density profile also causes the jet to
become more collimated and change its shape from the parabolic-like well
inside the recollimation point to cylindrical-like well
outside of the break in density.

Because X-ray observations
of density profile in M87 are consistent with a density jump
accompanying the density break at the Bondi radius
\citep{russelletal2015}, we also consider model A2x3. In addition to
the density break of model A2, it also features a jump in density by a
factor of $\chi=3$ at the position of the break, $r_t = 100 r_0$
(please see equation~\ref{rho_Ax}, the left
panel in Fig.~\ref{Model_schematics}, and Table~\ref{table0}). The third panel from the left
in Fig.~\ref{Final_density_modelsA} shows that the density jump causes
the recollimation to be more violent and significantly slows down the
jet compared with the A2-3D run. Indeed, as
Fig.~\ref{3D_figure} shows, whereas at small radii 
the jets maintain an ordered structure of
a tightly wound axisymmetric helix, beyond the recollimation
point the jet helix (i) develops significant asymmetries and irregularities
characteristic of the 3D magnetic kink instability and (ii) becomes
much less tightly wound. The two effects are related to each other:
the 3D instability dissipates the magnetic energy by reducing the
dominant toroidal magnetic field component and thereby 
also reducing the tightness of the helix (see also \citealt{omerandsasha16}).
The left panel of Fig.~\ref{LF_A2x3_and_A1} shows that the A2x3-3D jet reaches a lower Lorentz factor $\Gamma \sim 5$ 
before the recollimation point and a lower $\Gamma \sim 2$ beyond it,
as compared to model~A2. 
More importantly, as opposed to a density break, the combination of
the break and the jump causes the jet to slow down significantly and turn sub-fast magnetosonic. After 
the recollimation point, the jet accelerates and becomes super-fast
again. 

Note that in order for a jet to switch from super-fast to sub-fast
speeds it needs to go through a shock. Dissipation, associated with such a
shock, might in principle be observable. However, we found that in our
simulations, such shocks dissipate a very small fraction of
jet power, $\lesssim 5$\%, an order of magnitude less than the
internal magnetic kink instability; see Sec.~\ref{sec:sub-fast-super}.

In order to investigate the sensitivity of our results to variations
in jet power, we also carried out simulation A1-3D, which has the
density break at a smaller distance ($r_t = 10 r_0$) without a jump
($\chi=1$), see Tables~\ref{table0} and \ref{table3}, and
Fig. \ref{Model_schematics}. The smaller distance to the density break results in a jet that 
propagates in a denser ambient medium. Effectively, this corresponds to
a jet power that is
$10$ times weaker than in model A2. Because weaker jets propagate at
slower speeds, the smaller distance to the density break allows us to
keep the simulation costs down for the same physical distance covered
by the jet. It takes approximately $\sim 6$ times longer for the jet in
model A1 to reach the same physical distance as in model A2. 

The Lorentz factor of the A1-3D jet is presented in the right panel 
of Fig. \ref{LF_A2x3_and_A1}, which shows a recollimation 
point at $\sim 25 r_0$ and a much slower jet compared with the 
models discussed so far.  It also shows that essentially the entire jet 
remains sub-fast (except for small super-fast patches). 
This is not surprising, because as we saw for the 
previous models the location of the fast surface is at $z\sim 20r_0$, which
would fall beyond the break in density for model~A1.  
This is why in all our previous models we placed the break in density 
profile at $z\sim100r_0$, so that our jets accelerate to super-fast 
magnetosonic velocities by the time they reach the break in the ambient 
medium as we expect realistic AGN jets to do. Given that the A1-3D 
jet remains sub-fast, it describes a different astrophysical 
system, more akin to short- and long-duration GRB jets very close to their 
launching site than to AGN jets. We will investigate the differences
between the stability and other properties of 
sub-fast and super-fast jets in future work.

The rightmost
panel of Fig. \ref{Final_density_modelsA} and the right panel 
of Fig. \ref{LF_A2x3_and_A1} show that the weaker A1-3D jet 
inflates a cavity with clear small-scale and large-scale 
asymmetries that reflect the
development of small-scale, \emph{internal} kink modes, which operate
inside the jet, and global,
\emph{external} kink instability modes, which
operate at the interface between the jet and the ambient medium.
We have seen that these instabilities can be active at different
levels in the headed jets presented so far. We now investigate these instabilities in more detail.

\subsubsection{Deviations from axisymmetry and role of 3D instabilities}
\label{sec:convergence-role-3d}

As seen in Figs. \ref{Sequence-A}--\ref{LF_A2x3_and_A1}, 
the 3D models we have considered so far
remain mostly axisymmetric at distances smaller than 
the recollimation point, at which
the jets pinch toward the axis. However, beyond
the pinch some deviations from axisymmetry are evident, both in
density and Lorentz factor colour maps.  While these deviations are
small for model A, they increase substantially for other models
featuring an obstacle in the jet's way, e.g., in models A1 and A2 with a
density break and even more so in the model A2x3 with a density break and jump.

We quantify the degree of jet non-axisymmetry by measuring the
deviation of the jets from the rotation axis. To do this, we take the last snapshot
of each of our simulations and compute the total energy flux $\dot{E}$ (without 
the rest-mass contribution) weighted values of the transverse coordinates, $x$ and
$y$, as follows: 
\begin{equation}
\langle x \rangle = \frac{\int \! x \, \mathrm{d}\dot{E}}{\int \! \, \mathrm{d}\dot{E}}, 
\end{equation}
and similarly for the $y$ coordinate. In the case of an axisymmetric jet, we expect the total energy 
to be symmetric along the $-x$ and $+x$ (and also along the $-y$ and $+y$) 
directions, yielding $\langle x \rangle = 0$ ($\langle y
\rangle$=0). Thus, the deviations of $\langle x\rangle$ and $\langle
y\rangle$ from zero offer us a quantitative measure of jet deviations
from axisymmetry.

We calculate $\langle x \rangle$ and $\langle y \rangle$ for our model
A jet and plot these as a function of the radial coordinate $r$ in
the top panel of Fig.~\ref{Asymmetry_1}. We see that the jet remains mostly
axisymmetric, with very small deviations from the axial symmetry,
before it passes through the recollimation point.  Even after
the jet passes through it, its deviations from axisymmetry remain quite
small relative to the width of the jet, as seen in
Figs.~\ref{Sequence-A} and~\ref{Final_density_modelsA}. 

In the top panel of Fig.~\ref{Asymmetry_1}, we also show the same
diagnostics for model A2 whose only difference from model A is the
presence of a density break at $r_t=100 r_0$. 
The presence of the break causes the deviations from
axisymmetry to rapidly increase beyond the recollimation point and
most of the jet energy flux to shift away from the rotational axis. In
fact, the tip of the jet is displaced a distance from the rotational
axis comparable to the width of the jet. These deviations from
axisymmetry can be quite easily seen for both the density and Lorentz
factor colour maps of model A2 in Figs.~\ref{Sequence-A2} and~\ref{Final_density_modelsA}. 
This shows that even smooth changes in
the external medium can have a \emph{qualitative} effect on the
structure of the jets. 

We also compute the same diagnostics for the
jets in models A2x3 and A1 and show them in the bottom panel of
Fig.~\ref{Asymmetry_1}.  Noticeable deviations from axisymmetry are
found also for these models, with the strongest deviation in the
lowest-power model, A1. This is consistent with the findings of
\citet{omerandsasha16} who find that the lower the power of a jet, the
less stable it is to the external kink instability. 

\subsubsection{Comparison to 2D simulations}
\label{sec:comp-2d-simul}
In order to understand the role of 3D instabilities on the jet
structure and emission, we compare the 3D simulation results described
above to 2D simulations results. 
Because 3D instabilities are not present in 2D
simulations, the differences between 2D and 3D simulations tell us
about the effect of non-axisymmetric instabilities on the structure of
the jets, including internal dissipation. Furthermore, because we can
carry out convergence tests in 2D with rather little computational
expense, comparison to 2D simulations at different resolutions allows
us to quantify the effect of the resolution changes on our
results. This, of course, does not replace the proper 3D convergence
studies, which we also perform, as we discuss below. We present most
of our 2D simulations and 2D convergence tests in Appendix~\ref{sec:2d-models-headed}, but discuss
the main features of the 2D simulations results in the main text.
  
We have performed 2D simulations of our headed jets at different
resolutions, see Tables \ref{table1} and \ref{table3}.  All our 3D
models that interact with a change in the external medium (e.g., all
A models except A-3D) show much slower propagation of the jet head
compared with 2D models. Also, as shown in Section
\ref{sec:convergence-role-3d}, these 3D jets show deviations from
axisymmetry, whereas 2D jets are axisymmetric by design.  These
non-axisymmetric modes cause the jets to slow down relative to 2D
\citep{omerandsasha16,sashaandomer16}. The 3D instabilities also cause
additional dissipation in the jets; we will now compare to 2D
simulations to quantify this intrinsically 3D effect.
  
\subsubsection{Energy dissipation} \label{Dissipation}

A major goal of this study is to understand whether and how changes in
the external medium affect jet structure and cause internal energy
dissipation. For this, we plot the contributions of different energy
flux components to the
total energy flux, $\dot{E}$, as a function of radius: (i) the electromagnetic energy flux,
$\dot{E}_{\rm EM} = \iint 2p_{\rm mag} \Gamma^2 v_r\, {\rm d}A$, (ii) the
kinetic energy flux,
$\dot{E}_{\rm KE} = \iint \rho \Gamma (\Gamma-1) v_r \, {\rm d}A$, and
(iii) the enthalpy flux,
$\dot{E}_{\rm INT} = \iint (u_g+p_g) \Gamma^2v_r\, {\rm d}A$, where
$\rho$, $u_g$, $p_g \equiv (\gamma-1) u_g= u_g/3$, and $p_{\rm mag}$ are
density, internal energy, thermal pressure, and magnetic pressure, 
all measured in the fluid frame; $\Gamma$ is the Lorentz
factor and $v_r$ is radial velocity. The integration is carried out
over a sphere, and
${\rm d}A = r^2\sin\theta \, {\rm d}\theta \, {\rm d}\varphi$ is the
area element. At the
base of the jet essentially all of the energy flux is in the electromagnetic form. It
transforms to kinetic energy as the jet accelerates and to internal energy as
the jet dissipates.  
The distance at which the energy flux drops to zero marks
the location of the jet head.

We calculate these energy fluxes at the final time of our simulations,
when the head of the jet has reached a distance at least a factor of
$\sim 10$ times larger distance than where the change of the external medium
occurs. In order to isolate the role of 3D
instabilities in the energy dissipation in the jets, we will compare
the energy fluxes in 3D and 2D simulations at the same spatial
resolution, $N_r \times N_{\theta}$.

\begin{figure*}
\begin{tabular}{ccc}
\includegraphics[width=5.5cm, angle=0]{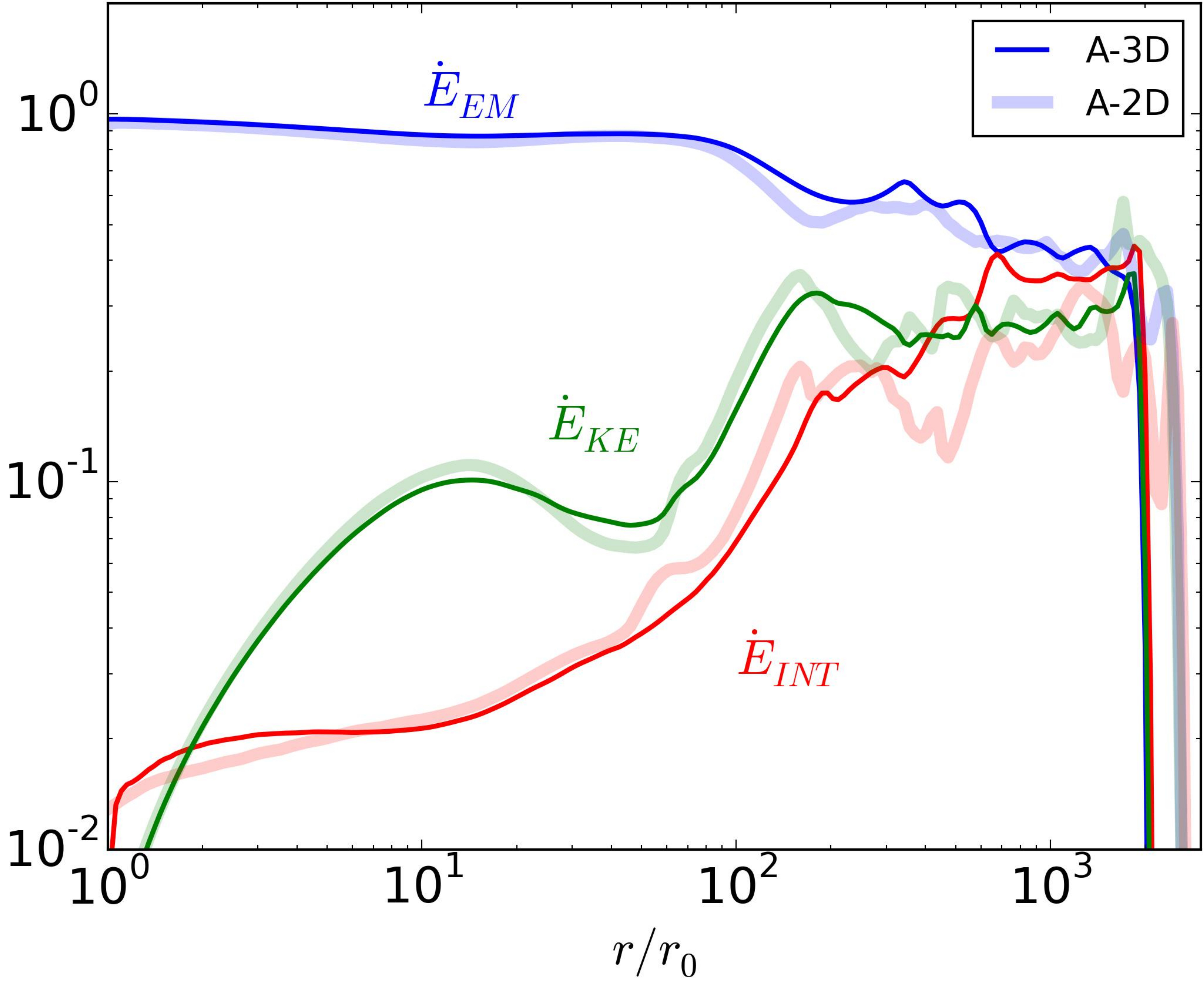} &
\includegraphics[width=5.5cm, angle=0]{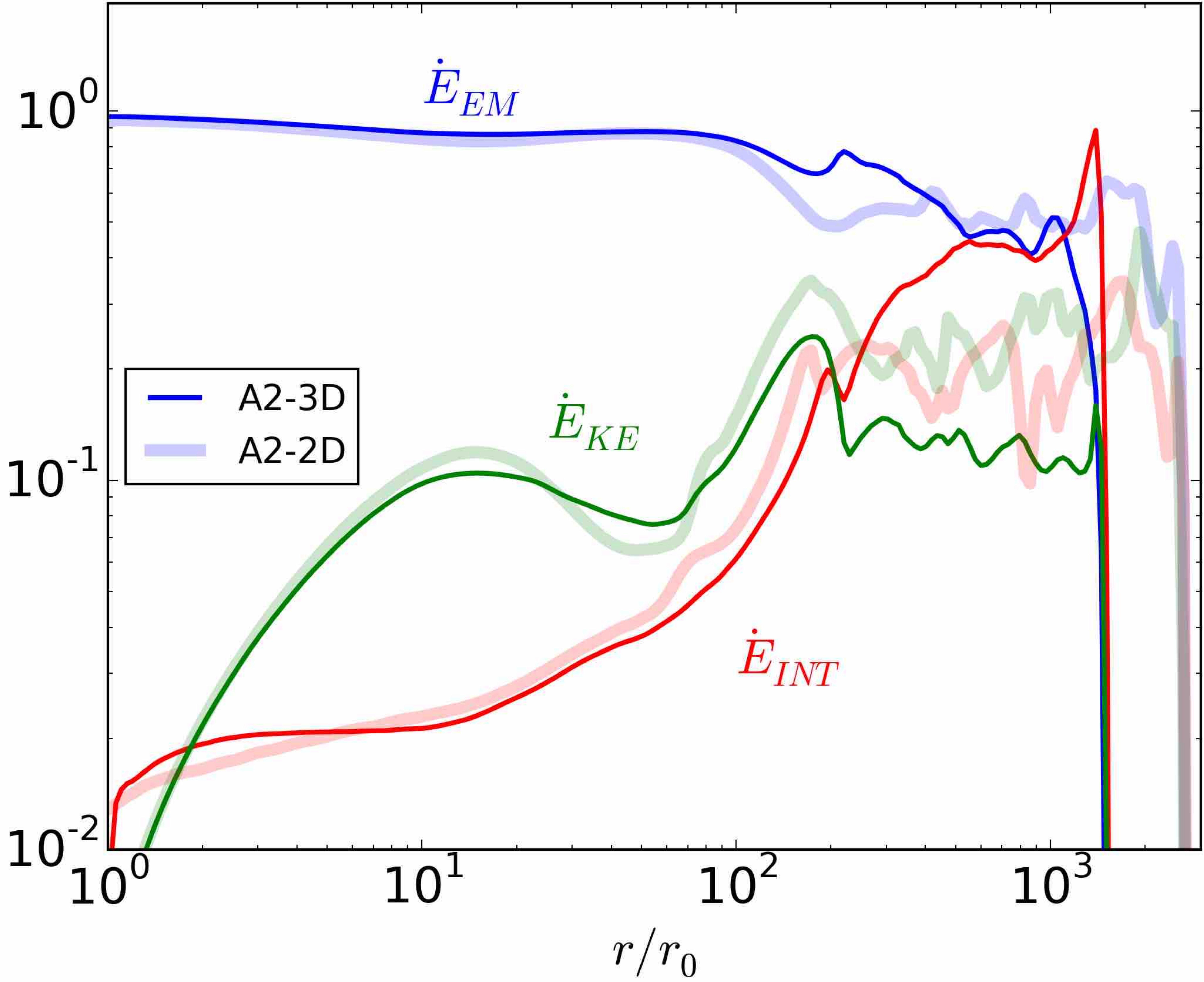} &
\includegraphics[width=5.7cm, angle=0]{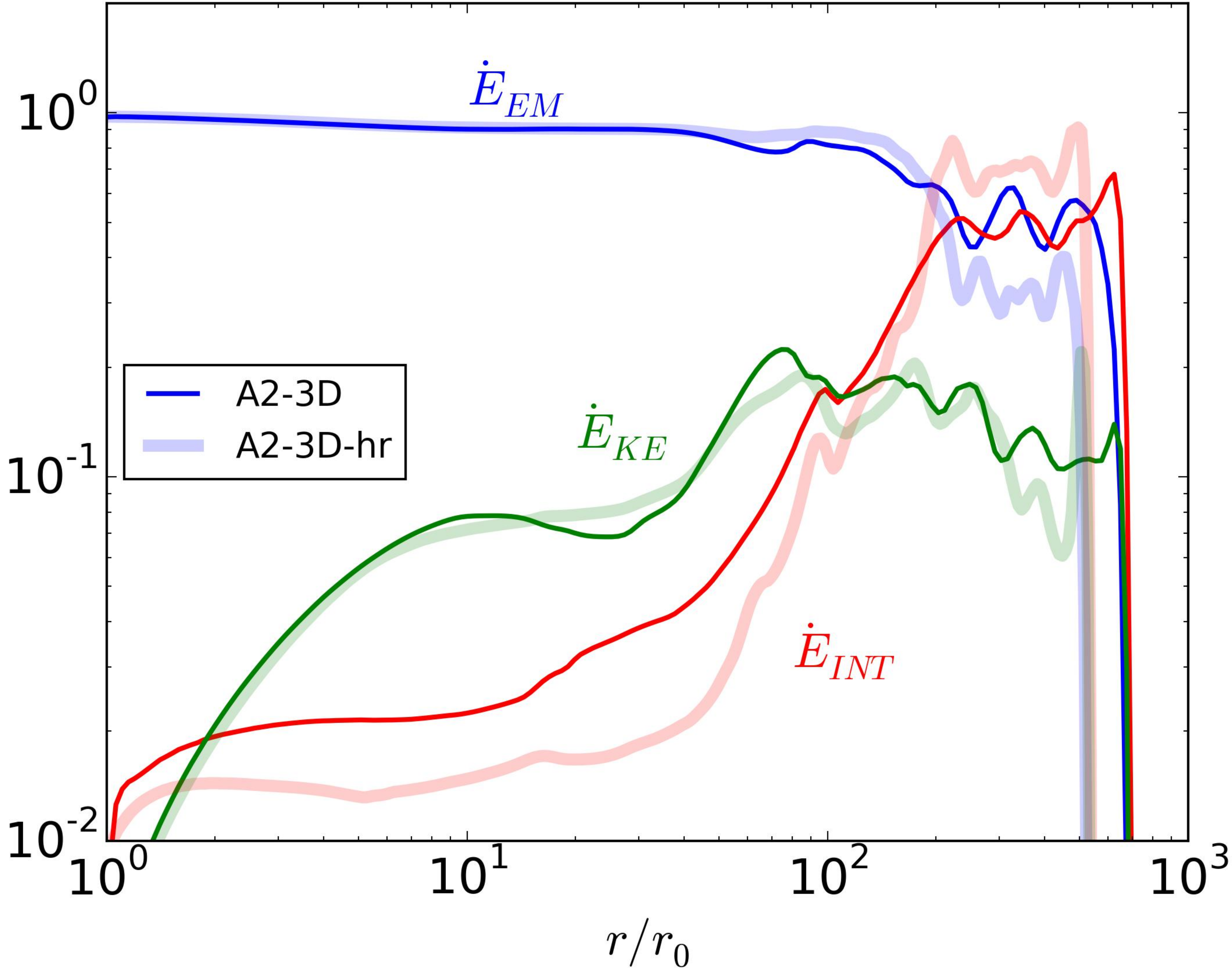} \\
\end{tabular}
\caption{Different components of the energy flux (see labels) 
for 3D and 2D simulations (see legends) of models A 
($t \approx 2700 r_0/c$) and A2 ($t \approx 3000 r_0/c$) in the 
left and middle panel, respectively. 
The energy fluxes have been normalized to the value 
of the total energy flux at $r=r_0$. The distance at which the energy flux 
drops to zero marks the location of the jet head.
In the right panel we present the results of our A2-3D-hr simulation compared
to those of A2-3D (at $t \approx 1500 r_0/c$). Model A shows a similar level of electromagnetic energy 
dissipation in 3D and in 2D (left panel). However, model A2 shows a large increase
in internal energy compared to what it is seen in 2D, due to the presence of 
3D instabilities (middle panel, for $\gae 300 r_0$). The A2-3D-hr run shows even more 
electromagnetic energy dissipation and higher internal energy than the A-3D run
implying that high resolution is required to follow the development of the MHD instabilities 
(right panel).}   
\label{Edot_A_first}
\end{figure*}

We first focus on model A and show its energy flux components in the left panel of
Fig.~\ref{Edot_A_first}. In the A-2D simulation, 
the electromagnetic energy decreases at $r \gae 100 r_0$, 
whereas the internal energy increases. 
However, at higher resolutions in 2D, 
the internal energy increase becomes suppressed.
This indicates that the most likely origin of this increase is 
numerical dissipation due to insufficient resolution 
(see Appendix~\ref{sec:2d-models-headed}). 
We draw the same conclusion regarding simulation A-3D, 
since its energy content is quite similar to the A-2D simulation.
In addition, given that the A-3D jet is
approximately azimuthally symmetric (see top panel of
Fig. \ref{Asymmetry_1}), 3D
instabilities, which could dissipate the electromagnetic energy and
make the jet wobble, are either not present or quite weak
for this model.

\begin{figure}
\includegraphics[width=8.30cm, angle=0]{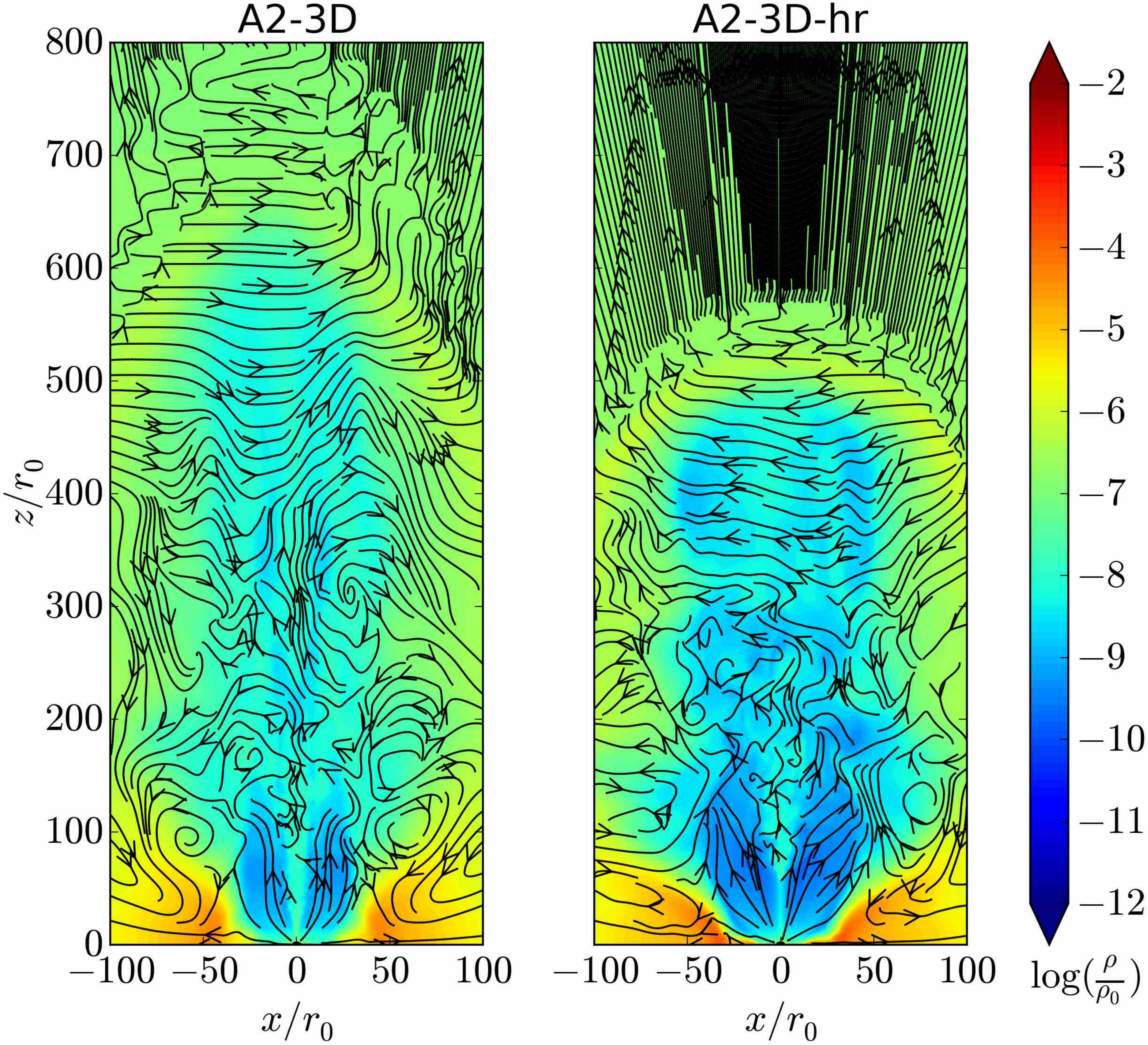}
\caption{2D cut of density contours of the A2-3D simulation (left
panel) and the A2-3D-hr one (right panel) at the final 
run time of the latter ($t \approx 1500 r_0/c$). 
Black arrows show magnetic field streamlines. 
The higher resolution A2-3D-hr simulation 
propagates much slower and inflates a bigger cavity, since it is 
able to better resolve the kink instability that slows 
down the jet propagation as the jet head wobbles.}
\label{Density_A2-3D-hr} 
\end{figure}

The middle panel of Fig.~\ref{Edot_A_first} shows various
contributions to the total energy flux in the simulation A2. Whereas the level of
dissipation in 2D is similar to the model A
without a break, in our 3D simulation A2-3D, the jet internal energy
exceeds that in 2D simulation by a factor of 2 beyond the break,
$r \gae 100 r_0 \sim r_t$. Additionally, the 3D jet 
shows clear deviations from axisymmetry, as seen in the top panel of
Fig.~\ref{Asymmetry_1}. This suggests that 3D instabilities are at
work in this jet.  In order to verify this, we have repeated the
simulation A2-3D at twice as high resolution in each direction. We
refer to this run as the simulation A2-3D-hr (see Table
\ref{table1}).\footnote{The cost of this simulation would be higher by a factor
of $16$ (a factor of $2^3$ from the $3$-dimensions, and an extra
factor of $2$ from twice as small time step). To save
computational time, we reduced the duration of the higher-resolution
simulation by a factor of $2$, resulting in the cost of
$\sim 6\times10^5$ CPU core-hours on the TACC Stampede supercomputer
(see Table \ref{table3}).} The right panel in
Fig.~\ref{Edot_A_first} shows that the higher-resolution simulation
A2-3D-hr exhibits a stronger decrease of electromagnetic energy flux at
$r \gae 100 r_0 \sim r_t$ accompanied by a more prominent increase of
internal energy flux than in the A2-3D simulation.
Fig.~\ref{Density_A2-3D-hr}  compares vertical slices through density
distributions in the two simulations.  The A2-3D-hr jet
propagates $30$\% slower than the A2-3D one, 
due to the better resolution of the external kink instability that slows down the jet
propagation \citep{omerandsasha16}.  

To summarize, the higher
resolution 3D simulation A2-3D-hr shows more energy dissipation than
our fiducial resolution simulation A2-3D. This confirms the robustness
of 3D instabilities in our headed jets and their role in dissipating
the internal energy. In fact, the high-resolution jet dissipates as
much as $70$\% of its energy flux into heat, more than sufficient to
account for the observed emission in jets. Higher resolution also
resolves better the global 3D instabilities and motions of the jet
head. These instabilities substantially slow the jet down as the jet
head wobbles.  We conclude that the presence of a break in the density
profile favors the development of 3D instabilities and considerably
strengthens the dissipation in headed jets.

\begin{figure}
\begin{center}
\includegraphics[width=5.5cm, angle=0]{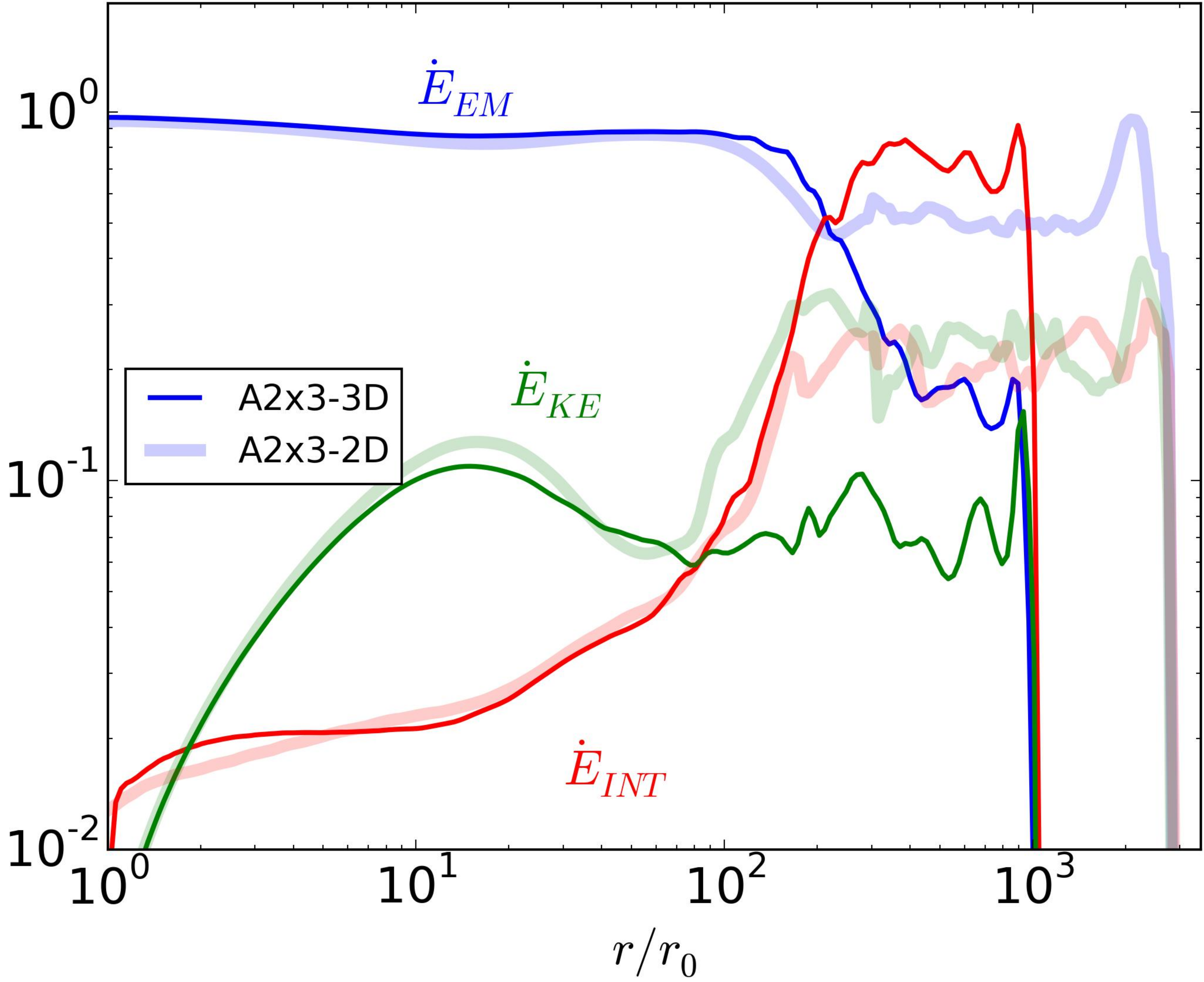}
\includegraphics[width=5.5cm, angle=0]{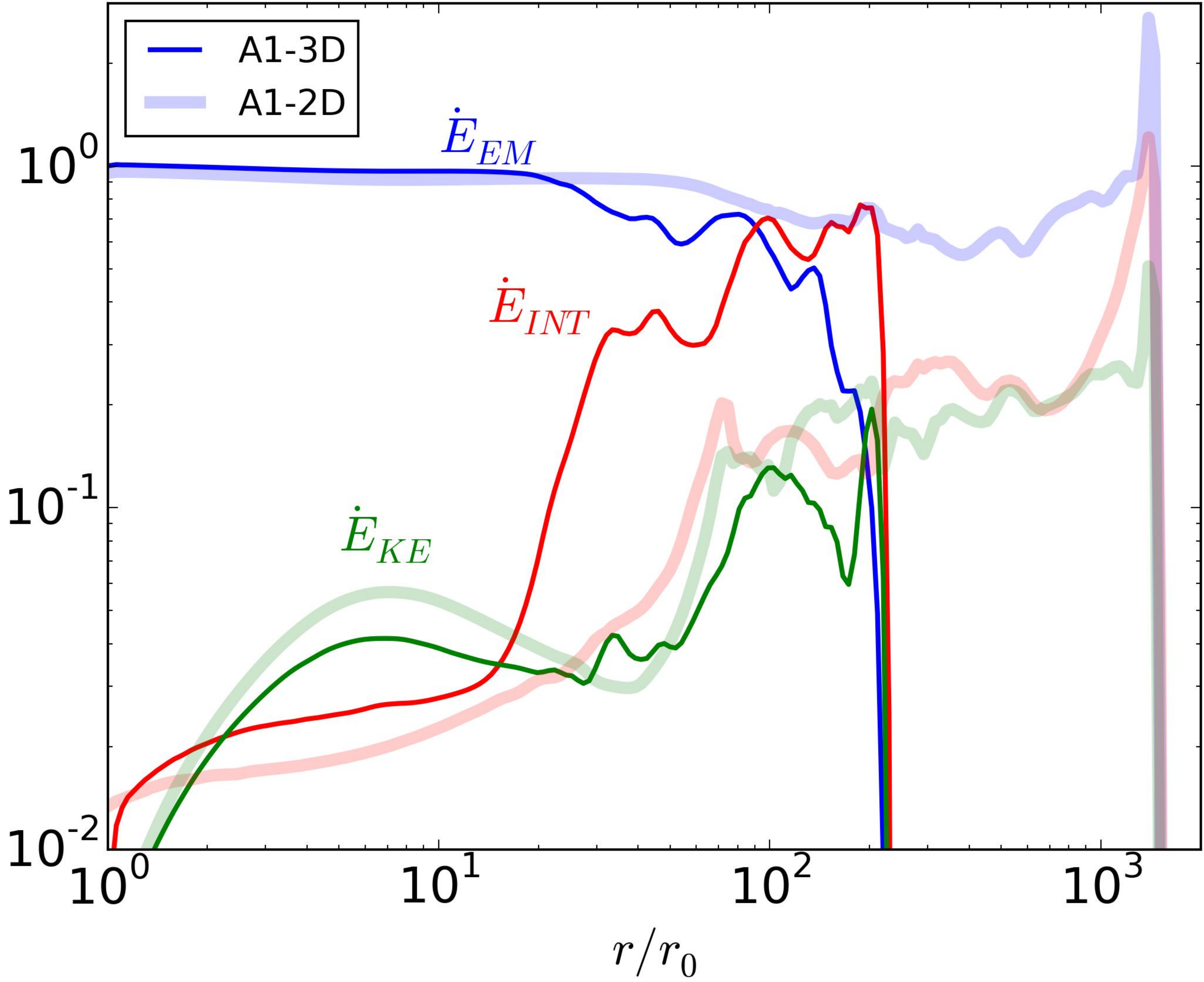}
\end{center}
\caption{Different components of the energy flux (see labels) 
for 3D and 3D simulations (see legends) of models A2x3 
($t \approx 3400 r_0/c$) and (A1 at $t \approx 2800 r_0/c$)
in the top and bottom panel, respectively. 
Both models show a decrease in electromagnetic energy flux accompanied 
by an increase in internal energy flux in 3D at the location of the density 
break. The energy dissipation is much more prominent than what is seen in 2D.}   
\label{Edot_A_second}
\end{figure}

So far we have considered models with jets propagating into continuous
ambient density profiles. The top panel of Fig.~\ref{Edot_A_second}
illustrates jet energy dissipation in model A2x3, in which a jet navigates a jump in 
the external density by a
factor of 3 in addition to the density break. Whereas the 2D version
of this model, A2x3-2D, shows a similar level of dissipation to the
simulation A2-2D (which does not have a jump), in the 3D version,
A2x3-3D, the presence of the jump
increases the fraction of electromagnetic energy converted into heat
substantially: as much as $80$\% of jet electromagnetic energy is
converted into heat in the A2x3-3D jet, which is much higher than the
$20$\% level of dissipation seen in the 2D models A2x3-2D and A2-2D
(this level of dissipation in 2D models is due to limited numerical
resolution, as we show in  
Appendix~\ref{sec:2d-models-headed}).
The high fraction of dissipated energy, coupled with the deviations
from axisymmetry seen for this model (see the bottom panel of
Fig.~\ref{Asymmetry_1}), makes a strong case for the presence of 3D
instabilities and their role in causing the dissipation of
electromagnetic into thermal energy in the headed jets that run into
an obstacle in the ambient medium. Additionally, the presence of a
density jump in the ambient medium is favorable for the growth of
instabilities and energy dissipation.

The bottom panel of Fig.~\ref{Edot_A_second} illustrates the energy
dissipation in model A1. This model
corresponds to a lower power jet propagating into a density
profile with a break (without a jump) at $r_t = 10
r_0$. For this model, the A1-2D 
simulation shows an approximately constant value of the electromagnetic 
energy flux as a function of radius. However, the A1-3D run shows a
decrease in the 
electromagnetic energy flux, which occurs right at the location of the density break 
$r_t \sim 10 r_0$, and is accompanied by an increase in the internal
energy flux at the same
location. At large radii, the internal energy flux is at least 50\% of the total jet power.
These features are not seen at any resolution in the 2D
simulations (see Appendix~\ref{sec:2d-models-headed}).
Thus, the dissipation and the high degree of 
asymmetry in the shape of the A1-3D jet (see bottom panel of Fig. \ref{Asymmetry_1}) 
strongly indicate the presence of 3D instabilities and their
contribution to the internal jet dissipation.  

\subsection{Sub-fast and super-fast regions and dissipation}
\label{sec:sub-fast-super}

As we discussed in Sec.~\ref{sec:model-ambi-medi}, 3D jets in models A and A2 
are super-fast magnetosonic. However, Figs.~\ref{Sequence-A}
and \ref{Sequence-A2}  show that a fraction of the jet around the polar axis
 remains sub-fast. This sub-fast region ``knows'' about the
impending collision with the ambient medium and develops an internal
3D kink instability that converts magnetic into thermal energy. 
Since energy flux in jets is edge-concentrated \citep{sashaetal2008}, these polar 
sub-fast regions contribute very little to the
internal energy flux, only about a few per cent of the total 
jet power. The jet in model A2x3 develops larger sub-fast
regions (see Fig. \ref{LF_A2x3_and_A1}) with
an internal energy flux that makes up $\sim 15$\% of the 
total jet energy flux; the super-fast regions carry $\sim 65$\% 
of the total energy flux, which together yield the $\sim 80$\% total 
internal energy flux seen at large radii in Section \ref{Dissipation} 
for this simulation. 
For the 3D jet in model A1, which 
remains sub-fast throughout except for small super-fast patches, 
the internal energy flux in the sub-fast regions is 
at the $\sim 10$ \% level of the total jet energy flux. 

In regions where the jets go 
from being super-fast to sub-fast magnetosonic (e.g., in model A2x3-3D, 
left-panel of Fig.~\ref{LF_A2x3_and_A1}), a shock 
forms and slows down the flow. More generally, we expect such a shock
to form
every time a super-fast flow rams into the ambient medium. 
We expect the shock to dissipate some energy; however, 
given the high magnetization at the location of the shock,
we expect the dissipation to be rather weak \citep[e.g.,][]{2010MNRAS.407.2501M}. There is evidence for the presence
of such shocks, for example, in the energy fluxes of the high-resolution 2D simulations of 
model A2x3, 
where a sharp increase in internal energy flux is observed at $r\sim
150 r_0$  (see Appendix~\ref{sec:2d-models-headed}).\footnote{The spike in the internal energy 
is also seen to a lesser degree in our 3D simulations, e.g.,
A2-3D and A2-3D-hr, in the right panel of Fig.~\ref{Edot_A_first}, at
$r \sim 100r_0$. Although
the shocks are present in both our 2D and 3D simulations, the
limited resolution of our 3D simulations prevents them from resolving the internal
energy down to the $\sim5$\% level accuracy necessary for quantifying the dissipation
at such shocks
for our headed jets.}
However, as expected, the internal energy flux at the post-shock
region is small, 
$\lae 5$\%. 
Even though energy dissipation at the shock is small, by slowing down
the flow the shock triggers the development of 
MHD instabilities downstream that can efficiently dissipate magnetic energy.

\begin{figure}
\begin{center}
\includegraphics[width=4.37cm, angle=0]{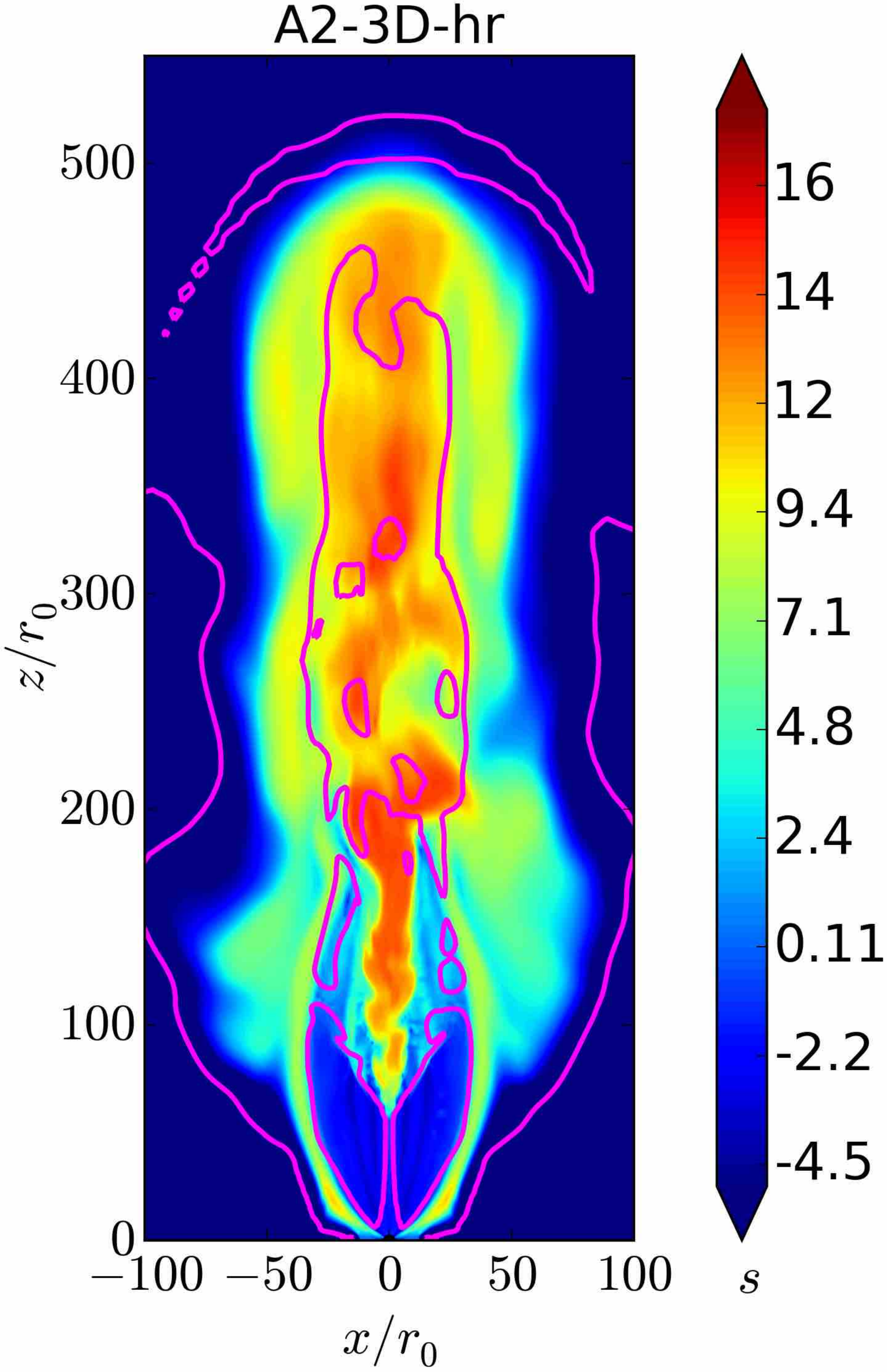}
\includegraphics[width=3.98cm, angle=0]{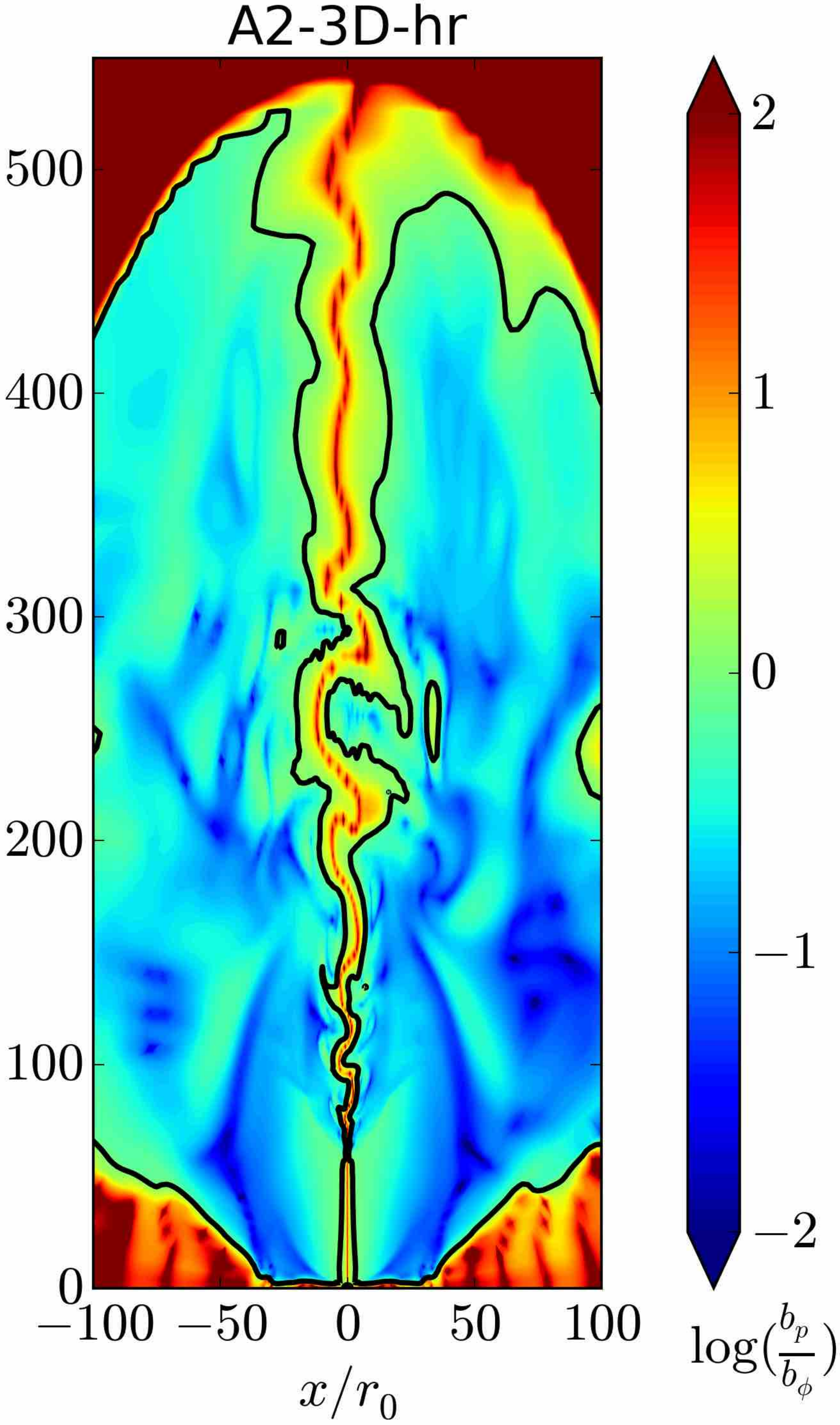}
\end{center}
\caption{{\it Left panel:} Vertical slice through entropy in the A2-3D-hr simulation 
at the final run time, $t \approx 1500 r_0/c$. The magenta lines show the 
position of the fast magnetosonic surface. At $z\gtrsim10r_0$, the jet
crosses the fast magnetosonic surface the first time and turns from 
sub- to super-fast magnetosonic. At $z\gtrsim 50 r_0$ the jet crosses
the fast surface the second time, this time turning from super- to
sub-fast magnetosonic. This second fast surface is a
fast shock at which the entropy sharply increases. The entropy
increases further beyond the shock, at $z\gtrsim100r_0$, due to the internal kink instability. {\it Right
  panel:} 2D cut of 
$\log_{10}(b_{\rm p}/b_{\phi})$ contours of the A2-3D-hr simulation 
at the final run time ($t \approx 1500 r_0/c$). The black lines show the 
surface where $b_{\rm p}/b_{\phi} = 1$ roughly dividing the regions 
where the jet is more ($b_{\rm p}/b_{\phi} < 1$) and less ($b_{\rm p}/b_{\phi} > 1$) 
susceptible to the magnetic kink instability. As the magnetic energy is dissipated by the 
kink instability and converted to internal energy, the jet turns more stable to this 
instability.}
\label{Entropy_A2-3D-hr} 
\end{figure}

The presence of a shock can also be identified by the variation of 
entropy, 
\begin{equation}
s = \frac{1}{\gamma - 1} \ln \left( \frac{p_g}{ \rho^{\gamma}} \right),
\end{equation} 
where $p_g = (\gamma - 1) u_g = u_g/3$ is the thermal pressure. This
diagnostic is an alternative to the inspection of energy fluxes
discussed above (which are angle-integrated quantities and thus are
not well-suited for identifying dissipative features that are extended
in radius, such as conical shocks). We show a colour map of entropy in
a vertical slice through the jet axis of model A2-3D-hr in the left panel of
Fig. \ref{Entropy_A2-3D-hr}. The low-entropy flow (since the jet is
initially cold) is super-fast at the end of the simulation and turns
sub-fast around $z \sim 80 r_0$. Interestingly, just as the flow turns
sub-fast, the entropy increases abruptly (the fast magnetosonic
surface and the entropy increase match precisely), indicating the
presence of a shock as discussed above. This entropy plot very clearly
highlights the various dissipation regions in our headed jets
discussed above: the dissipative cone at small radii along the jet
axis and the internal kink at larger radii. Both of these are regions that can
also be identified in, e.g.,~Fig.~\ref{Sequence-A2}.

As mentioned in \cite{omerandsasha16}, the magnetic kink instability is triggered 
as the flow passes through the recollimation region. They derived an instability criterion 
that they applied to their particular scenario of mildly relativistic, 
$\Gamma \approx 1$, and initially weakly collimated jet that runs into
an ambient medium. Our jets move relativistically and 
are highly collimated. The stability criterion that describes our jets remains to 
be studied in detail. However, in analogy with the work of \cite{omerandsasha16}, the growth 
time-scale of the magnetic kink instability is proportional to the magnetic pitch, which is the 
ratio of the poloidal to toroidal magnetic field strength in the comoving frame, 
$b_{\rm p}/b_{\phi}$ (see, \citealp[e.g.,][]{2000A&A...355..818A}). We show a colour map of 
$\log_{10}(b_{\rm p}/b_{\phi})$ for the A2-3D-hr jet in the right panel of Fig.~\ref{Entropy_A2-3D-hr}, which can be 
easily compared to the entropy colour map in the left panel.  The narrow jet, which extends 
from $x \sim - 50 r_0$ to $x \sim 50 r_0$ shows $b_{\rm p}/b_{\phi} < 1$ at small radii, 
which suggests a rapid growth of the magnetic kink instability. At larger radii, the region 
with $b_{\rm p}/b_{\phi} > 1$ grows in size, suggesting that the jet turns stable at large radii
(see \citealp{omerandsasha16}). This happens because the toroidal component 
of the magnetic field, $b_{\phi}$, which dominates at small radii and turns the jet kink unstable, is 
dissipated away as the magnetic energy is converted to internal energy.  After this occurs, 
$b_{\phi}$ decreases making the jet less susceptible to the kink instability. This can be also 
clearly seen in Fig. \ref{3D_figure}, where the irregular bends and asymmetries of the magnetic 
field lines become dissipated away. This causes the field to 
become less tightly wounded and more ordered at large radii. 

Summing up, the presence of obstacles in 
the ambient density, such as a factor of few jump in density, or even just a break in the
density slope, can have a dramatic effect on jets in  3D. These features
can lead to the development of 3D instabilities and
prominent dissipation, up to $\gtrsim50$\% of
the total jet flux (see Figs. \ref{Edot_A_first} and
\ref{Edot_A_second}).  This dissipation might power bright features in
jets and the blazar zone as we discuss in Secs.~\ref{sec:discussion}
and \ref{sec:conclusions}.

\subsection{Headless jet interacts with the external medium}
\label{sec:headl-jet-inter}
\subsubsection{Modeling the external medium}

In Sec.~\ref{sec:introduction}, we discussed the distinction between
jets of two types different by whether they drill through the ambient
medium: \emph{headed} and \emph{headless} jets.
In Sec.~\ref{sec:headed-jet-interacts}, we considered how \emph{headed} jets,
which drill through the ambient gas, are affected by the changes in
the external medium density profile.  Here, we perform a similar
analysis for \emph{headless} jets, which do not have to drill through the
ambient medium because, e.g., they have a pre-existing funnel to
propagate through.  We expect such jets in several astrophysical
scenarios (see Sec.~\ref{sec:introduction}).  In the AGN context, 
we would like to model a jet that propagates outward through
essentially an empty
(or very low-density) funnel and collimates against the surrounding accretion
disc outflow. In this case, we can think of the walls of the funnel as
providing the
confining medium for the jet and setting the jet shape.

In order to model such jets, we use a density distribution of the form
\begin{equation} \label{rho_B}
\rho_{\rm B} = \rho_0 \left(\displaystyle\frac{r}{r_0}\right)^{-\alpha} \sin^{\delta} \theta,  \qquad \text{(Model B)}
\end{equation}
where $\theta$ is the polar angle (the angle measured with respect to
the jet axis). As in model A, we adopt $\alpha = 3$ and, similarly, choose 
$\rho_0=4500$ and $B_0=(4\pi)^{1/2}$, the initial density and magnetic
field strength at $r_0$, respectively. The $\sin^{\delta} \theta$ term effectively
evacuates the density along the rotational axis and essentially
prescribes the shape of the walls of the funnel\footnote{For the
  chosen value of $\rho_0$, we find that $\delta \approx 6$ yields a
  super-fast magnetosonic jet early on, by the time that the jet head
  reaches $\sim 100 r_0$.}. We will refer to this as model~B, see
Table~\ref{table0} and the middle panel in
Fig.~\ref{Model_schematics}. This density mimics what is seen in
simulations of thick accretion discs (e.g.,
\citealp{devilliersetal2005, mckinney2006, sashaetal2011}); a more
detailed comparison will be left for a future study.

\begin{figure}
\includegraphics[width=8.3cm, angle=0]{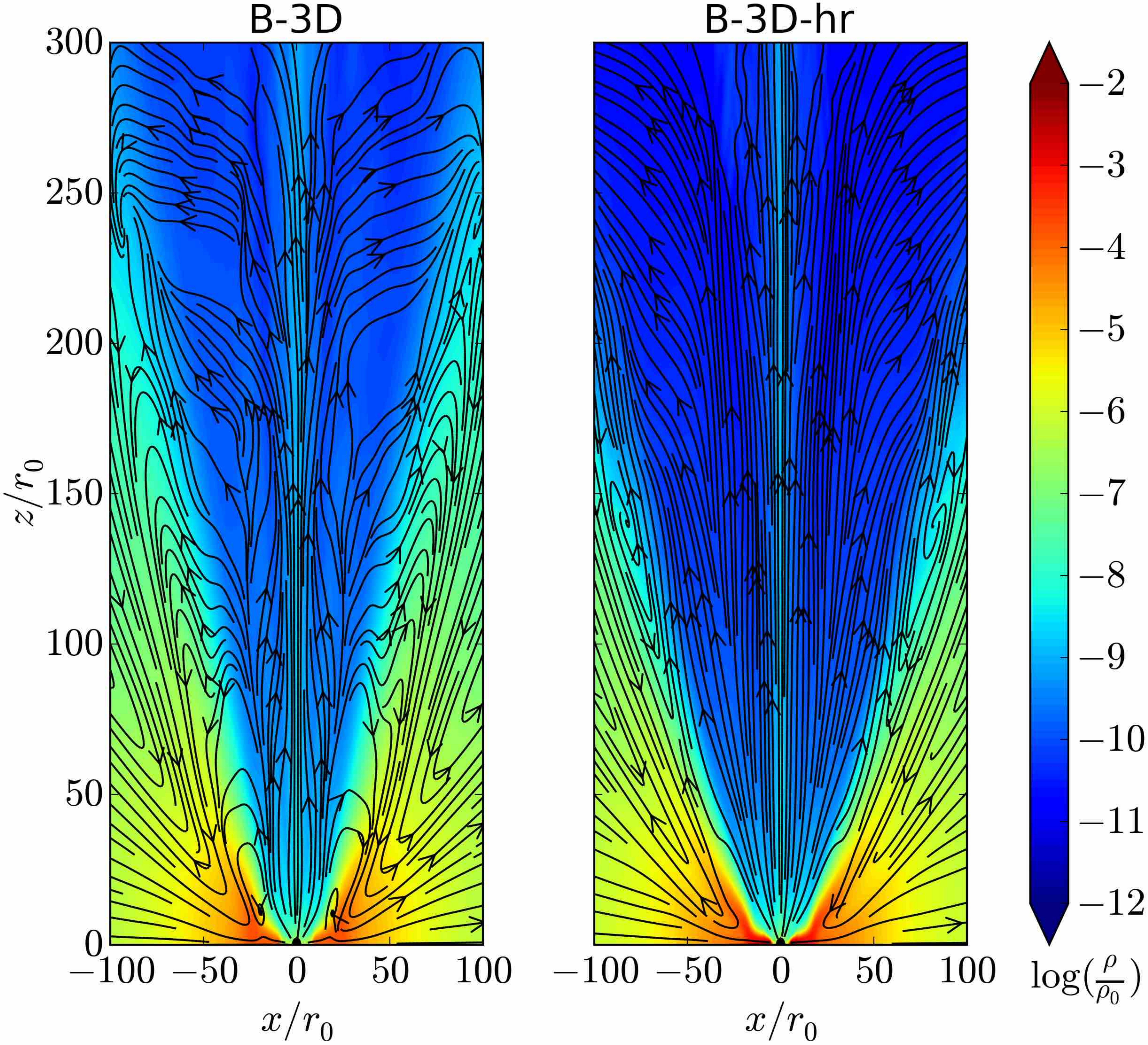} 
\caption{2D cut of density contours of the B-3D simulation (left
  panel) and the B-3D-hr one (right panel) at the final 
  run time of the latter ($t \approx 1200 r_0/c$). Black arrows show
  magnetic field streamlines. The jet propagates
  along the funnel as it expands sideways.  The B-3D and B-3D-hr simulations are
  quite similar. We have zoomed-in on a region $z < 300 r_0$ even though the 
  jet head is at $\sim 1200 r_0$. There are no significant deviations from axisymmetry in 
  this model. }
\label{Density_B} 
\end{figure}

We refer to our 3D simulation of model B as simulation B-3D (see
Table~\ref{table3} for more details). In the left panel of
Fig.~\ref{Density_B}, we show the colour map of density in a vertical
slice through the jet axis.  We see that the jet propagates along
the funnel as it expands sideways, maintaining roughly a parabolic
shape.  At large radii, the jet attains a
Lorentz factor of $\sim 7$.

\begin{figure}
\includegraphics[width=8.3cm, angle=0]{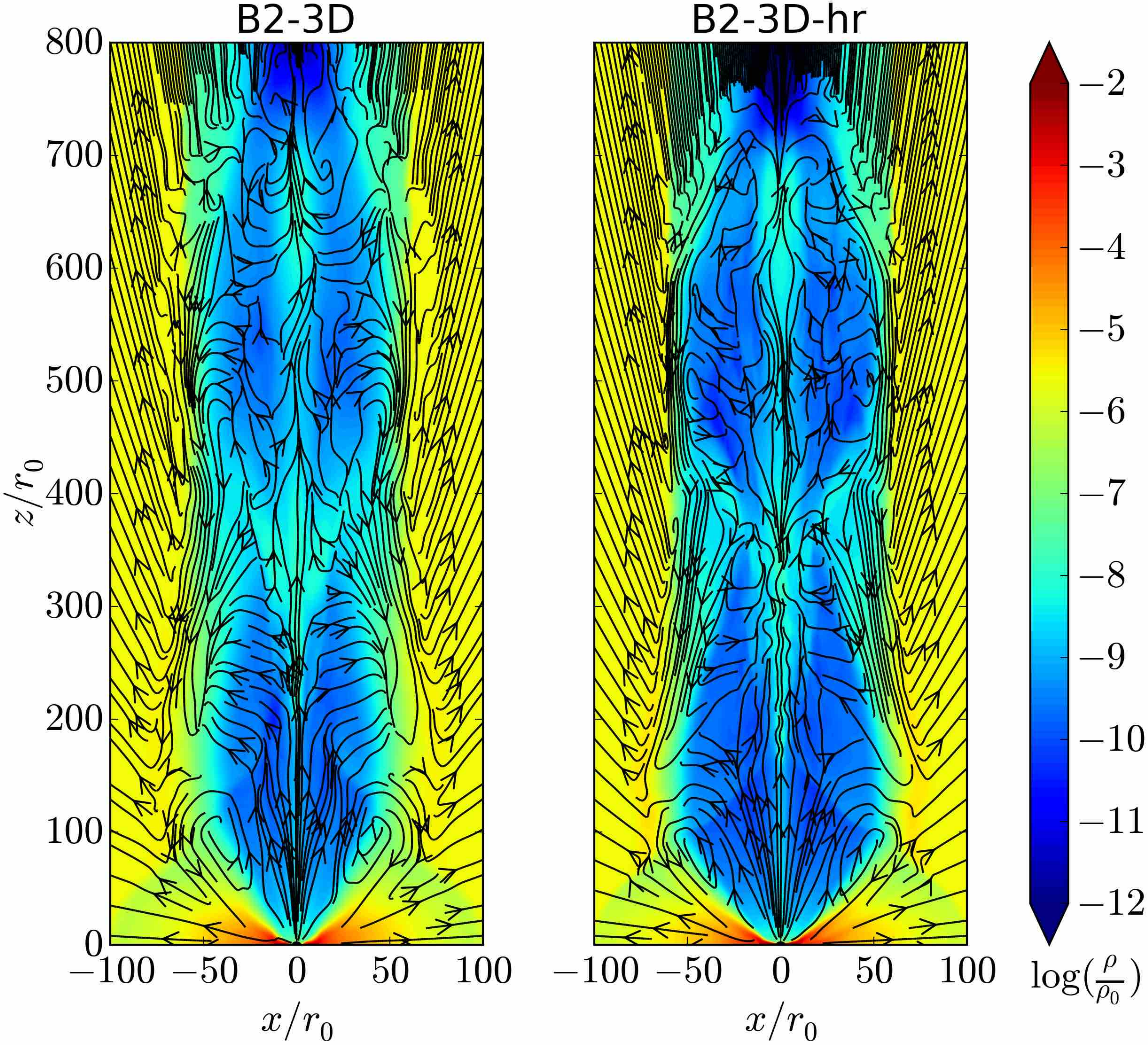} 
\includegraphics[width=8.3cm, angle=0]{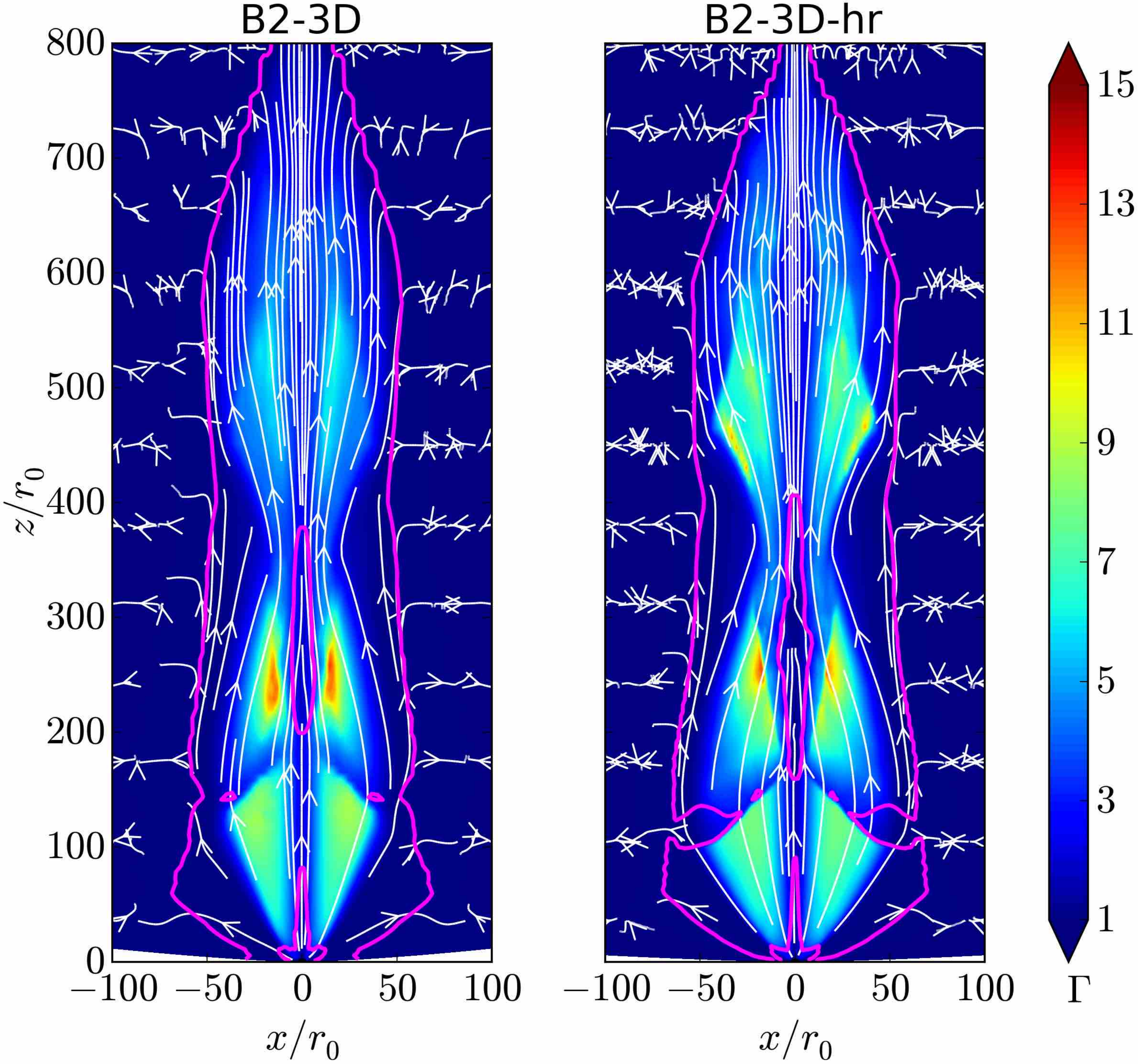} 
\caption{2D cut of density contours (top panels) and Lorentz factor 
contours (bottom panels) of the B2-3D simulation (left
panels) and the B2-3D-hr one (right panels) at the final 
run time of the latter ($t \approx 700 r_0/c$). Black arrows in the 
top panels show magnetic field streamlines. White arrows 
in the bottom panels show velocity streamlines and magenta
lines show the location of the fast-magnetosonic surface
(most of the jet, colored regions, is super-fast, except 
for a small thin region along the jet-axis; this will be discussed 
in Section \ref{Dissipation_headless}). 
The jet is confined to propagate in a pre-existing funnel that becomes cylindrical
at $z = 80 r_0$. The presence of conical shocks is evident by inspecting the Lorentz 
factor plots (as coloured shapes with a sharp boundary). We see a conical 
shock at $z \sim 150 r_0$ and another one between $z \sim 200-300 r_0$. 
The B2-3D-hr simulation shows a more defined structure
between $z \sim 400 r_0$ and $600 r_0$ and the presence of yet another 
conical shock in this region.}      
\label{Density_LF_B2} 
\end{figure}

Now that we have simulated a jet confined by an accretion disc, we
would like to modify the set up so that the jet eventually interacts with
the ambient medium. We model this interaction by adjusting the shape
of the funnel to abruptly become cylindrical. Because the funnel is
essentially empty, the jet does not have to drill through the ambient
medium and remains headless. 
Namely, we consider the following density distribution 
\begin{equation} \label{rho_Bx} 
\rho = \begin{cases} 
\rho_{\rm *} (z)  & |z| \le z_0 \\
\rho_{\rm *} (z_0) & |z| > z_0, 
\end{cases}
 \qquad \text{(Model B\#)}
\end{equation}
where $z = r \cos \theta$ is the distance along the jet axis of
rotation, and $z_0$ is the distance beyond which we ``freeze" the
lateral profile of the external medium at the value of $z=z_0$, that
is, the funnel becomes cylindrical. In order to ensure that the funnel
walls are heavy enough to prevent the jet from displacing them, we
use 
\begin{equation} 
\rho_{\rm *} = \begin{cases} 
\rho_{\rm B}(r) & r \le r_t \\
\chi \left(\displaystyle\frac{r}{r_t}\right)^{-\beta} \rho_{\rm B}(r=r_t,\theta=\pi/2) & r > r_t,
\end{cases}
\end{equation}
with $\beta =1$, $r_t = 100 r_0$ and $\chi = 3$. 
In equation (\ref{rho_Bx}) we choose $z_0 = 80 r_0$, which leads to a cylindrical 
funnel of radius $\simeq 60 r_0$. We refer to this setup as model B2, see 
Tables~\ref{table0} and \ref{table3} and the right panel in 
Fig.~\ref{Model_schematics}. 

We show a snapshot of our 3D simulation, B2-3D, in Fig.
\ref{Density_LF_B2} (left panels), where we present the slices through
the density and Lorentz factor contour maps.  The cylindrical funnel
confines the jets and forces them into a cylinder. We find that, at the location where the funnel transitions
to being cylindrical, the jet undergoes a conical shock as evident by
the sharp feature in the Lorentz factor seen in the
bottom-left panel of Fig.~\ref{Density_LF_B2}, especially the features
at $(x,z) \sim (\pm 40 r_0,150 r_0)$. There is also a second conical 
shock, between $z \sim 200-300 r_0$, whose surface is close to the jet axis. We discuss these
conical shocks below.

Our jets in models B and B2 propagate close to the speed of light,
because they propagate along an evacuated funnel and 
do not have to do work against the external medium. 
This is also why they easily
accelerate to super-fast magnetosonic velocities before they encounter
changes in the ambient medium, just like jets in nature. Indeed,
the bottom left panel of Fig.~\ref{Density_LF_B2} shows that the
jets cross the fast surface at around $z\sim 10r_0\ll z_0$. This is
important, as we discussed previously, because if this were not the
case, the ambient medium would be able -- in an unphysical way -- to
affect the conditions of the jets near the central compact object.

\subsubsection{Deviation from axisymmetry and role of 3D instabilities}
\label{sec:devi-from-axisymm}

We have calculated the deviations from axisymmetry for models B-3D and B2-3D
by calculating $\langle x \rangle$ and $\langle y \rangle$ as
described in Section \ref{sec:convergence-role-3d}. Only extremely small 
deviations of the order of $\langle x \rangle/r_0 \sim \langle y \rangle/r_0 \sim 1$
are observed for these simulations at the final times of the runs.
For models B-3D and B2-3D, this
result shows that headless jets tend to maintain their axisymmetry,
even when they navigate abrupt changes in the jet
shape. Indeed, even after passing through recollimation shocks, the jets do
not develop global 3D instabilities, although this is perhaps not so
surprising given that the shape of the jet is held fixed by the hard
wall. This is in agreement with \citet{omerandsasha16}, who found that
headless jets are more stable than their headed counterparts. The
absence of global 3D stabilities of headless jets
enables us to use 2D simulations for studying most of the
large-scale dynamics of these jets.

\subsubsection{Numerical convergence of jet morphology}
\label{sec:numer-conv-jet}  

To ensure that our simulations are numerically-converged, we
perform 2D and 3D simulations of models B and B2 at various resolutions (see Tables
\ref{table1} and \ref{table3}). For model B, 3D simulations at various 
resolutions are quite similar: both maintain roughly a parabolic shape, as seen in
Fig.~\ref{Density_B}. The jets propagate at a very similar velocity.
The jet of model B-3D-hr is a bit wider due to the increase of $N_{\theta}$-resolution 
and the ability to resolve better the interface between the jet and
the pre-carved funnel. Our 2D simulations reproduce the main features
seen in 3D simulations reflecting the near absence of 3D instabilities
in headless jets (see Appendix~\ref{sec:2d-simulations-model}).

Likewise, Fig.~\ref{Density_LF_B2} shows that the simulations of model
B2 at different resolutions show very similar jet shapes and jet
propagation velocities. The jets show conical shocks, seen as sharp
jumps in density and Lorentz factor. In
Appendix~\ref{sec:2d-simulations-model}, we show that these features
are present in both 2D and 3D simulations of model B2 and therefore
are not inherent to 3D.  Because the shocks are oblique, most of the
jet flow remains super-fast magnetosonic even after passing through
the shocks.  The first set of shocks at $z \sim 100r_0$ occurs as the
jet accelerates to the point that it drops out of lateral causal
contact, i.e., its side-ways expansion becomes super-fast
magnetosonic. The shock slows down the jet and brings it back into
lateral causal contact with the ambient medium. The jet then
collimates off the ambient medium, converges and accelerates into the
jet axis.  Note that this acceleration occurs due
to jet sideways \emph{contraction}, which is the opposite of the
standard jet acceleration due to sideways \emph{expansion}
\citep{2006MNRAS.367..375B,komietal2007,sashaetal2008,2009ApJ...699.1789T},
and we discuss a possible reason for this unexpected behaviour 
in Sec.~\ref{sec:discussion}.

Once the motion toward the axis becomes
super-fast magnetosonic, a second set of conical shocks develops, as
seen at $z\sim250r_0$, and slows down the transverse motion to
sub-fast magnetosonic speeds. In fact, a small central part of the jet (closest
to the axis) slows down so much that its \emph{net} velocity becomes
sub-fast magnetosonic. This region, marked by the magenta line, shows
signs of 3D magnetic instabilities, as
suggested by the irregular streamline shape seen at $x\simeq0$, $z\simeq300r_0$ in the
bottom-right panel of Fig.~\ref{Density_LF_B2}.  As we discuss in
Sec.~\ref{3D_convergence_headless}, these 3D instabilities dissipate a
very small fraction of jet magnetic energy into heat.

\begin{figure}
\includegraphics[width=4.15cm, angle=0]{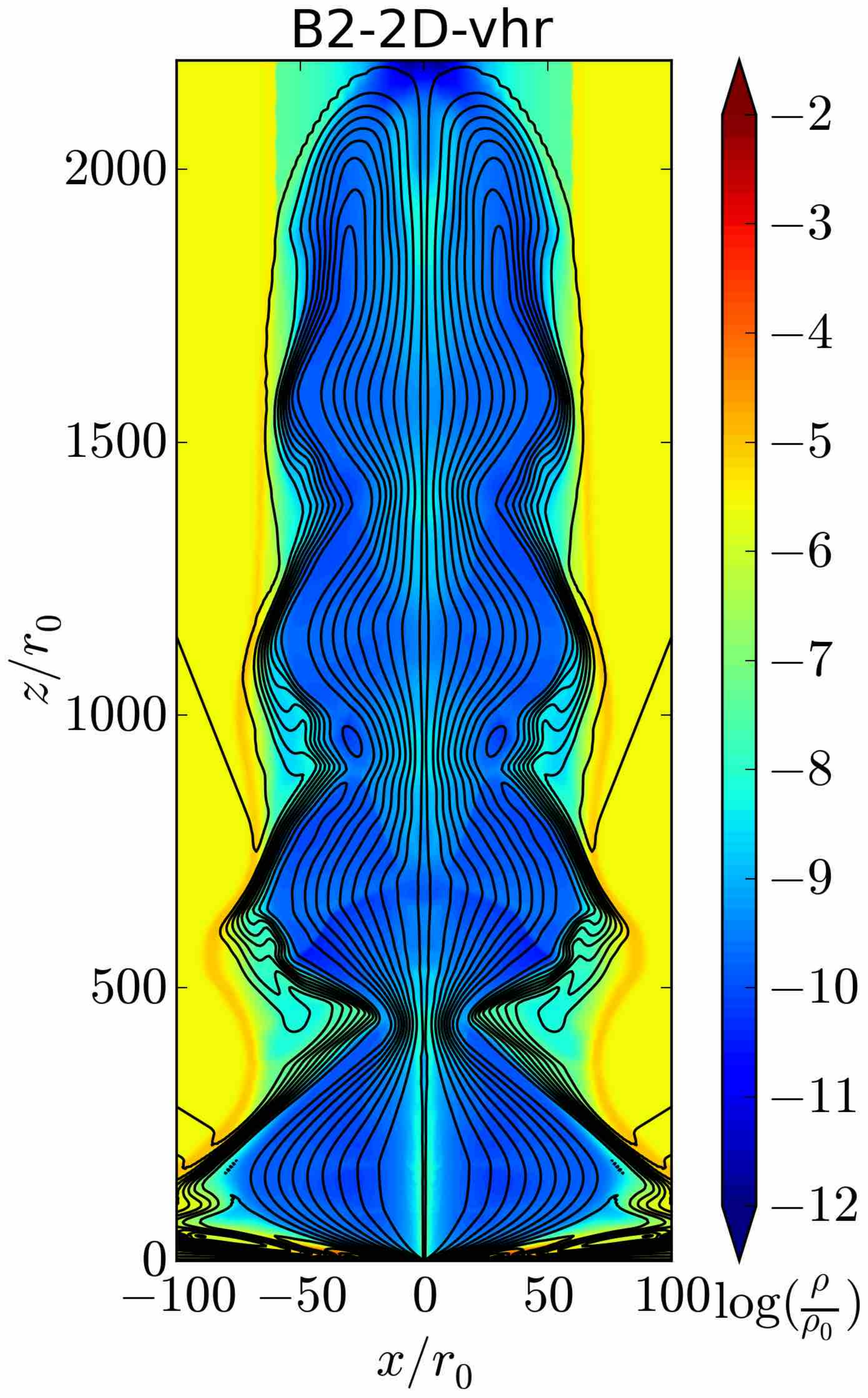} 
\includegraphics[width=4.15cm, angle=0]{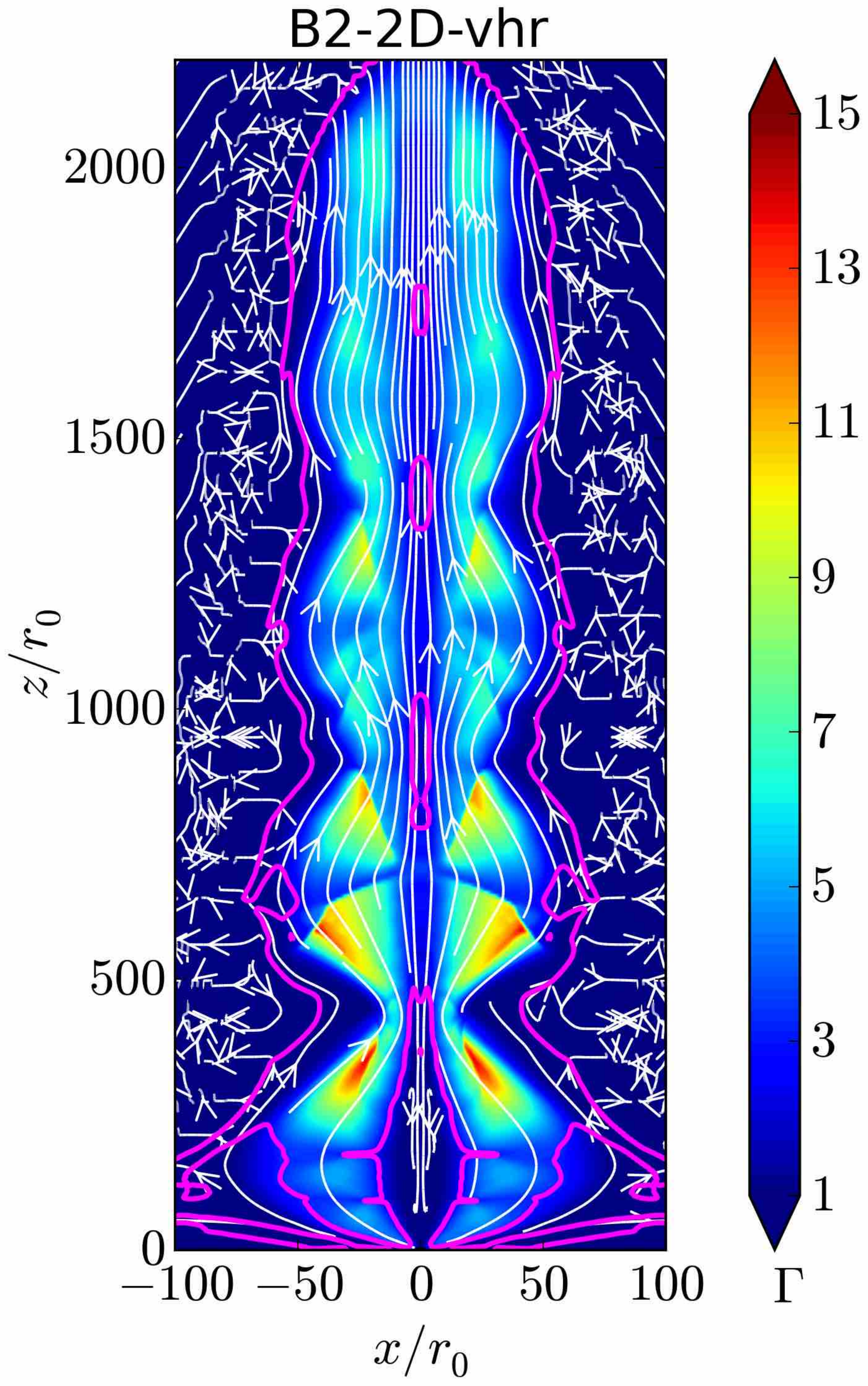} 
\caption{Density contour (left panel) and Lorentz factor 
contour (right panel) of the B2-2D-vhr simulation at the final 
snapshot, $t \approx 2200 r_0/c$; note that the vertical scale is compressed 
relative to the horizontal. Black lines in the 
left panel show poloidal magnetic field streamlines. White arrows 
in the right panel show velocity streamlines and magenta
lines show the location of the fast-magnetosonic surface.
A series of conical shocks is seen throughout the cylindrical funnel, 
although the ones farther away from the central object are less 
well-resolved.}      
\label{Density_LF_B2-2D-vhr} 
\end{figure}

After the second set of shocks, the jet bounces back and accelerates, and the shock
structure repeats itself.
In fact, at a higher resolution, a third set
of conical shocks emerges around $z \sim 500 r_0$ for the 
B2-3D-hr jet, as seen in the
bottom right panel of Fig.~\ref{Density_LF_B2}. Even higher
resolutions and longer run times, which we are able to achieve in 2D,
point to the emergence of a regular, periodic shock structure formed
by the bouncing jet that cannot come to terms with being confined by
the cylindrical funnel, see Fig.~\ref{Density_LF_B2-2D-vhr}.

The result is a wavy jet shape, with a
pronounced ``sausage'' or ``breathing'' ($m=0$) mode.
Related jet oscillations have been observed in the context of
magnetized jets confined by a single power-law flat pressure profile
medium analytically \citep{lyubarski2009} and numerically
\citep{mizunoetal2015,
  komietal2015}. We compare and contrast these results to ours in Sec.~\ref{sec:discussion}. 

\subsubsection{Energy dissipation} \label{Dissipation_headless}

\begin{figure*}
\begin{tabular}{cc}
\includegraphics[width=5.5cm, angle=0]{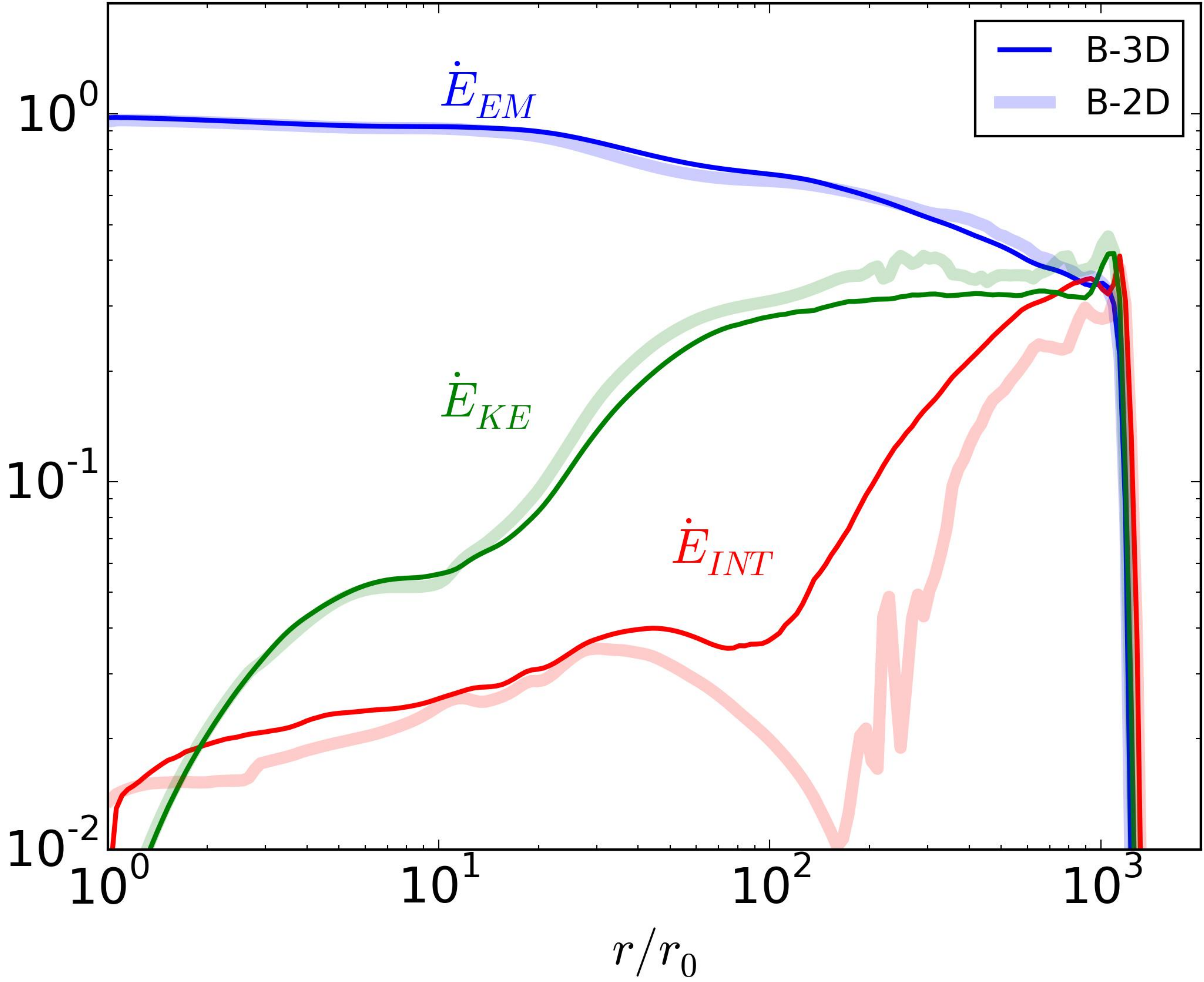} &
\includegraphics[width=5.5cm, angle=0]{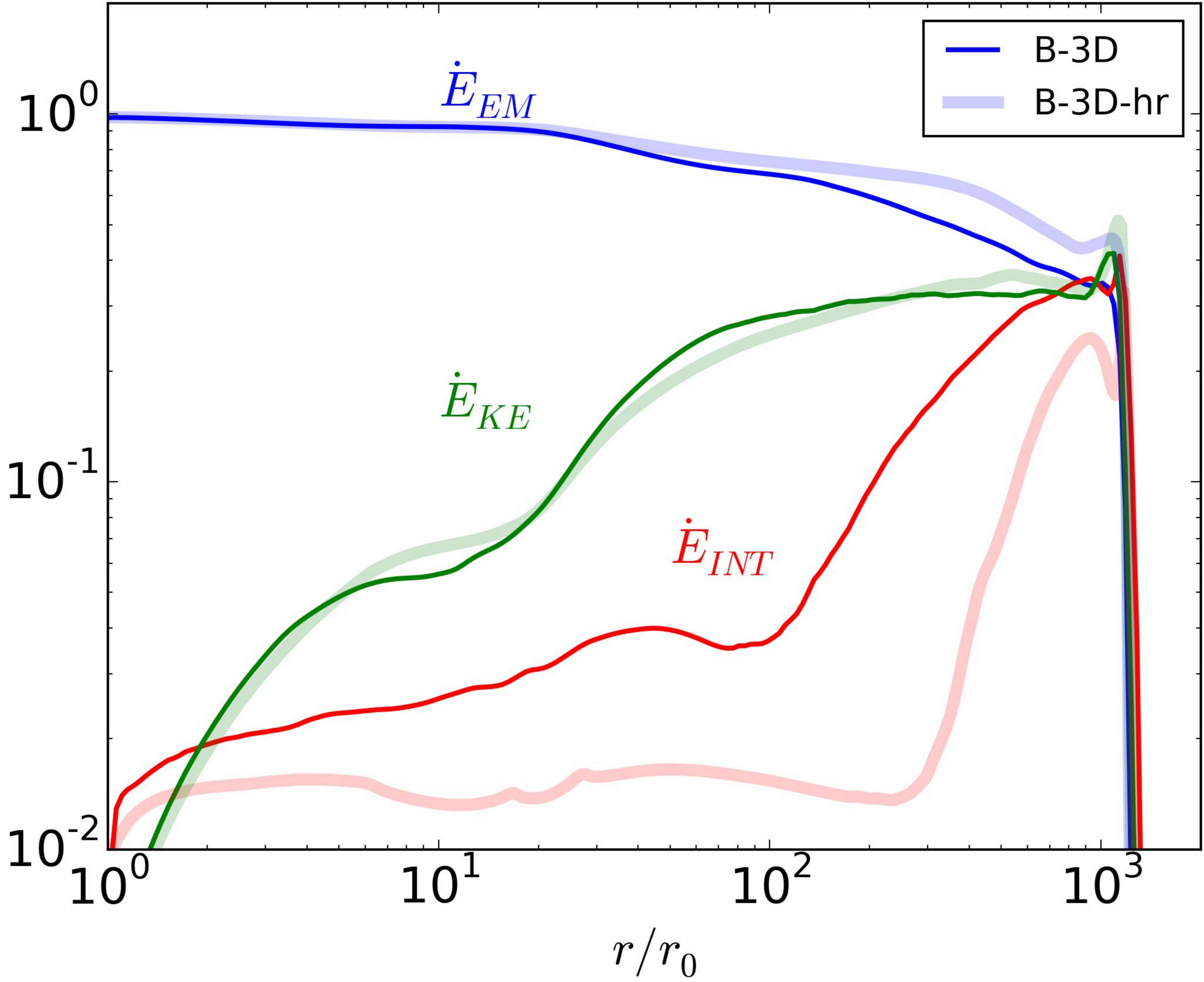} \\
\includegraphics[width=5.5cm, angle=0]{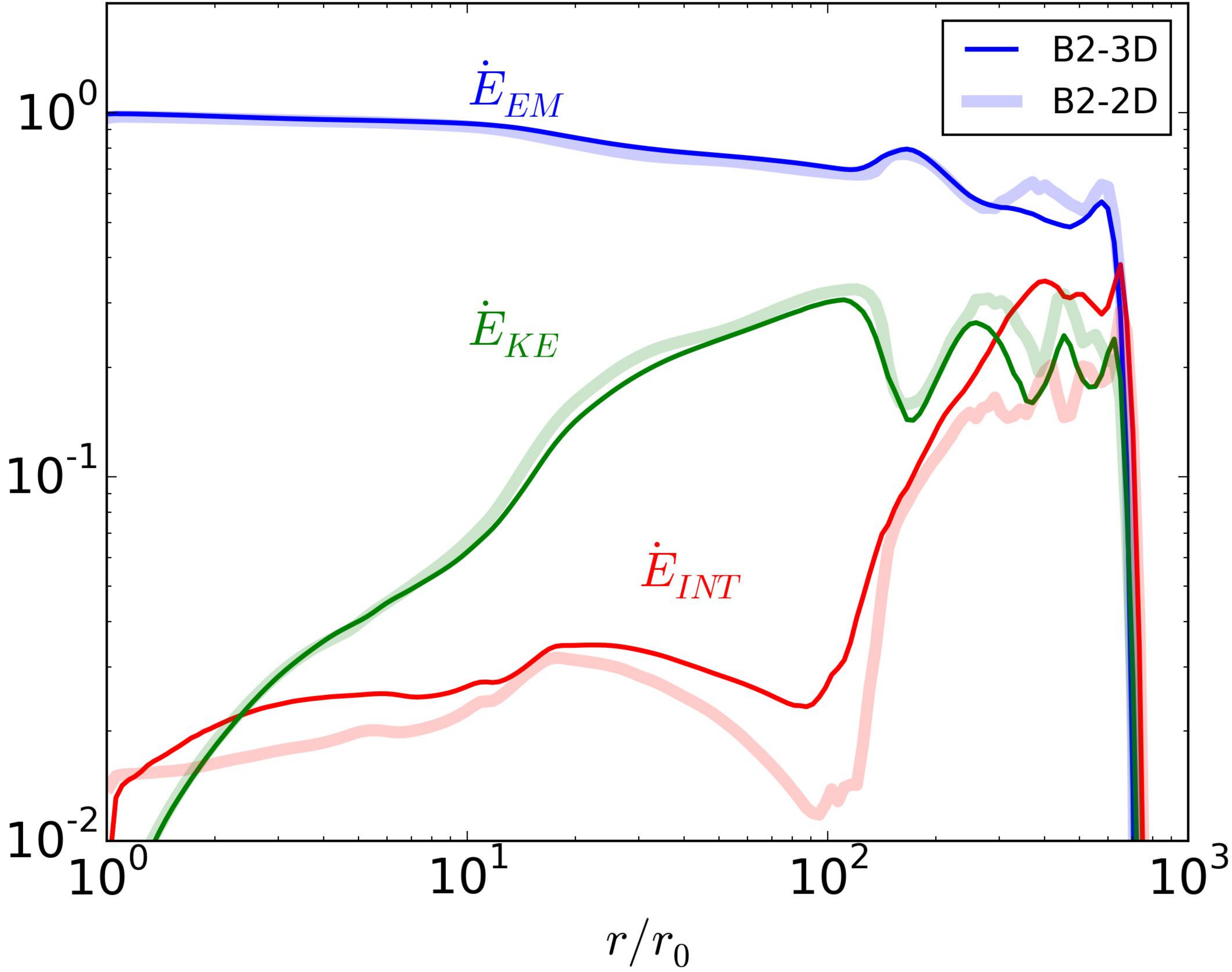} &
\includegraphics[width=5.5cm, angle=0]{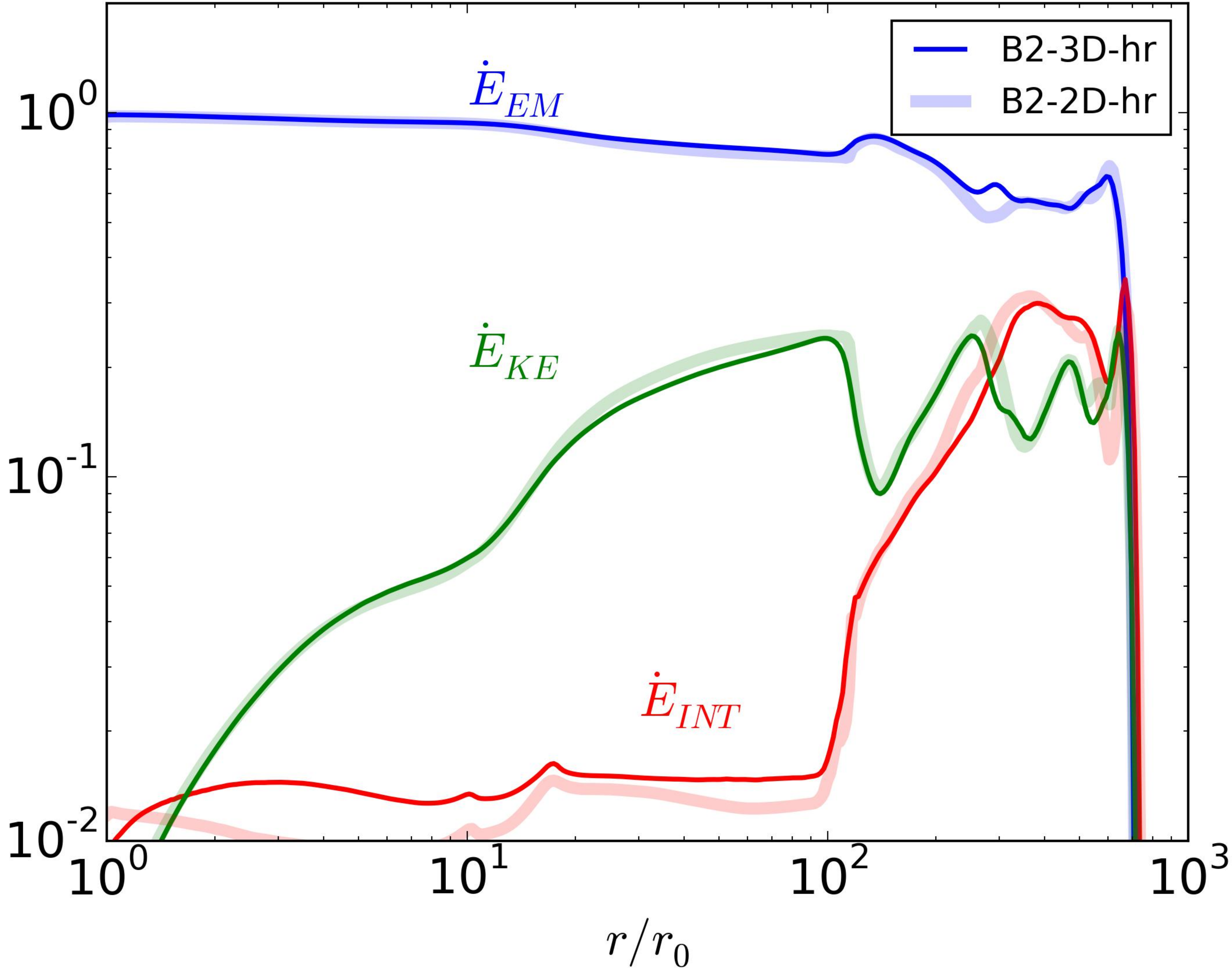} \\
\end{tabular}
\caption{Different components of the energy flux (see labels) for 3D and 2D
simulations (see legends) of model B ($t \approx 1200 r_0/c$)
and B2 ($t \approx 700 r_0/c$) in the top and bottom panels, respectively. 
The top-left panel shows that model B exhibits similar levels of electromagnetic
energy dissipation both in 2D and 3D, with an increase in internal
energy close to $r\sim 100 r_0$. However, the top-right panel shows
that this increase goes away at higher 3D resolution, indicating that
the jets in model B show no dissipation (see also the main text and
Appendix~\ref{sec:2d-models-headed}), but only a decrease of electromagnetic 
energy in order to accelerate the jet (not to heat it up).  Model B2 exhibits
oscillations in the kinetic energy due to the presence of
multiple conical shocks, which slow down the jet, decrease the
kinetic energy and convert it into electromagnetic and internal
energy: see, e.g., the increase of $\dot{E}_{\rm EM}$ and
$\dot{E}_{\rm INT}$ at $\sim 150 r_0$ (bottom-left panel). The 
bottom-right panel shows the jet energy content for higher 
resolution runs. By comparing the bottom-left and bottom-right 
panels, we find that simulations B2-3D and B2-3D-hr show quite 
similar features in their energy content, indicating that they are numerically 
converged. Also the energy contents of the jets in B2-3D-hr and
B2-2D-hr simulations (shown in the bottom right panel and carried
out at the same $N_r \times N_{\theta}$ resolution) are
almost identical, indicating that the contribution of 3D instabilities
to dissipation in these jets is subdominant. }
\label{Edot_B}
\end{figure*}

In this section, we analyze our headless jets in terms of their energy
flux content (for a similar analysis of headed jets, see Section
\ref{Dissipation} and Figs.~\ref{Edot_A_first} and
\ref{Edot_A_second}). 
To isolate the role of 3D effects, we compare the simulations B-3D and
B-2D, in the top-left panel of Fig.~\ref{Edot_B}. In both 2D and 3D,
we see similar behaviour of the different energy flux components.
There is a steady increase in the kinetic energy at the expense of the
electromagnetic energy. This is expected, as the headless jets
accelerate easier than the headed ones, and reach higher
velocities. The internal energy shows a stronger increase starting at
$r\sim 100 r_0$. However, as the top-right panel of
Fig.~\ref{Edot_B} shows, at higher resolution in 3D, this
internal energy increase becomes substantially suppressed (see also
Appendix~\ref{sec:2d-models-headed} for a detailed 2D
study). This indicates that this increase is mostly due to low
numerical resolution, which shows up as dissipation at large
distances. There are no 3D instabilities evident in this model, and
the jet shape shows azimuthal symmetry: neither the jet head nor the
body wobbles or is perturbed. This is consistent with our analysis of
global instabilities in Sec.~\ref{sec:devi-from-axisymm}.

The change in the shape of the jet-confining wall into a cylinder in
model B2 leads to qualitatively different jet behaviour, as seen in
bottom-left panel of Fig.~\ref{Edot_B}. Initially, the electromagnetic
energy is converted to kinetic energy as the jet accelerates, just as
we saw in model B. However, the situation changes at the break in the
funnel shape.  The bottom panels of Fig.~\ref{Edot_B} show that at
different resolutions both our 2D and 3D simulations display a similar
increase in the internal energy at the position of the break,
$r\gae 100 r_0$. This conversion to internal energy occurs at the
conical shocks, which are seen as jumps in the density and Lorentz
factor in Fig.~\ref{Density_LF_B2} (see also
Sec.~\ref{sec:numer-conv-jet}). The shocks slow down and heat up the
jet, and compress the magnetic field. The bottom-left panel of
Fig.~\ref{Edot_B} shows that in simulation B2-2D the kinetic energy
decreases at the first shock at $r\sim 100 r_0$ (a ``dip'' in the
curve), while the electromagnetic energy increases (a ``bump") 
and the internal energy quickly rises. An oscillating
pattern of the kinetic energy flux as a function of distance from the 
central object, as the jet encounters the second and third shock (less
well-resolved), is also evident in the bottom panels of 
Fig.~\ref{Edot_B}. The B2-3D jet shows the same features, while
exhibiting azimuthal symmetry (Sec.~\ref{sec:devi-from-axisymm}). This
robustly points to energy dissipation via conical shocks, which slow
down the jets by converting the kinetic energy into heat and
compression of the magnetic
field. This is in contrast to headed jets, where the dissipation
occurs due to 3D instabilities, as evident by not only the increase in
internal energy, but also by the wobbling/asymmetric nature of the
jets and the emergence of the irregular magnetic field structure and
the current sheets.

To quantify the energy dissipation in headless jets, we consider the
variation of entropy. In
steady state, absent dissipation, entropy remains constant; it
increases in the presence of dissipation.
Fig.~\ref{B2-3D_streamline} shows the radial profiles of entropy and
Lorentz factor along a representative streamline in our model
B2-3D-hr. The streamline is shown in the figure inset.  At small
radii, the jet is cold and strongly magnetized. The application of
density and internal energy ``floors" in this region leads to deviations
from constancy in the entropy profile. By $r\gtrsim20r_0$, the
magnetization drops due to the bulk flow acceleration by the magnetic
fields, the density and internal energy floors are no activated,
and the entropy profile settles to a constant, $s\approx-3.5$ that
corresponds to essentially a cold jet. At the location of the shock,
$r\sim100r_0$, the entropy rises sharply and the Lorentz factor drops.
Beyond the shock, the jet re-accelerates and the Lorentz factor
smoothly increases.

The jet sharply decelerates at $r \sim 200 r_0$ as it passes through
the second conical shock. The entropy additionally increases at this
shock, reflecting the shock heating experienced by the jet. By
$r\sim 250r_0$, the Lorentz factor decreases to essentially
non-relativistic values, $\Gamma\lesssim1.5$. This occurs when the
streamline enters a sub-fast magnetosonic region, indicated by the
magenta line, as seen on the figure inset. The flow in this small region is
unstable to the internal 3D magnetic kink instability, which
dissipates magnetic energy into heat. This dissipation is evidenced by
the increase in the entropy of the streamline. Once the streamline
exits the dissipative region at $r\sim350r_0$, the entropy slightly
decreases. While we would not expect such a decrease along a
streamline in the steady state, we suggest that the time-dependent
dissipation and irregular flow in the sub-fast region
could cause this behaviour.

\subsubsection{3D instabilities and heating in headless jets} \label{3D_convergence_headless}

\begin{figure}
\includegraphics[width=8.2cm, angle=0]{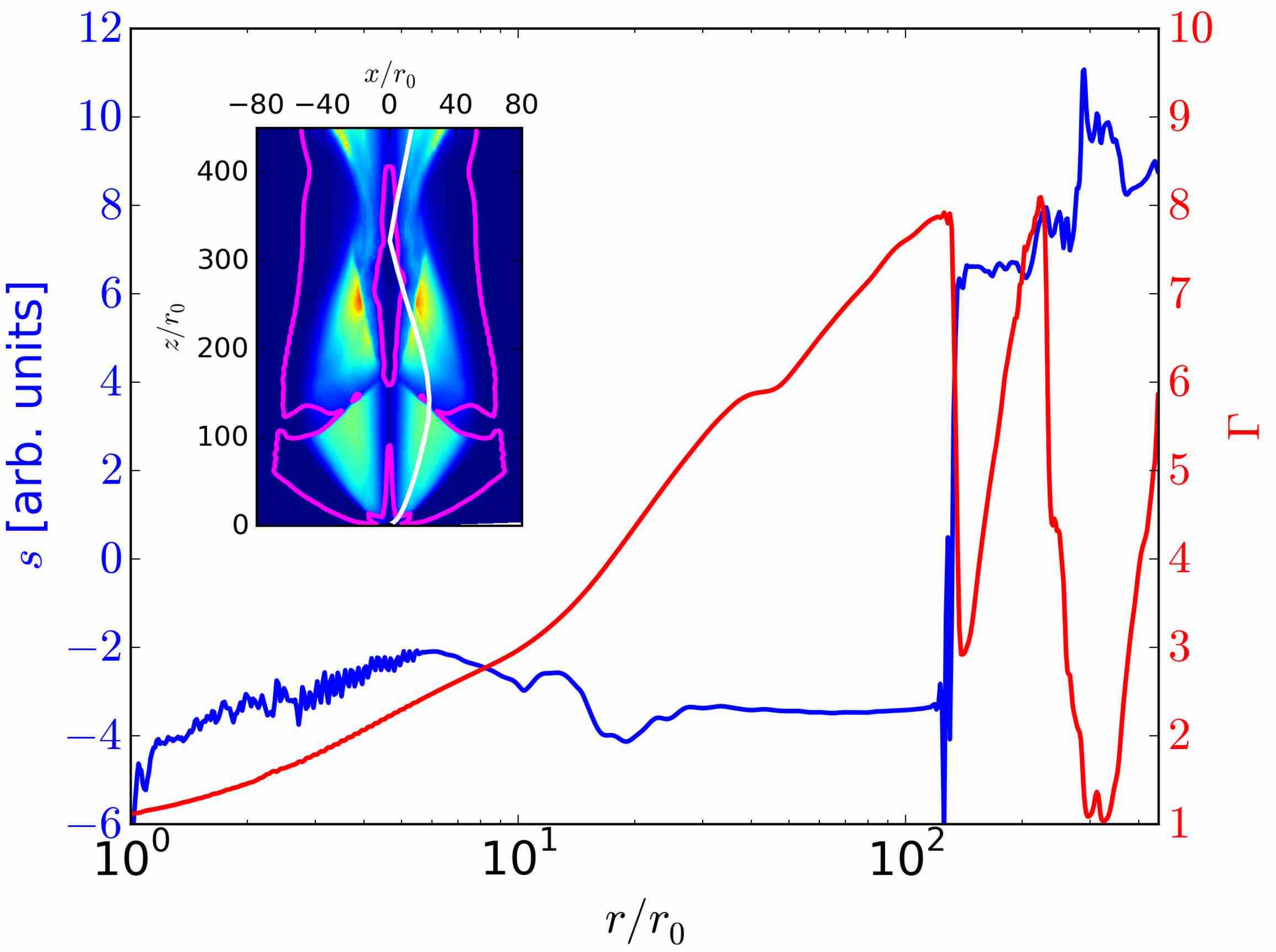} 
\caption{Entropy (blue) and Lorentz factor (red) along a velocity
streamline of the jet in the B2-3D-hr simulation (note different
y-axes). The inset shows the bottom-right panel of
Fig.~\ref{Density_LF_B2}, and the white line shows the particular
chosen velocity streamline. Entropy increases and velocity decreases
at $r \sim 100 r_0$ and $\sim 200r_0$, the locations of the first
and second conical shocks. Outside the shocks, the Lorentz factor
smoothly increases, reflecting the magnetic acceleration of the
jets. The sharp jumps in the entropy and Lorentz factor reveal
and confirm the presence of conical shocks. Note that at $r\sim 250r_0$, the
streamline enters a sub-fast region in which 3D magnetic
instabilities dissipate magnetic energy into thermal energy, as
evidenced by a temporary increase in the entropy of the streamline 
(see text for more details).}
\label{B2-3D_streamline} 
\end{figure}

\begin{figure}
\includegraphics[width=8.2cm, angle=0]{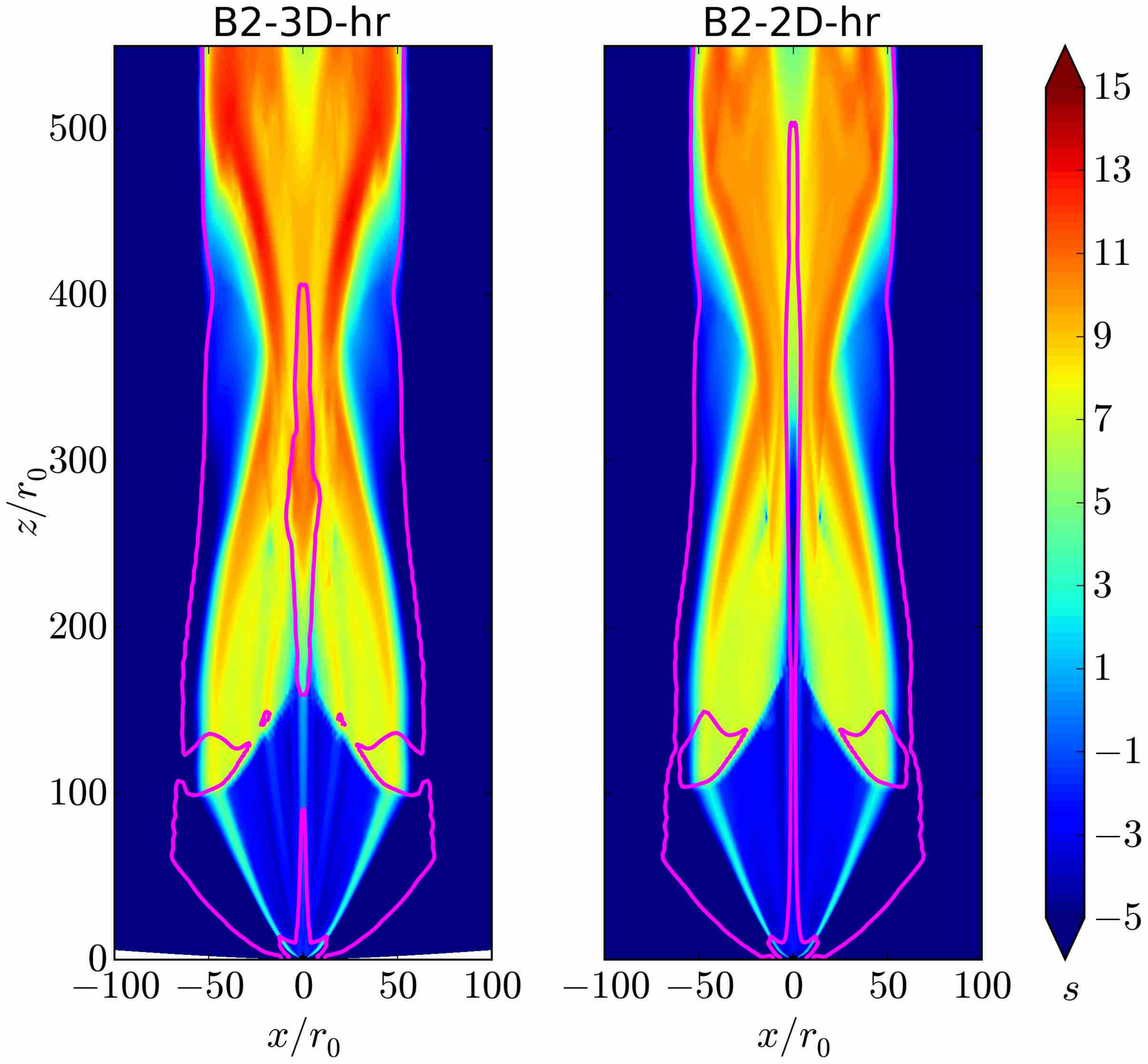} 
\caption{2D entropy contours (arbitrary units)
of models B2-3D-hr (left panel) and B2-2D-hr (right panel) 
at the final snapshot of the B2-3D-hr simulation ($t \approx 700 r_0/c$). 
The magenta lines show the position of the fast magnetosonic surface in both panels.
The entropy maps of these two models are almost identical. This shows that 
energy dissipation in the 2D and 3D runs is essentially the same, so there is 
no large contribution from 3D instabilities.  There is a small fraction of energy that could be 
dissipated in the sub-fast region that develops along the jet axis a few $\times 10^2 r_0$ 
away from the central object in the 3D-hr run (left panel) and also the 3D run. 
Since most of the jet power travels along the jet edges, energy
contribution of the sub-fast region is quite small. }      
\label{B2-zoom_in} 
\end{figure}

Another way to study jet dissipation is through a map of
entropy in the jets, shown in Fig.~\ref{B2-zoom_in} for our
high-resolution simulations of model B2.  Because the jets start out
cold, the entropy near the origin is low both in the left panel,
showing the 3D simulation, B2-3D-hr, and in the right panel, showing
the 2D simulation at the same resolution, B2-2D-hr. The entropy
sharply increases in both panels at the
first conical shock at $r\sim100r_0$ (as seen in Fig.~\ref{B2-3D_streamline}), 
reflecting the fact that the dissipation at the conical shocks is 
inherently a 2D phenomenon. 

Even though most of the energy dissipation in our model
B2 occurs via conical shocks, in certain cases 3D instabilities
might still play a role. Note that in models B2-3D and B2-3D-hr there
is a small sub-fast region that develops around the jet axis at
$z\sim300r_0$, as seen in Fig.~\ref{Density_LF_B2}.  In analogy with
headed jets, because it is sub-fast, this region `knows' about the
impending catastrophe -- that the jet is running into an obstacle, in
this case the walls of the cylinder -- and tries to compress in the
anticipation of the collision.  This compression is what drives the
jet to be 3D magnetic kink unstable. As we also saw for headed jets,
these unstable regions can turn dissipative. Indeed, our results for
model B2 (see Sec.~\ref{Dissipation_headless}), suggest the presence
of additional dissipation in this region. 
In the left panel of Fig.~\ref{B2-zoom_in} one can clearly see that  
the narrow sub-fast region centred around $x\simeq0$, $z\simeq300r_0$
shows high-entropy that is indicative of dissipation.
Because this sub-fast region in our 2D simulation is low-entropy, as seen in the right
panel of Fig.~\ref{B2-zoom_in}, this suggests that internal 3D instabilities are
responsible for the dissipation of energy in the sub-fast region.
However, since most of the jet power travels along the jet edges, 
we do not expect this region to be the dominant contribution to the overall jet dissipation. 
In fact, the internal energy flux in the sub-fast region of the B2-3D-hr jet
is at the level of $\sim2$\% of the total energy flux.
Future work will focus on studying this additional source of
dissipation in the sub-fast regions of headless jets. 

Note the existence of high-entropy regions in both panels of Fig.~\ref{B2-zoom_in}, 
near the jet edges at $z\gtrsim 200r_0$. They most likely reflect the numerical
dissipation in the sheet of current returning on the surface of the
jet and they emerge due to insufficient numerical resolution
across the jet, where the jet becomes narrow and difficult to resolve
transversely. Indeed, we find spatial alignment of these regions
and the current sheet at which the poloidal magnetic flux reverses. 
Properly capturing dissipation near the jet edge is
inherently difficult for numerical methods such as ours that employ
simple Riemann solvers such as the local Lax-Friedrichs (LLF) solver: in fact,
dissipation near a density discontinuity, such as near the jet edge,
might not converge away with increasing resolution, necessitating the
use of more advanced Riemann solvers such as
Harten-Lax-van-Leer-Discontinuities (HLLD) solver to properly capture
it \citep{ressleretal2017}. We leave this to future
work. 

\section{Discussion}
\label{sec:discussion}

How relativistic jets dissipate a large fraction of their energy and
convert it to radiation remains an important unsolved problem in high-energy
astrophysics.  Magnetized relativistic jets are prone to various
instabilities including the magnetic kink instability, which is
thought to be the primary candidate for dissipation in the jets
\citep[e.g.,][]{spruitetal2001, 2003astro.ph.12347L, gianniosandspruit2006, 
molletal08, 2011MNRAS.416.2193N,sironietal2015}. Jets in nature propagate
over as many as 10 orders of magnitude in distance. This gives plenty
of room for the instabilities to develop. Studying their development
is an inherently 3D multi-scale problem that involves jet
acceleration, collimation and interaction with the ambient medium.
Different elements of this problem have been studied extensively via numerical simulations
\citep[e.g.,][]{nakamuraetal2007, mckinneyandblandford2009, 
mizunoetal2012, mignoneetal2013,
  2014ApJ...781...48G, porthandkomi2015, omerandsasha16,
  sashaandomer16, singhetal2016}.

To make numerical studies of jets feasible, the standard approach
is to reduce the dynamic range covered in simulations.  Primarily
because of this, the development of magnetic instability in jets has
been extensively studied via 3D numerical MHD simulations that inject
relativistic jets at large distances from the central compact object
(\citealp[e.g.,][]{2014ApJ...781...48G}).  
A crucial parameter that determines jet stability is the magnetic
pitch, or the ratio of longitudinal to transverse magnetic field
components in the jet, $B_p/B_\varphi$ \citep[see,
e.g.,][]{2000A&A...355..818A,2014ApJ...781...48G}. Unlike many
previous studies, where the magnetic pitch is a free parameter set by
the ``jet injection'' boundary condition (\citealp[e.g.,][]{nakamuraetal2007}), 
we follow a different
approach and initiate the jets the way nature does it, via the
rotation of the central magnetized compact object
(\citealp[e.g.,][]{blandfordandznajek1977, 2006ApJ...641..103H,mckinneyandblandford2009,komietal2007,2009ApJ...699.1789T,2010ApJ...711...50T,2010NewA...15..749T,sashaetal2011,
omerandsasha16, sashaandomer16}). This fixes the magnetic pitch
to its natural value and allows us to study jet stability from first
principles.

The kink instability triggers magnetic reconnection, which in turn leads to magnetic energy
dissipation that could readily explain the bright multi-wavelength
emission attributed to jets in various astrophysical systems ranging
from short- and long-duration GRBs to AGN, tidal disruption events, and pulsar wind
nebulae \citep{omerandsasha16}. The focus of this work is on extending
the simulated dynamic range and, for the first time in the context of
AGN jets, following the jets from their birth at the surface of the
central compact object out to their \emph{interaction} with the external
medium (ISM).  

In the first attempt at addressing this important problem, we adopt
several simplifying assumptions that make the current-generation of
our simulations
computationally feasible: (i)~we reduce the range of jet acceleration
and collimation zone from $5{-}6$ orders of magnitude in nature down
to $2$ orders of magnitude; (ii)~we consider powerful jets that
propagate the fastest in order to reduce the computational cost;
(iii)~we ignore gravity because it is not important at large distances
where the instabilities take place and over the short time-scales
simulated; (iv)~we assume a monopolar magnetic field geometry at the central
object; and (v)~we approximate ambient medium via an initially prescribed
density profile.  While this work focuses on relativistic AGN jets,
our results can be extended to other astrophysical systems with
relativistic jets.

In this paper we studied relativistic jets that interact with the
ambient medium in order to understand how and whether this interaction
triggers jet energy dissipation. We did this via 2D and 3D
relativistic MHD simulations with different external medium density
distributions (see Tables~\ref{table1}--\ref{table3} and
Fig. \ref{Model_schematics}).  Considering different types of ambient
medium allowed us to study two distinct classes of jets: headed and
headless jets.  Headed jets drill through and push the ambient
medium. Headless jets propagate through a previously evacuated
funnel. Although this distinction is not always clear-cut (since
realistic astrophysical systems could involve a combination of these
two types of jets), considering the two limiting cases helps us to
elucidate the jet physics in the simplest circumstances.

\subsection{Headed jets} 

Headed jets were found to be unstable in previous 
3D MHD simulations of
mildly relativistic jets that followed them from the central object 
\citep{omerandsasha16, sashaandomer16}. These simulations
did not set the magnetic pitch as an initial condition: the pitch was self-consistently 
determined by the rotation of the central object. These authors found that
a stable jet that interacts with a density profile,
$\rho \propto r^{-\alpha}$ with $\alpha \lae 2$, tends to turn unstable
as it propagates farther away from the compact object
\citep{sashaandomer16}. However, if this jet interacts with a profile
with $\alpha \gae 2$, then it would remain stable (see also
\citealt{porthandkomi2015} who considered stability of headless jets
as a function of the power-law index of the ambient pressure profile). These conclusions 
were found for mildly relativistic jets,
$\Gamma\sim1$.  

We found that headed jets that run into a density break from steep to
flat density profile (we adopt for the flat profile $\alpha=1$, as suggested
observationally, \citealt{russelletal2015}) switch from parabolic to
cylindrical shape. Recent observations of jets
in the M87 galaxy \citep{asadaandnakamura2012} and several other AGN
\citep{2016arXiv161104075A,tsengetal2016} also indicate a 
switch in the geometry of the jet close to the Bondi radius.
We note that whereas observations prefer a conical jet shape
beyond the Bondi radius, our simulated jets tend to collimate into
cylinders. It is possible that this difference comes from the neglect of
gravity in our simulations: without gravity, the hot jet exhaust flows
back along the jet and equilibrates the confining pressure. 
Because the pressure is uniform along the jet, it collimates the
jet into a cylinder. The effect of gravity would be to
gravitationally stratify the confining gas. This would lead to pressure
decreasing with increasing distance from the source and possibly 
resulting in a finite opening angle of the jet.

We also found that at the transition between the two geometries the jets recollimate,
i.e., abruptly bend on to the jet axis. At this recollimation point,
the jets exhibit a `pinch' or a `waist' (see also
\citealt{omerandsasha16}). 
Such places in the jet are natural sites for the development of
  the kink instability: because the instability growth time-scale is
  essentially the sound crossing time across the jet, the natural
  place for the instability to develop is where the jet cross-section
  is the smallest, i.e., near the recollimation point.
Indeed, we find that at the recollimation point, our jets develop an internal magnetic kink
instability that twists the internals of the jet on to itself and
generates current sheets that dissipate $\gtrsim50$\% of jet magnetic energy into
heat over a factor of few in distance.  

Dissipation in jets occurs on the small, resistive scales.  These
  dissipative scales in jets in nature are small compared to the jet
  size. In fact, they are so much smaller than the size of the
  numerical cells in our simulations that it is unrealistic to hope to
  capture them in a global simulation. In fact, while we would like to
  include non-ideal effects into our simulations explicitly, we would
  not be able to do so realistically: any dissipative scale we would
  include would still be much larger than in nature. Because of this,
  we do no include explicit resistivity into our simulations. The
  numerical scheme instead relies on the fact that when two oppositely
  oriented magnetic field lines get squished into a single numerical
  cell, they annihilate and their energy gets converted into heat. So,
  our simulations have a dissipative scale length comparable to the
  grid cell size. The hope is that if the numerical resolution is high
  enough, the dissipative scale becomes so much smaller than the
  global scales involved in the problem that it no longer matters. We
  studied the convergence of our simulations with numerical resolution
  in 2D and 3D by doubling the resolution in every dimension. We found
  that in 3D a higher resolution leads to a \emph{larger} amount of
  energy dissipated by the kink instability
  (Sec.~\ref{Dissipation}). This indicates that the dissipation
  process is robust and the fraction of magnetic energy that is
  dissipated into heat can be of order unity. As more computational
  resources become available, in the future we will attempt to carry
  out simulations at even higher resolutions. This will help us to
  answer an important question of whether the simulation results
  converge to a well-defined answer in the infinitely high resolution
  limit that is closest to reality in the sense of the dissipation
  length scale being small. Local kinetic
  \citep[e.g.,][]{sironietal2015,2016MNRAS.462...48S,2016arXiv160505654L}
  and resistive MHD \citep[e.g.,][]{2016MNRAS.460.3753D} simulations
  could inform us about the best way of including physical dissipation
  in global jet models.

The recollimation point moves much slower than the jet fluid.
It is tempting to associate the recollimation point and the associated
dissipation with
quasi-stationary or ``slow" moving features in the core of the M87 jet
(e.g., HST-1 and other knots; \citealp{birettaetal99,meyeretal2013}) and 
jets in other systems \citep{jorstad_agn_jet_2005, 
listeretal2013, cohenetal2014}.
At larger radii, our jets develop large-scale bends and asymmetries
characteristic of the external kink instability and powerful \citet{1974MNRAS.167P..31F}
type II AGN such as Cygnus~A \citep[see also][]{sashaandomer16}.

An important new element of our work is the presence of acceleration
and collimation zone extending from the central compact object to the
ISM. This allows our jets to reach super-fast
magnetosonic velocities. Because of this the jet outruns the fast
magnetosonic waves, ensuring that no signals can propagate
backwards to the central compact object, just as expected in nature
(see Sec.~\ref{sec:model-ambi-medi} for discussion). Once such a super-fast
jet runs into the ISM, it develops a shock or a series of shocks,
which decrease its velocity.
We find that the fraction of electromagnetic energy flux
dissipated via such shocks is rather low in
headed jets, $\lesssim 5$\% (see
Appendix~\ref{sec:2d-models-headed}). 
Nevertheless, 
the shocks slow down the flow
substantially, allowing for the instabilities to proceed in the
first place (see Section \ref{Dissipation}).
Most of the energy flux,
$\gtrsim50$\%, is dissipated through magnetic reconnection in the
current sheets formed by the internal magnetic kink
instability ; there is evidence in Figs. \ref{Final_density_modelsA} 
and \ref{Density_A2-3D-hr} that magnetic field lines are perturbed, which reflects both the 
emergence of irregular magnetic fields in the jets and the large-scale deviations 
of the jets out of the image plane, both caused by the 3D magnetic kink instability 
and ideal locations for magnetic reconnection to occur. 

Interestingly, in addition to dissipation co-spatial with the
recollimation feature, our headed jets develop an unstable, sub-fast
magnetosonic region near the jet axis at distances comparable to
the location of the density break in the ambient density (see
Fig.~\ref{Sequence-A2} and Section \ref{Dissipation}). 
It is tantalizing that this sub-fast region 
can even in some cases extend to distances smaller than the location of 
the density break, and can appear even in the cases when there is no density break (see
Fig.~\ref{Sequence-A}). This could provide a potential source 
for energy dissipation along the central region of the jet
\citep{omerandsasha16}. Even though the amount of energy flowing
through and dissipated in this region is small, the axial
dissipation associated with it could power the low-frequency jet radio
emission and contribute to the formation of the flat radio spectrum
of jetted AGN. Intriguingly, this feature appears to be robust, as  we
find an indication for a similar feature in our headless
jets, as we discuss below. Our simulations suggest that our jets 
are initially slow at the spine and fast at the edge, and are 
surrounded by a slower sheath. With increasing distance, as the jet accelerates, 
the relative fraction of jet cross-section occupied by the central slow spine shrinks, 
which might lead to a more uniform transverse luminosity profile of the jet.
Our findings might have relevant connections with the observational evidence 
of limb brightening in Markarian 501, M87 and other jets 
\citep[e.g.][]{girolettietal2004,kovalevetal2007}.

\subsection{Headless jets} 

To model headless jets, we carved out in the
ambient gas a very low-density smooth polar escape route for a jet, as seen
in the middle panel of Fig.~\ref{Model_schematics} (model B). This
allows the jets to propagate freely along the polar region without the need
to push any external medium, in contrast to our study on headed
jets. Our headless jets easily accelerate to high Lorentz factors and
maintain a parabolic-like shape. The jets remain essentially
axisymmetric. We find that the electromagnetic energy transforms to
kinetic energy, and the jets accelerate to a Lorentz factor
$\Gamma\sim10$ over several orders of magnitude in distance, without undergoing
significant internal energy dissipation. There is no indication of 
magnetic kink instabilities in these jets. This is consistent with 
the findings of \cite{mckinneyandblandford2009}, who found 
their jets also to be stable in a similar scenario.

We introduced the external medium for headless jets by changing the
shape of our funnel into a dense cylinder, as seen in the right panel
of Fig.~\ref{Model_schematics} (model B2).  We made this change
approximately two orders of magnitude away from the central object.
As the jet expands and propagates, it accelerates and loses lateral causal 
contact before it encounters the cylindrical part of the
funnel. Once the jet 
hits the cylindrical funnel walls, it develops a conical shock that brings the 
jet into lateral causal contact and increases the toroidal magnetic
field strength. The stronger toroidal magnetic field causes the jet to
pinch and
recollimate toward the jet axis, where it goes through another conical
shock
before expanding again. 
This process repeats, leading to a wavy jet 
featuring many recollimation points with a rather regular spacing 
(see Fig.~\ref{Density_LF_B2-2D-vhr}). The recollimation points propagate 
very slowly, with $v \lae 0.1c$. These points move because the funnel walls, 
not completely rigid, are pushed by the jet and move extremely slowly, making the 
situation not completely static and thus allowing the recollimation points to also 
move, albeit with a small velocity. The jet remains essentially axisymmetric. 
At the expansion and contraction of the jet between
the recollimation points, two or more conical shocks are observed. 

\cite{lyubarski2009,komietal2015} studied a 2D evolution of a
magnetized jet confined by a medium of a flat pressure profile. They
found a breathing ($m=0$) mode in the jet similar to what we observe
in our model B2 jet. These works considered jets in lateral causal
contact with the ambient medium: their Lorentz factor smoothly
increased (decreased) in response to expansion (contraction) of the
jet radius.  Our simulations are the first 3D study of jet breathing
mode. They show that this mode can
naturally develop as a result of jet interaction with an
\emph{obstacle} in the ambient medium. Unlike the above works, our
jets drop out of lateral causal contact by the time they become
reconfined by the obstacle in the ambient medium. As a result, jet
radius oscillations are accompanied by \emph{conical shocks} that keep
bringing the jet into lateral causal contact and can even lead to a 3D
kink instability near the jet axis (see below).  In contrast to previous
works, we find that the jet accelerates \emph{both} at the expansion
and at the contraction.  We suggest that the acceleration occurs in
both cases due to the sensitivity of the Lorentz factor to the
poloidal curvature of the magnetic field lines \citep{sashaetal2008}.
\citet{mizunoetal2015} found a related breathing mode in their 2D
numerical simulations of over-pressured jets. As their jets adjusted to
the ambient medium, they developed shocks and rarefactions (for details 
of rarefaction acceleration, see, \citealp[e.g.,][]{aloyandrezzolla06, mizunoetal08, 
2010NewA...15..749T, komietal2010}). However, \citet{mizunoetal2015} 
assume that initially the relativistic jet is kinematically dominated for all values of the 
magnetic field they consider. In contrast, our jet is initially magnetically dominated.  
Future work will explore the physics of jet acceleration in our simulations.

Recently, a scenario in which a jet encounters an abrupt flattening of
the pressure profile at some radius, which would be somewhat analogous
to our model B2, has been presented by
\cite{levinsonandglobus2017}. These authors find strong jet radius
oscillations upon transition from the steeper to the flatter pressure
profile, which we also find as our funnel transitions to a cylindrical
shape.  However, as mentioned before, in our 2D and 3D numerical
simulations we find jet acceleration both at the expansion and
contraction (instead of only at the contraction) and resolve the
presence of conical shocks, which efficiently decelerate the jet.
This highlights the importance of accounting for discontinuities and
shocks in the jet, which can significantly affect the jet Lorentz factor.

In contrast to headed jets, we find that headless jets are stable to
3D kink instability.  The main reason is that headed jets show a
larger toroidal to poloidal magnetic field ratio, a steep poloidal
field profile, and they are slower, whereas headless jets show a
smaller toroidal to poloidal magnetic field ratio, a shallower
poloidal field profile, and they are faster since they do not have to
push against a large ambient density (e.g., \citealp{mizunoetal2012,
  omerandsasha16}). These factors affect the growth of kink
instability, making it grow very slowly for headless jets.

The lack of 3D instabilities does not mean that energy dissipation
does not occur.  We find that headless jets that propagate in a smooth
low-density funnel convert very little magnetic and kinetic energy to internal energy
(model B). However, when we abruptly change our funnel to be
cylindrical (model B2), the jets develop conical shocks at which internal
energy increases.  A conical shock abruptly slows down
the jet, decreasing its kinetic energy.  This energy is almost equally
shared towards an increase in electromagnetic energy, since the shock
compresses the magnetic field, and towards an increase in internal
energy.  The level of energy dissipation is modest, $\sim 10$ per cent
of the total jet power at the first conical shock. This level is
lower than for headed jets, where the kink instability dissipates
more energy, and will likely be even lower for jets at higher
magnetization than we considered \citep[e.g.,][]{2010MNRAS.407.2501M,2011MNRAS.416.2193N}.

Although we have been able to resolve numerically only a few conical
shocks, we expect that many more of these conical shocks develop if we
follow the jet out to larger scales, as suggested by Fig.~\ref{Density_LF_B2-2D-vhr}.  A sequence of many such
shocks would result in a higher dissipation efficiency.  It is
tempting to associate a sequence of such dissipative shocks with the
sequence of essentially equidistant knots seen in the M87 jet starting at 
the HST-1 knot \citep[e.g.,][]{meyeretal2013}.  In contrast to headed jets, our
cylindrical headless jets can dissipate energy over a large range of
distance so long as they maintain their cylindrical shape. We note that 
our jet is aligned relative to the cylindrical funnel. A misaligned jet (or a non-axisymmetric
external medium) would likely prevent the development of a long sequence of shocks 
over distance and may develop additional instabilities.

Somewhat similar to the headed jets, in addition to energy dissipation
at the location of conical shocks, there is a hint that some
dissipation could be occurring along the jet axis via internal 3D
magnetic kink instabilities in a sub-fast region (see Section \ref{3D_convergence_headless}).
This region does not develop for model B, since its jet interacts with
a very low-density ambient medium, but it develops for model B2, where the 
funnel turns cylindrical and causes conical shocks to form. 
This unstable region could be a side-product of the conical shocks, which
slow down the jet fluid and increase the strength of the toroidal magnetic
field, thereby encouraging the MHD instabilities to develop and
additionally dissipate the electromagnetic flux.  
Furthermore, we see indications that at very late times this unstable sub-fast region could 
even extend to distances a bit smaller than the location of the cylindrical transition, 
which could provide a potential site for energy dissipation along the central 
region of the jet. Although small jet energy flux (internal energy flux at a level of a percent 
of the total jet energy flux) travels through the unstable region 
along the jet axis (most of the energy flows along the jet edges), the 
additional axial dissipation could be a promising mechanism for forming the
low-frequency radio emission of flat radio spectrum jetted AGN.

\section{Conclusions}
\label{sec:conclusions}

We have performed a suite of fully 3D relativistic MHD simulations at various resolutions 
that follow two types of jets from the compact object: (i) headed jets, which interact
with the ambient medium and push it out of the way in order to propagate, and 
(ii) headless jets, which propagate freely through a previously evacuated 
very low-density funnel \citep{omerandsasha16}. The jets are launched 
self-consistently via the rotation of a magnetized compact object and are 
followed for about 3 orders of magnitude in distance. 

We find that  as the headed jets encounter a shallower
density profile, they recollimate and pinch toward the axis and
succumb to a magnetic kink instability ($m=1$ mode). This instability triggers strong, concentrated energy dissipation localised within an order of magnitude in
distance of the location 
of the density profile break. The result is a slowly propagating
recollimation point accompanied with dissipation, 
which could be associated with the slowly moving or stationary
features observed in AGN jets. 

Headless jets that encounter a sudden change in the shape of the
low-density funnel through which they propagate,
do not exhibit strong magnetic kink instabilities. Instead, they dissipate
energy at essentially equidistant conical shocks that form as the jet develops a violent
breathing ($m=0$) mode in an attempt to come to terms with the new shape of
the confining funnel. In contrast to headed jets, dissipation in headless jets, although
less efficient, can occur over a large range of distances so long as the 
funnel retains its shape. Such a series of equidistant shocks 
could correspond to the series of equidistant knots seen in the M87
jet \citep[e.g.,][]{meyeretal2013}. 

In both types of jets, a sub-fast unstable narrow region along the jet
axis develops close to (and in some cases before) the location of the
density profile break. This instability could light up helical
structures near the jet axis and account for several, apparently helical
threads in the jet of M87 upstream of HST-1 seen in a space-VLBI image
with RadioAstron and a large global-VLBI array taken at 18cm
(Savolainen et al., in preparation).

Among the simplifications of our numerical simulations, 
we have considered high-power jets, since they propagate 
the fastest in order to reduce the computational cost.  In the future,
we plan to 
extend our work to lower jet powers, and to increase the spatial range
and duration of our simulations. 
We will also study the dependence of our results on the value of 
the jets' initial magnetization, currently the same for all our simulations, which 
could have important consequences for the development of the magnetic 
kink instability and the recollimation structures.

\section*{Acknowledgments}

We thank Omer Bromberg and Tuomas Savolainen for useful discussions, 
and Alan Marscher for comments on the manuscript. 
RBD dedicates this work to Jessa and Lucas Barniol. We acknowledge support
from NASA through grant NNX16AB32G issued through the Astrophysics
Theory Program.  This work used the Extreme Science and Engineering
Discovery Environment (XSEDE) computational time allocation
TG-AST100040 on NICS Darter, TACC Stampede and Ranch. XSEDE is
supported by National Science Foundation grant number ACI-1053575.
Resources supporting this work were also provided by the NASA High-End
Computing (HEC) Program through the NASA Advanced Supercomputing (NAS)
Division at Ames Research Center (for the simulations of the models
without the ambient medium, models A and B).  This research was supported in
part through computational resources provided by Information
Technology at Purdue, West Lafayette, Indiana, and also by computing
time granted by UCB on the Savio cluster. 
Support for AT was provided by the TAC fellowship and NASA through Einstein Postdoctoral
Fellowship grant number PF3-140131 awarded by the Chandra X-ray
Center, which is operated by the Smithsonian Astrophysical Observatory
for NASA under contract NAS8-03060.

\bibliographystyle{mn2e}
\bibliography{references}

\appendix

\section{Numerical convergence for 2D models of jets}
\label{sec:2d-models-headed}

Whereas in the main text we focused on the numerical convergence of 3D
models, in this Appendix we present the results of numerical
convergence for the 2D models. This allows us to compare the levels of
dissipation between 2D and 3D simulations and thereby to determine the
dissipation regions that likely represent physically-motivated
dissipation sites. 

\begin{figure*}
\begin{tabular}{ccc}
\includegraphics[width=5.5cm, angle=0]{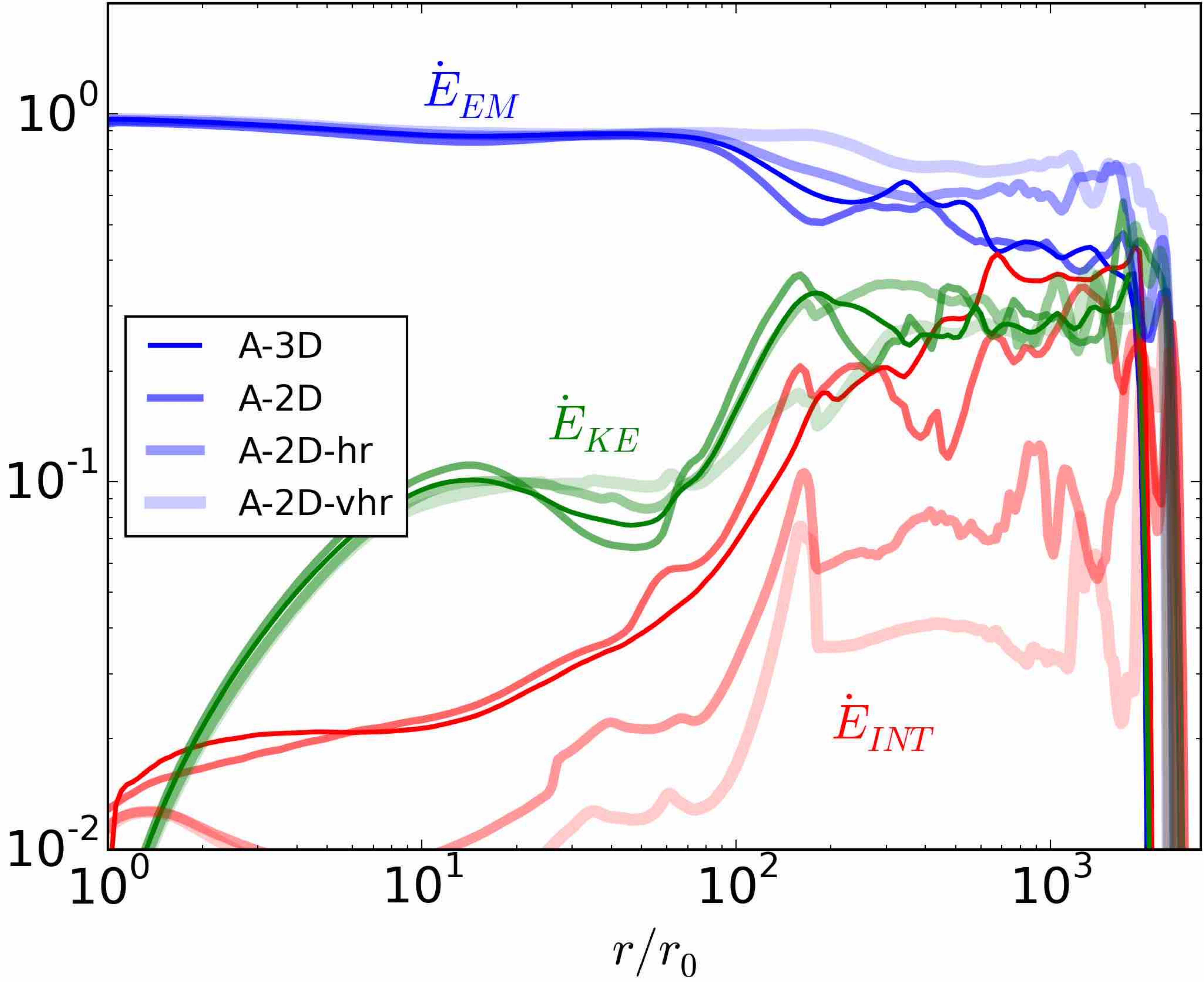} &
\includegraphics[width=5.5cm, angle=0]{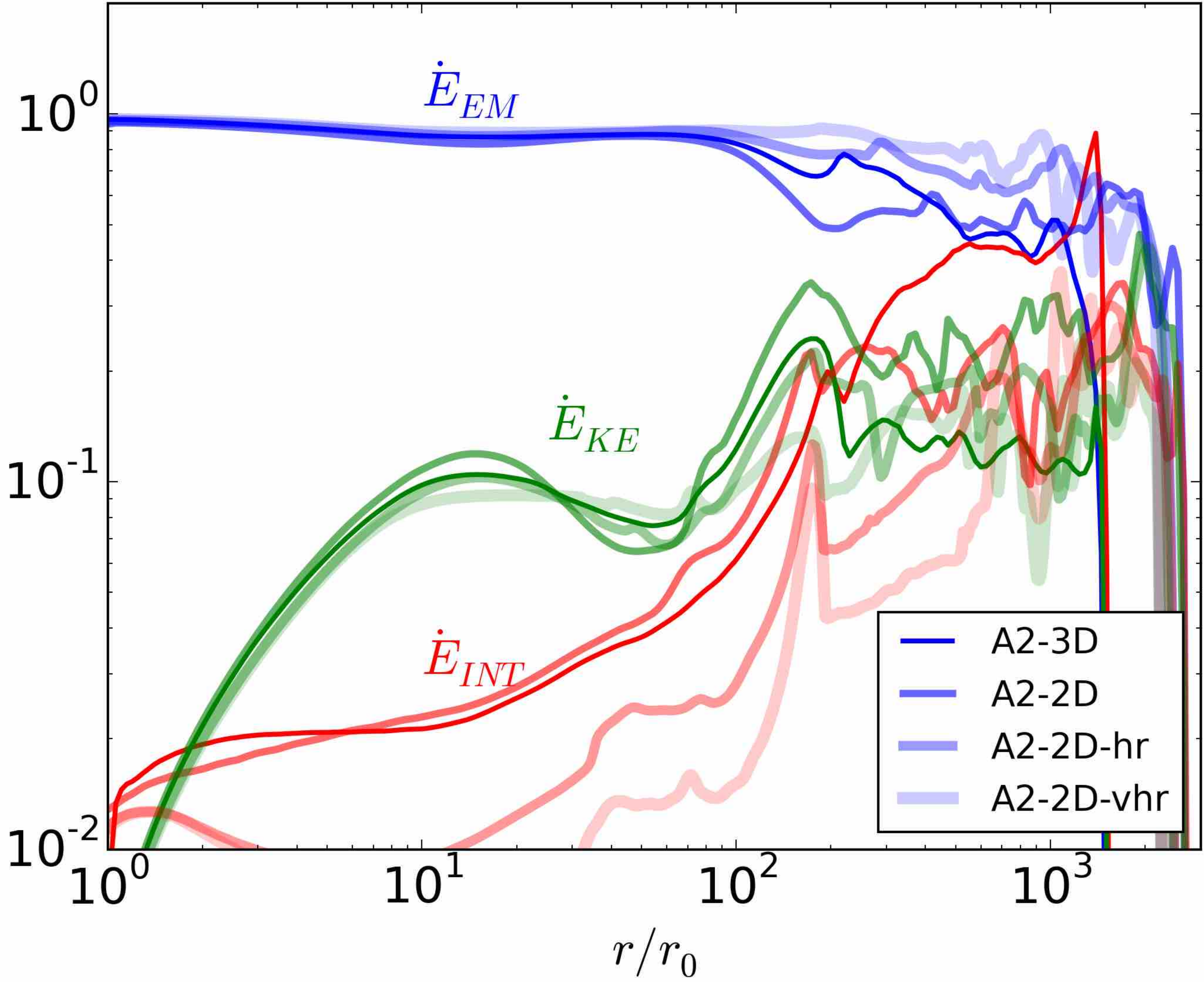} &
\includegraphics[width=5.5cm, angle=0]{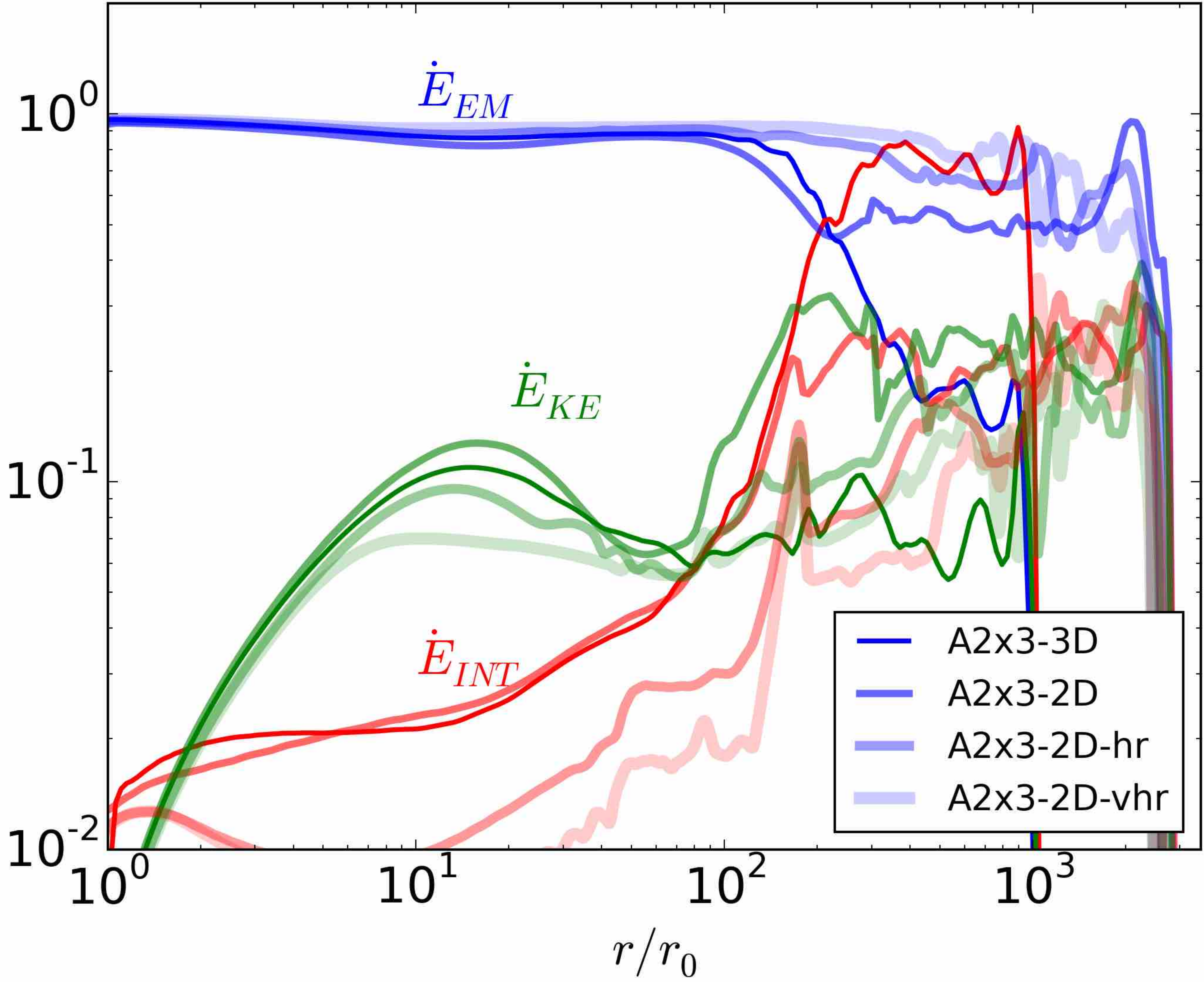} \\
\includegraphics[width=5.5cm, angle=0]{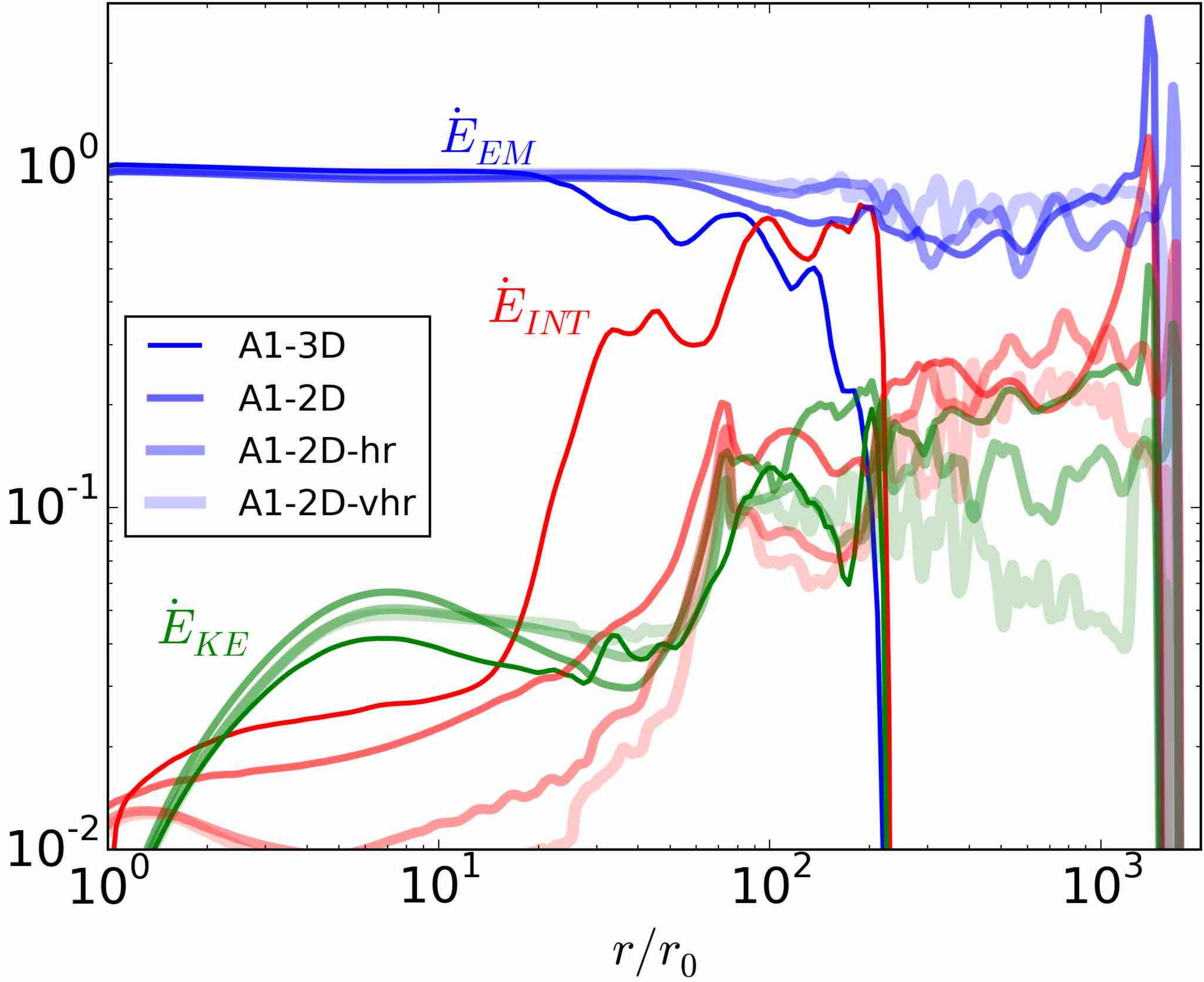} &
\includegraphics[width=5.5cm, angle=0]{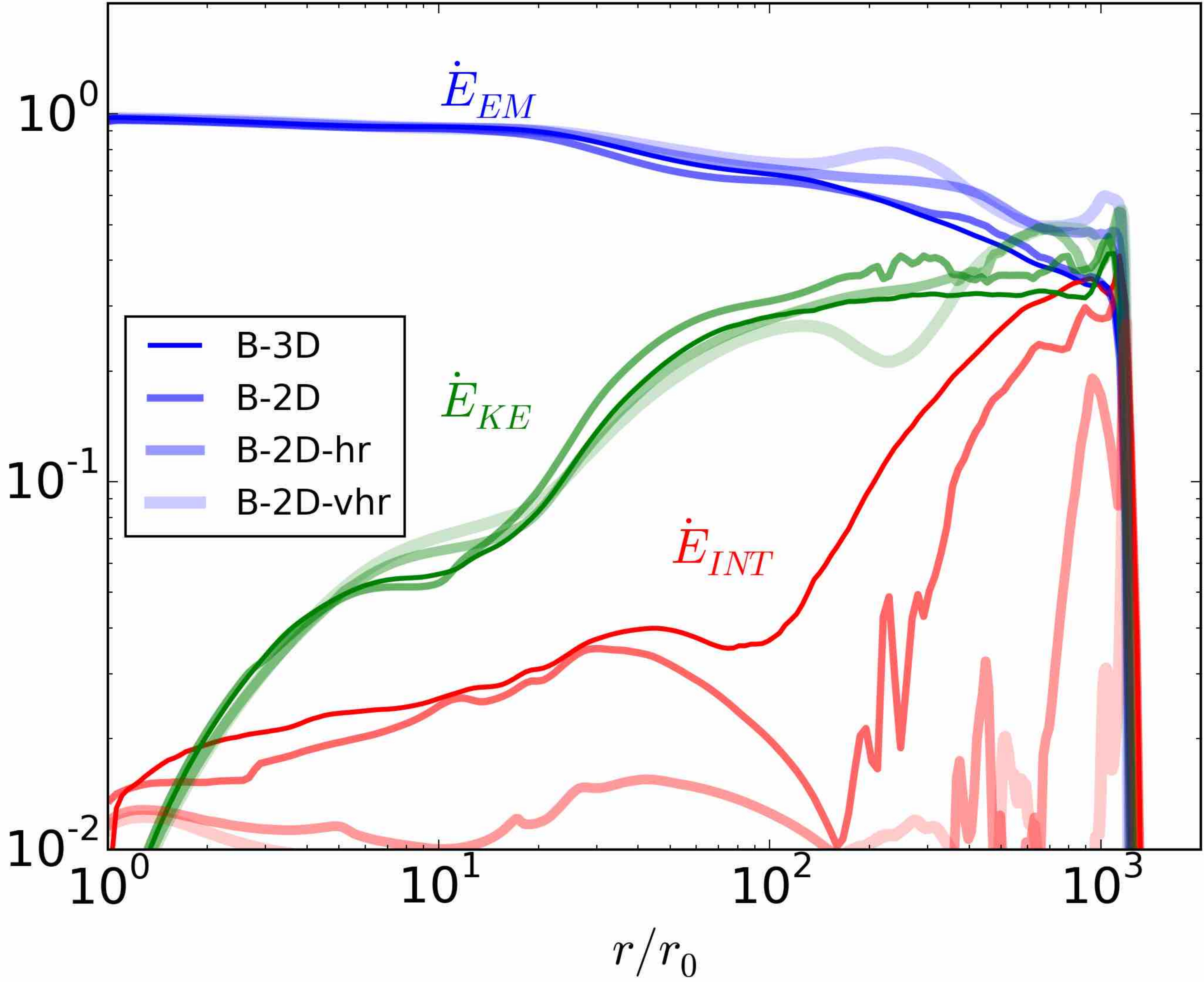} &
\includegraphics[width=5.5cm, angle=0]{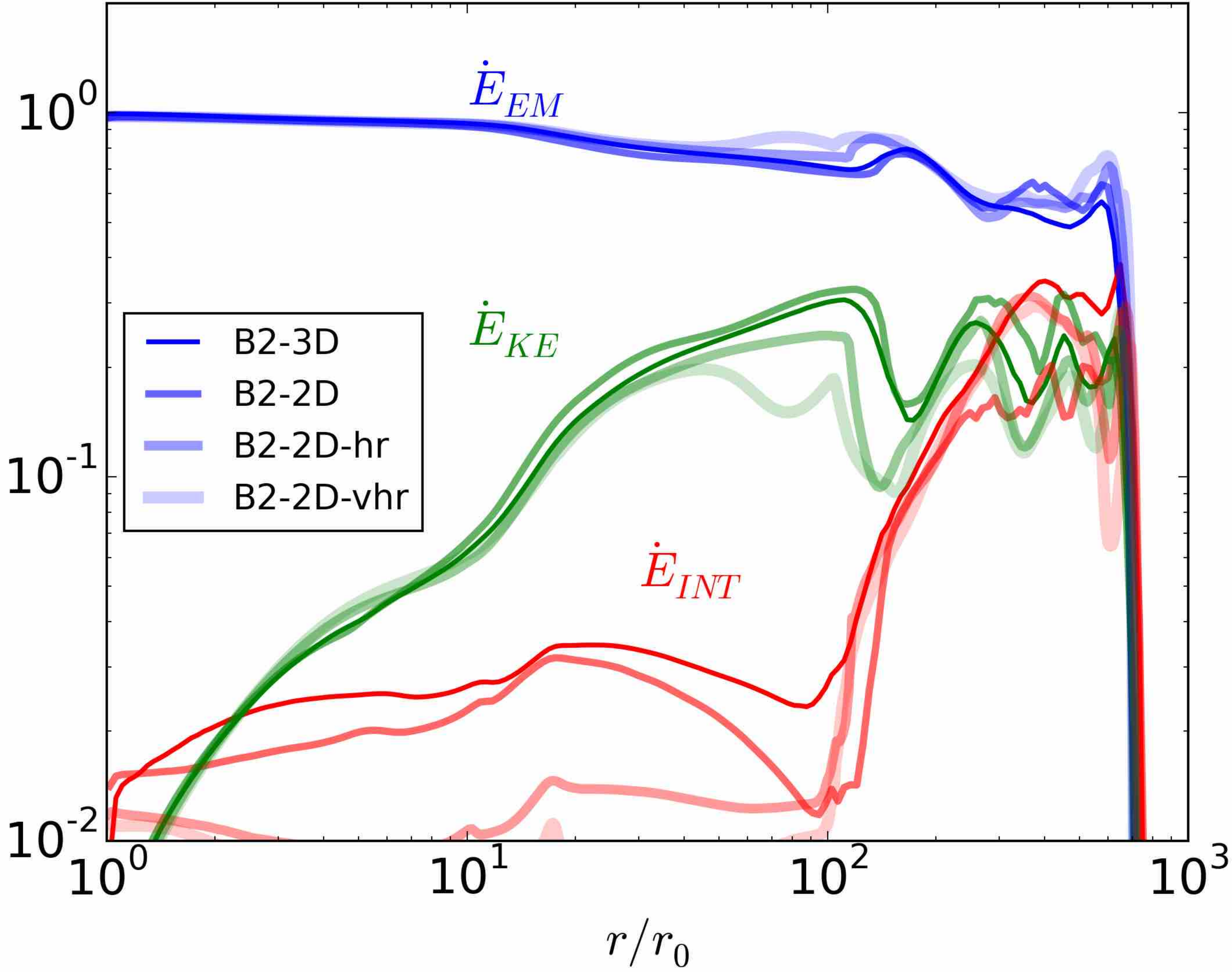} \\
\end{tabular}
\caption{Different components of the energy flux (see labels) 
for 3D and 2D simulations of model A (top-left, $t \approx 2700 r_0/c$), 
A2 (top-middle, $t \approx 3000 r_0/c$), A2x3 (top-right , $t \approx 3400 r_0/c$), 
A1 (bottom-left, $t \approx 2800 r_0/c$), B (bottom-middle, $t \approx 1200 r_0/c$) 
and B2 (bottom-right, $t \approx 700 r_0/c$). All quantities 
are calculated at the final snapshots of the simulations, except for models B 
and B2, where we have used the final snapshot of the 3D-hr simulations, respectively
(see Table \ref{table3}).
The energy fluxes have been normalized to the value 
of the total energy flux at $r=r_0$. The distance at which the energy flux 
drops to zero marks the location of the jet head. We ascribe the energy dissipation 
and internal energy flux increase seen in models A-3D and B-3D due to 
insufficient numerical resolution. However, the internal energy flux increase 
seen in models A2, A2x3 and A1 exceeds the level seen in 2D runs. Model B2-3D
shows a nice convergence with 2D runs. For 3D convergence studies see Figs. 
\ref{Edot_A_first} and \ref{Edot_B}.}   
\label{Edot_final_total}
\end{figure*}

In Fig.~\ref{Edot_final_total} we present the different energy fluxes as 
described in Section \ref{Dissipation} for all our models.  Instead of presenting 
only 3D and 2D simulations as we did in Figs. \ref{Edot_A_first}, 
\ref{Edot_A_second} and \ref{Edot_B}, in Fig.~\ref{Edot_final_total}
we also present the results from our higher-resolution 2D simulations, 
2D-hr and 2D-vhr.  

As can be seen in Fig.~\ref{Edot_final_total}, in model A (top-left), as we increase 
the 2D resolution, the internal energy flux decreases. Since in the A-3D 
simulation the internal energy flux is at the level of the A-2D simulation, we ascribe 
the energy dissipation seen in the A-3D simulation to insufficient resolution.  
For models A2 (top-middle), A2x3 (top-right) and A1 (bottom-left), the internal energy flux  
decreases as we increase the 2D resolution. However, the level 
of internal energy flux in the 3D runs of these simulations exceeds that of 
all 2D runs at different resolutions. This suggests that in these 3D models 
the dissipation is physical. In fact, 
a 3D convergence study, shown in the right panel of Fig.~\ref{Edot_A_first}, demonstrates 
that at a higher-resolution 3D simulation A2-3D-hr
shows an even larger internal energy flux than the fiducial resolution
simulation A2-3D. This indicates that the dissipation is
robust and consistent with being due to the 3D magnetic kink instability. 

In addition to energy dissipation in the current sheets produced by
the internal kink instability, we identify another potential
dissipation site: one or several shocks through which our super-fast
magnetosonic jets learn about the presence of the ambient medium. They
have to do so through shocks because by the time our jets reach the
ambient medium, they move at a super-fast magnetosonic velocity and
therefore do not know anything about the impending collision. The
shocks are accompanied by a spike in the internal energy. They are
easiest to identify in the highest resolution simulations, in which
the value of the internal energy is the most accurate: at $r\sim 100r_0$ in
the right-most panel in Fig.~\ref{Edot_A_first}, at $r\sim150 r_0$ in
simulations A-2D-hr/A-2D-vhr, A2-2D-hr/A2-2D-vhr, and
A2x3-2D-hr/A2x3-2D-vhr in the top row of panels in
Fig.~\ref{Edot_final_total}. As the resolution increases, the fraction
of energy carried in the form of heat (or internal energy) decreases
in the post-shock region, becoming as low as $\lesssim 5$\% at the highest
resolutions.  This demonstrates that the dissipation by these shocks is
sub-dominant to the dissipation by the internal kink instability,
which converts as much as $50$\% of the total energy flux into
heat. In future work, we will investigate how the fraction of the total
jet power dissipated at these shocks depends on the jet magnetization
and on the dynamical range accessible to the jets in the collimation
and acceleration region:
it is plausible that if the jets had a chance to accelerate to higher
values of the Lorentz factor, they would have reached the shock at a
lower value of magnetization and would be more readily able to heat up 
\citep[e.g.,][]{2010MNRAS.407.2501M}.

For model B in the bottom-middle panel of Fig.~\ref{Edot_final_total}, it is 
clear that increasing the 2D resolution essentially reduces the internal 
energy flux to zero. Thus, the internal energy flux increase seen in B-3D 
is due to insufficient resolution. Also, a 3D convergence 
study showed a significant decrease in the internal energy flux 
for the simulation B-3D-hr (top-right panel of Fig.~\ref{Edot_B}). Finally, as 
seen in the bottom-right panel of Fig.~\ref{Edot_final_total}, 
model B2 shows almost identical levels of energy fluxes for 
different 2D resolution runs, indicating that all our simulations 
show convergence. We found the same conclusion with our 
3D convergence study for this model in the bottom-right panel 
of Fig.~\ref{Edot_B}.

\section{2D simulations of models B and B2}
\label{sec:2d-simulations-model}

\begin{figure*}
\includegraphics[width=13.5cm, angle=0]{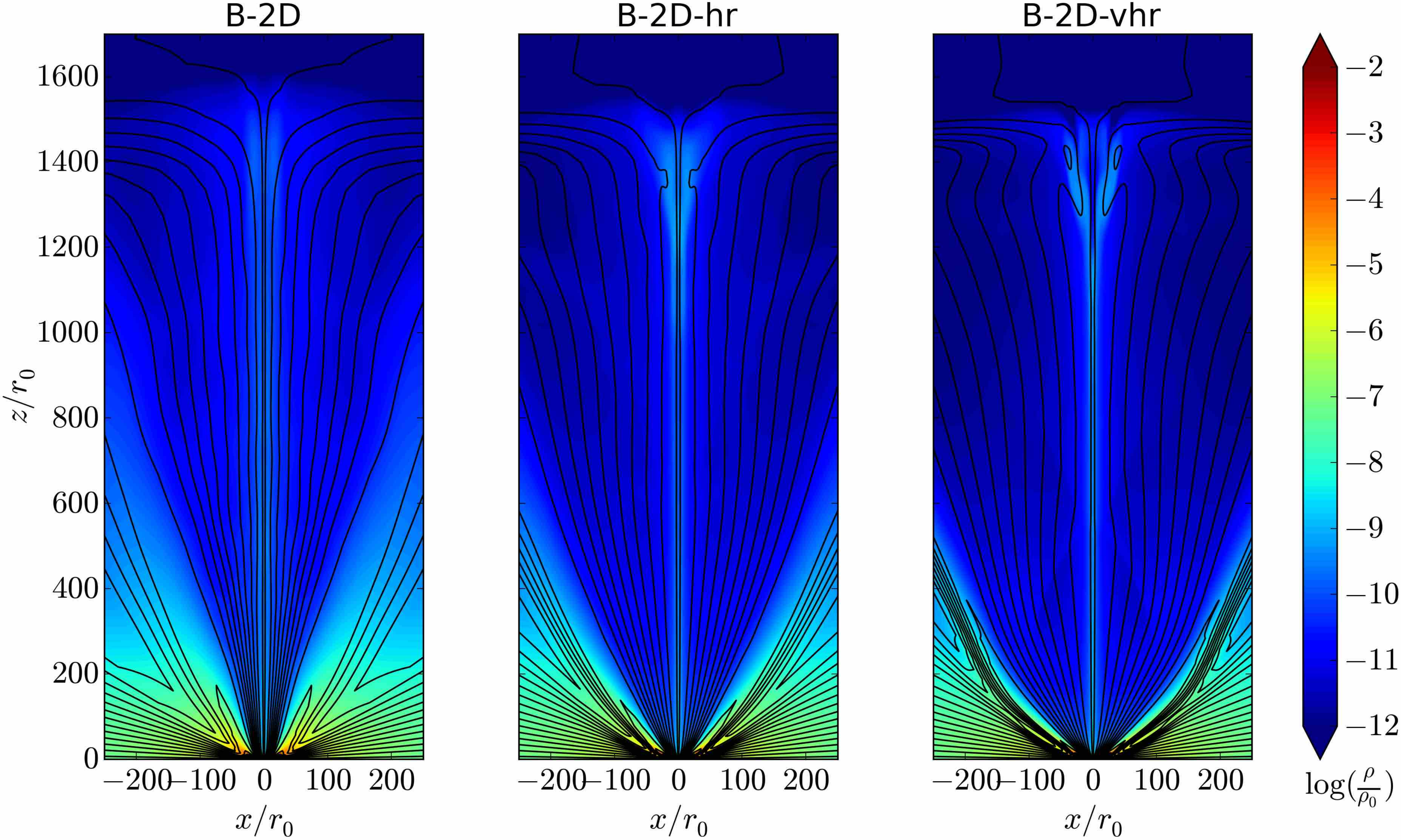} 
\caption{Density contours of 2D simulations of model B: B-2D (left), B-2D-hr (middle) and B2-vhr (right panel)
at the final snapshot ($t \approx 1500 r_0/c$; same scales 
and colorbar in all panels). Black lines show magnetic flux surfaces. 
The jet in model B propagates freely in a pre-existing funnel and it expands as it travels.
Different 2D resolutions show similar jet propagation velocities and similar jet shapes, 
although as we increase 2D resolution the jet becomes wider due to the increase 
of the $N_{\theta}$-resolution and our enhanced ability to resolve the interface
between the jet and the pre-existing funnel.}      
\label{Density_LF_B} 
\end{figure*}

\begin{figure*}
\includegraphics[width=13.5cm, angle=0]{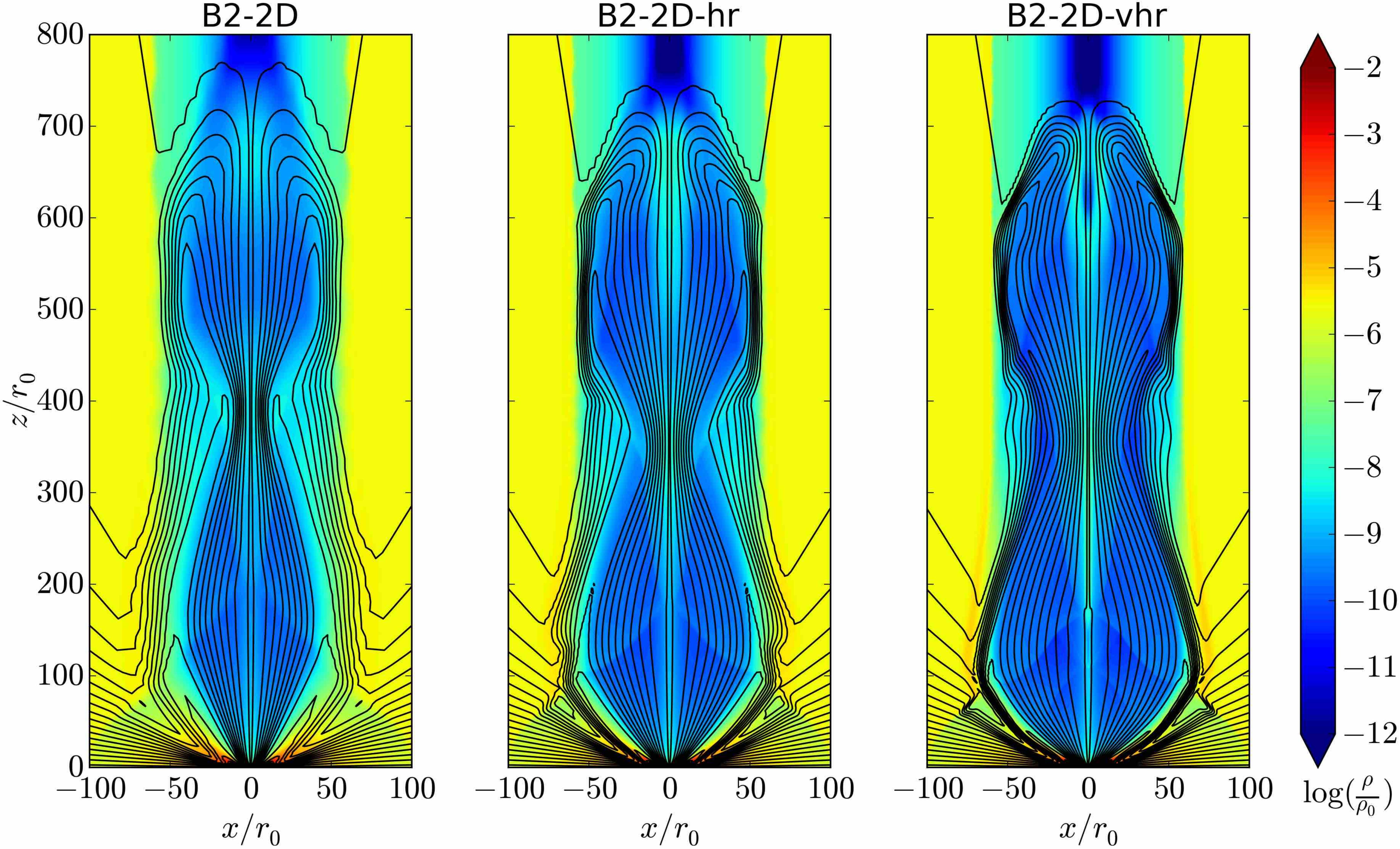} 
\includegraphics[width=13.5cm, angle=0]{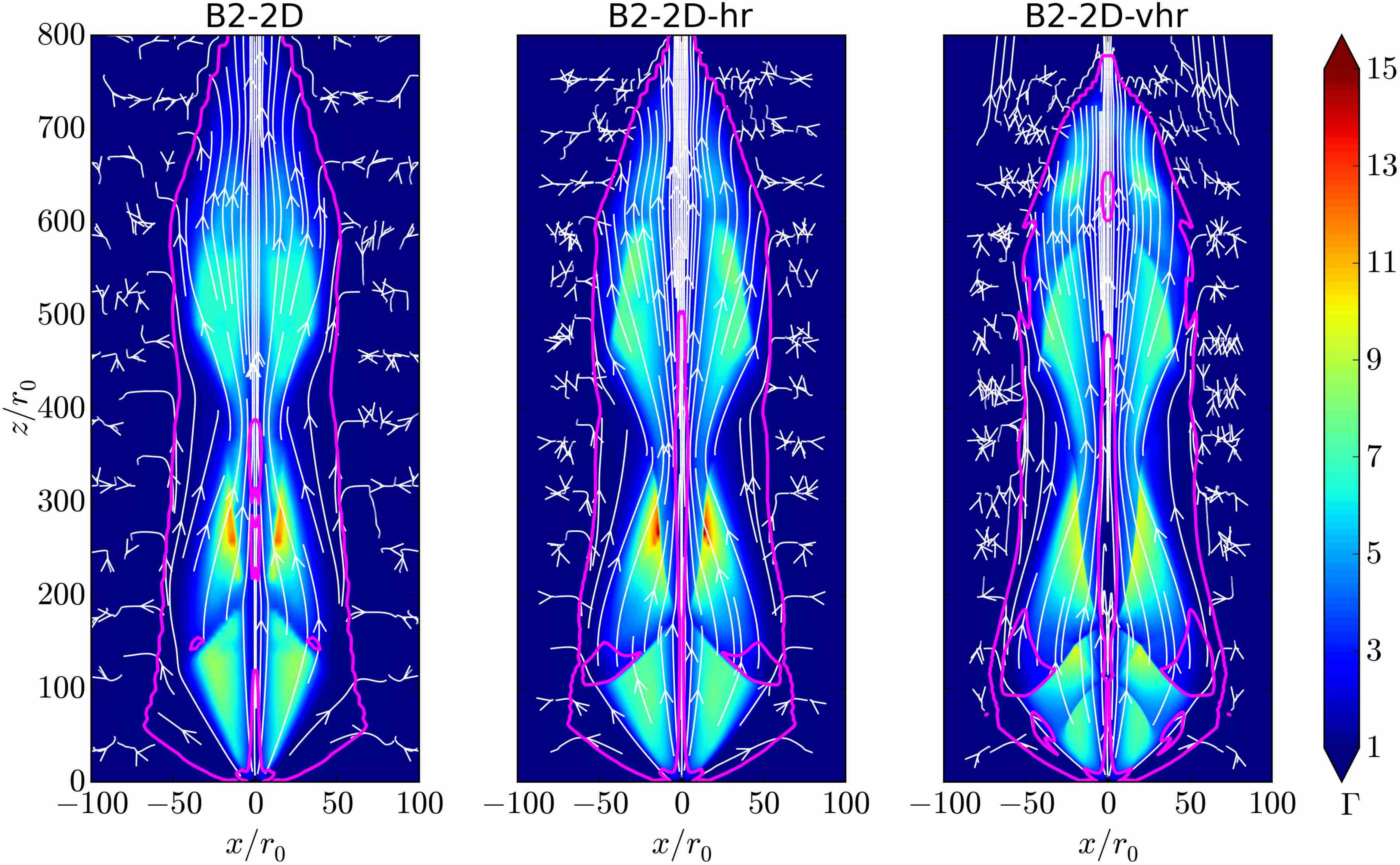} 
\caption{Density contours (top panels) and Lorentz factor contours (bottom panels) 
of model B2: B2-2D (left), B2-2D-hr (middle) and B2-vhr (right panel)
at the same time ($t \approx 700 r_0/c$; same scales in all panels, 
same density and Lorentz factor colorbars for the top and bottom panels, respectively). 
Black lines show magnetic flux surfaces. 
The jet in model B2 propagates in a pre-existing funnel that becomes cylindrical
at $z = 80 r_0$. The presence of shocks is evident by inspecting the Lorentz 
factor plots.  2D simulations show multiple shocks, which become better resolved at higher resolution.
Different 2D resolutions show similar jet propagation velocities and similar jet shapes, 
although as we increase 2D resolution the jet becomes wider due to the increase 
of the $N_{\theta}$-resolution and our enhanced ability to resolve the interface
between the jet and the pre-existing funnel.}      
\label{Density_LF_B2_appendix} 
\end{figure*}

We show 2D runs at different resolutions of model B and B2, 
see Figs. \ref{Density_LF_B} and \ref{Density_LF_B2_appendix}, respectively.
The 2D jets propagate at similar velocities at different 2D resolutions and their 
shapes appear similar, although as we increase the $N_{\theta}$-resolution 
we are able to better resolve the interface between the jet and the pre-existing 
funnel and thus the jet appears a bit wider. The 2D jets appear also very 
similar to the 3D jets of these models seen in Figs. \ref{Density_B} and \ref{Density_LF_B2}.
As also seen in our 3D convergence study (Fig. \ref{Density_LF_B2}), 
when we increase the 2D resolution of our model B2 runs, we are able to better 
resolve multiple shocks.

\end{document}